\documentclass[fleqn,usenatbib]{mnras}

\usepackage{newtxtext,newtxmath}

\usepackage[T1]{fontenc}

\DeclareRobustCommand{\VAN}[3]{#2}
\let\VANthebibliography\thebibliography
\def\thebibliography{\DeclareRobustCommand{\VAN}[3]{##3}\VANthebibliography}

\usepackage{graphicx}	
\usepackage{amsmath}	

\newcommand{\orcid}[1]{\href{https://orcid.org/#1}{\includegraphics[width=10pt]{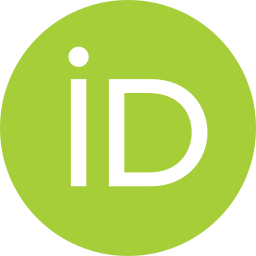}}}

\title[SILCC -- IX. The Low-Metallicity ISM]{SILCC -- IX. The multi-phase interstellar medium at low metallicity}

\author[V. Brugaletta et al.]{Vittoria Brugaletta$^{\orcid{0000-0003-1221-8771}}$$^{1, 2}$\thanks{E-mail: brugaletta@ph1.uni-koeln.de} , Stefanie Walch$^{1}$, Thorsten Naab$^{2}$, Tim-Eric Rathjen$^{1}$, Philipp Girichidis$^{3}$, \newauthor Daniel Seifried$^{1}$, Pierre Colin Nürnberger$^{1}$, Richard Wünsch$^{4}$, Simon C. O. Glover$^{3}$, Sanjit Pal$^{1}$ \newauthor and Lukas Wasmuth$^{1}$  
\\
$^{1}$ I. Physikalisches Institut, Universität zu Köln, Zülpicher Str. 77, 50937 Köln, Germany \\
$^{2}$ Max Planck Institute for Astrophysics, Karl-Schwarzschild-Str. 1, 85748 Garching, Germany\\
$^{3}$ Universit\"at Heidelberg, Zentrum f\"ur Astronomie, Institut f\"ur Theoretische Astrophysik, Albert-Ueberle-Str. 2, D-69120 Heidelberg, Germany\\
$^{4}$ Astronomical Institute of the Czech Academy of Sciences, Boční II 1401, 141 00 Prague, Czech Republic\\
}

\date{Accepted XXX. Received YYY; in original form ZZZ}

\pubyear{2025}

\begin{document}
\label{firstpage}
\pagerange{\pageref{firstpage}--\pageref{lastpage}}
\maketitle

\begin{abstract}
The gas-phase metallicity affects heating and cooling processes in the star-forming galactic interstellar medium (ISM) as well as ionising luminosities, wind strengths, and lifetimes of massive stars. To investigate its impact, we conduct magnetohydrodynamic simulations of the ISM using the FLASH code as part of the SILCC project. The simulations assume a gas surface density of 10 M$_\odot$ pc$^{-2}$ and span metallicities from 1/50 Z$_\odot$ to 1 Z$_\odot$.
We include non-equilibrium thermo-chemistry, a space- and time-variable far-UV background and cosmic ray ionisation rate, metal-dependent stellar tracks, the formation of HII regions, stellar winds, type II supernovae, and cosmic ray injection and transport. With the metallicity decreasing over the investigated range, the star formation rate decreases by more than a factor of ten, the mass fraction of cold gas decreases from 60\% to 2.3\%, while the volume filling fraction of the warm gas increases from 20\% to 80\%. Furthermore, the fraction of H$_\mathrm{2}$ in the densest regions drops by a factor of four, and the dense ISM fragments into approximately five times fewer structures at the lowest metallicity. Outflow mass loading factors remain largely unchanged, with values close to unity, except for a significant decline at the lowest metallicity. Including the major processes that regulate ISM properties, this study highlights the strong impact of gas phase metallicity on the star-forming ISM.  
\end{abstract}




\section{Introduction}

The ISM consists mainly of hydrogen and helium, with a small fraction of heavier elements (i.e., metals). This metal abundance, called metallicity, can play a very important role in shaping the ISM. Cooling in low-metallicity environments is inefficient, primarily due to the lack of major coolants like C$^+$, O, and dust \citep[see e.g.][]{Tielens2005, Drainebook2011}. Moreover, the chemical abundances found at low metallicity cannot simply be explained by scaling solar-metallicity abundances, as the involved chemical reactions for the creation and destruction of species change with metallicity \citep{Guadarrama2022}, and because of a different enrichment by stellar nucleosynthesis \citep{Bisbas2024}. Metal-poor environments exhibit lower dust-to-gas ratios \citep{Issa1990, Walter2007}, which allow dissociating UV radiation to enter deeper into molecular clouds, as dust shielding is reduced. This affects the star formation activity in dense clumps, possibly reducing the star formation efficiency at low metallicity \citep{Tanaka2018, Fukushima2020}, and impacting the structure of giant molecular clouds \citep{Maloney1988, Elmegreen1989, McKee1989, Rubio1993, Papadopoulos2002, Pelupessy2006, Bolatto2008, leroy2009, Hughes2010, Chevance2016, Jameson2018}. Low-metallicity environments also seem to experience a lower degree of fragmentation \citep{Omukai_2005, Omukai_2010, Walch_2011, Glover2012c, Moran2018, Bate2019, Whitworth2022}. Therefore, metallicity could have an effect on the galaxy-wide stellar initial mass function \citep{jerabkov2018, Sharda2023, Bate2025}.

Stellar feedback is an important source of energy and momentum to modulate star formation in the ISM, and this is affected in part by metallicity. For example, metal-poor stars are brighter and hotter than metal-rich ones with the same initial mass. Metal-poor massive stars have less wind mass loss (\citealt{maeder2000}), resulting in a slower chemical enrichment of the ISM and weaker stellar winds. In addition, for massive stars, metallicity can influence the ratio of blue to red supergiants (\citealt{langer1995}), and determine their final fate \citep[][]{Heger_2003}. All of these effects together will modify the impact of stellar feedback on the ISM. 

We can find metal-poor environments on several scales and cosmic distances from us. For instance, massive star-forming disc galaxies, including the Milky Way, show different metallicity gradients along the galactocentric radius \citep{Aller1942, Searle1971, Shaver1983, balser2011, Belfiore2017, mendez2022, piatti2023, Sextl2024}. At large radii, a negative metallicity gradient with increasing galactocentric radius is observed to be almost exponential \citep{Wyse1989, Zaritsky1992}. Therefore, disc galaxies host more low-metallicity gas in their outskirts. In the case of our Galaxy, \citet{Lian2023} report that the integrated stellar metallicity profile in the Milky Way shows a positive gradient between the centre and 6.9 kpc, with a slope of 0.031~$\pm$~0.010~dex~kpc$^{-1}$, and a negative gradient from 6.9 kpc to the edge, with a slope of -0.052~$\pm$~0.008~dex~kpc$^{-1}$. Less massive galaxies, however, can have flatter gradients \citep[see e.g.][]{Mingozzi2020, Ju2025}. Additionally, low mass galaxies are known to be metal-poor environments according to the mass-metallicity relationship \citep{Tremonti2004}. Metal-poor environments are also characteristic of high-redshift galaxies, where the chemical enrichment owing to the presence of a stellar population had less time to pollute the ISM with metals \citep{Savaglio2005}. With the launch of \textsc{JWST}, low mass galaxies have raised general interest, as we are now able to observe them with unprecedented resolution, allowing us to better investigate metal-poor environments \citep[e.g.][]{Cameron2023,Curti2023, Heintz2023, Tacchella2023, Doan2024}.

A complete theoretical study by means of simulations of the multiphase ISM in metal-poor environments, involving a detailed description of the stellar feedback (in the form of stellar winds, supernovae, radiation and cosmic rays), has not yet been performed. \citet{Kim2024} have run simulations similar to ours, for a range of metallicities, however they focus more on the validity of their dynamical equilibrium theory for star formation. We aim to fill this gap by providing state-of-the-art simulations of representative parts of the ISM of a galaxy with the same initial conditions except for the gas metallicity and the gas-to-dust mass ratio.

We discuss seven different initial metallicities representative of the solar neighbourhood, a randomly selected patch of the Milky Way (MW), the Large and Small Magellanic Clouds (LMC, SMC, respectively), the irregular dwarf galaxy Sextans A, the globular cluster NGC1904, and the blue compact dwarf galaxy I Zwicky 18 (I~Zw~18). In particular, metallicities comparable to those of the LMC and SMC environments are useful because of the richness of observational studies to which we can compare our results. The Magellanic Clouds have been the subject of numerous studies regarding their structure and the kinematics of the gas (e.g. \citealt{Marel2004, Fujii2014, Weiss2023, Grishunin2024} for the LMC, \citealt{Goldman2007, Jameson2018, Smart2019, Murray2019} for the SMC), their star formation history (e.g. \citealt{Baumgardt2012, Meschin2013} for LMC, \citealt{Sabbi2009, Rubele2009, Ramachandran2019} for SMC), and the episodes of star formation triggered by the interaction between the two \citep{Bekki2003}. 

On the other hand, the dwarf galaxy I Zwicky 18 \citep[][]{Zwicky1966} is one of the most metal-poor galaxies in our surroundings, with a distance of around 18~Mpc \citep{Aloisi2007, ContrerasRamos2011, Musella2012}. The structure of this galaxy has been analysed by \citet{Dufour1996}, who found a secondary partner associated with the main body of I Zwicky 18. \citet{ContrerasRamos2011} suggest that the very low metallicity of I Zwicky 18 could be explained by the accretion of primordial or very metal-poor gas, which diluted its ISM, or the removal of metals due to galactic winds. Moreover, the nature of its star formation history has been debated. \citealt{Garnett1997, Legrand2000, Legrand2001} assume a continuous and low star formation rate for a time equal to the Hubble time, to explain the observed abundances. On the other hand, \citealt{Aloisi1999} claim that there has also been starburst activity in addition to a continuous star formation phase. \citet{Bortolini2024} observe the presence of stars of all ages; however, they agree on the low star formation rate scenario for epochs older than 1~Gyr, which could be the origin of its low metal abundance.

The structure of our paper is as follows. In Section~\ref{sec:numerical_methods}, the numerical methods and the setup of the simulations are described. Our results are analysed in Sec.~\ref{sec:results}, and we discuss the main caveats of our work in Sec.~\ref{sec:discussion}. Finally, our summary and conclusions can be found in Sec.~\ref{sec:conclusions}.

\section{Numerical methods and simulation setup}
\label{sec:numerical_methods}
To describe the structure and evolution of the low-metallicity ISM, we perform magnetohydrodynamic (MHD) simulations with the adaptive mesh refinement code \textsc{Flash} version 4.6 (\citealt{Fryxell_2000, Dubey_2008,dubey2009}). The setup generally follows the stratified box simulations carried out in the SILCC Project (\citealt{silcc1, silcc2, silcc3, silcc4, silcc5, silcc6, silcc7, silcc8, Brugaletta2025}), which we briefly describe below. 

\subsection{Structure}
Our simulation domain is an elongated box with size 500~pc~$\times$~500~pc~$\times$~$\pm$ 4~kpc, which represents a small portion of a galaxy centered around the galactic midplane. The region is assumed to be far away from a central black hole. Due to the vertical elongation, we may study the development of outflows from the galactic disc. To better resolve the evolution of the gas near the midplane, where massive stars are forming, we fix our resolution $\Delta$$x$ to be uniform and equal to around 3.9~pc for $|z| <$~1~kpc, and up to 7.8~pc for $|z| >$~1~kpc. We adopt periodic boundary conditions in the x- and y-directions, while we allow for outflow along the z-direction. However, the gas exiting the box in the z-direction is not allowed to fall back in.

The total gravity in our setup consists of four different contributions: an external static dark matter potential following the NFW dark-matter profile from \citealt{NFW1996}, with $R_\mathrm{vir}$~=~200~kpc, distance from galactic centre $R$~=~8~kpc and concentration parameter $c$~=~12; an external static potential due to the presence of an old stellar population with stellar surface density $\Sigma_*$~=~30~M$_\odot$~pc$^{-2}$ and vertical scale height $H_*$~=~300~pc; the self-gravity of the gas computed on-the-fly using the tree-based algorithm from \citealt{wunsch2018}; and the gravity due to the presence of star cluster particles followed in the simulation \citep{Dinnbier2020}. 

At the beginning of each simulation the gas is distributed according to a Gaussian distribution centered on the midplane and with a scale height of 30~pc, assuming a gas surface density of 10~M$_\odot$~pc$^{-2}$. The Gaussian distribution is cut off at a height for which the disc density is equal to the uniform background density $\rho_{\rm b}$~=~10$^{-27}$~g~cm$^{-3}$.

We also employ a magnetic field oriented initially in the x-direction with a strength varying in the z-direction as
\begin{equation}
B_x(z) = B_{x,0} \sqrt{\rho(z)/\rho(z = 0)},     
\end{equation}
with $B_{x,0}$ = 6~$\mu$G and $\rho(z)$ the density at a height $z$ from the midplane.
To avoid sudden gravitational collapse at the beginning of the runs, we stir the gas for the first 20~Myr of evolution. This is achieved by injecting kinetic energy in the computational domain on the largest scale in the x- and y- directions, such that the mass-weighted root mean square of the velocity of the gas is 10 km~s$^{-1}$.

\subsection{MHD and cosmic rays}
The evolution of the gas is modelled by solving the ideal MHD equations. We also add cosmic rays protons, which can propagate in the ISM via anisotropic diffusion and advection \citep{GirichidisEtAl2016a,GirichidisEtAl2018a}. For each supernova event (see Sec.~\ref{sec:method_star_formation_feedback}), we inject 10\% of its energy (10$^{50}$~erg, \citealt{Ackermann2013}) in cosmic rays, which are treated as a separate relativistic fluid and are coupled to the MHD equations adding a pressure gradient and an additional energy source term $Q_\text{cr}$. The latter takes into account the injection energy due to supernovae and the cooling of cosmic rays via hadronic and adiabatic losses (\citealt{Pfrommer2017, Girichidis2020}). The modified MHD equations read:

\begin{equation}
    \frac{\partial \rho}{\partial t} + \nabla \cdot (\rho \boldsymbol{v}) = 0,
\end{equation}
\begin{equation}
    \frac{\partial \rho \boldsymbol{v}}{\partial t} + \nabla \cdot \Big( \rho \boldsymbol{v} \boldsymbol{v}^T - \frac{\boldsymbol{B} \boldsymbol{B}^T}{4 \pi} \Big) + \nabla P_\mathrm{tot} = \rho \boldsymbol{g} + \dot{\boldsymbol{q}}_\mathrm{sn},
 \end{equation}

\begin{align}
    \nonumber  \frac{\partial e}{\partial t} + & \nabla \cdot \Big[\ (e + P_\mathrm{tot})\boldsymbol{v} - \frac{\boldsymbol{B} (\boldsymbol{B} \cdot \boldsymbol{v})}{4\pi} \Big]\  \\ & = \rho \boldsymbol{v} \cdot \boldsymbol{g} + \nabla \cdot (\boldsymbol{\mathrm{K}} \nabla e_\mathrm{cr}) + \dot{u}_\mathrm{chem} + \dot{u}_\mathrm{sn} + Q_\mathrm{cr},
\end{align}

\begin{equation}
    \frac{\partial \boldsymbol{B}}{\partial t} - \nabla \times (\boldsymbol{v} \times \boldsymbol{B}) = 0,
\end{equation}
\begin{equation}
    \frac{\partial e_\mathrm{cr}}{\partial t} + \nabla \times (e_\mathrm{cr} \boldsymbol{v}) = - P_\mathrm{cr} \nabla \cdot \boldsymbol{v} + \nabla \cdot (\boldsymbol{\mathrm{K}} \nabla e_\mathrm{cr}) + Q_\mathrm{cr},
\end{equation}
with $\rho$ the density, $\boldsymbol{B}$ the magnetic field, $\boldsymbol{v}$ the velocity of the gas, $P_\text{tot} = P_\text{thermal} + P_\text{magnetic} + P_\text{cr}$ the total pressure, with $P_\text{cr}$ the pressure due to cosmic rays, $\boldsymbol{g}$ the gravitational acceleration, $e = \frac{\rho v^2}{2} + e_\text{thermal} + e_\text{cr} + \frac{B^2}{8 \pi}$ the total energy density, $\boldsymbol{\dot{q}}_\text{sn}$ the momentum input of unresolved SNe, $\boldsymbol{K}$ the cosmic ray diffusion tensor, $\dot{u}_\text{chem}$ the change in thermal energy due to heating and cooling processes, $\dot{u}_\text{sn}$ the thermal energy input from resolved supernovae, $Q_\text{cr} = Q_\text{cr, injection} + \Lambda_\text{hadronic}$. The latter term is assumed to be \citep{Pfrommer2017}

\begin{equation}
    \Lambda_\text{hadronic} = -7.44 \ \times \  10^{-16} \times \Big( \frac{n_e}{\text{cm}^{-3}} \Big) \times \Big( \frac{e_\text{cr}}{\text{erg cm}^{-3}} \Big) \text{ erg s}^{-1} \text{cm}^{-3}.
\end{equation}
In the diffusion tensor we adopt constant diffusion coefficients of $K_\parallel$~=~10$^{28}$~cm$^2$~s$^{-1}$ parallel to the magnetic field lines and $K_\perp$~=~10$^{26}$~cm$^2$~s$^{-1}$ perpendicular to them \citep{Strong2007, Nava2013}.

\subsection{Star formation and stellar feedback}
\label{sec:method_star_formation_feedback}
We model star formation in our simulations assuming that stars can form in star clusters, which are treated as collisionless sink particles (\citealt{bate1995, Federrath_2010, silcc3, Dinnbier2020}). To form a sink particle in a cell, the gas must be denser than a threshold density $\rho_\mathrm{sink}$~=~2~$\times$~10$^{-21}$~g~cm$^{-3}$, Jeans unstable, in a gravitational potential minimum, and found in a converging flow. Once a sink particle is created, for every 120~M$_\odot$ of gas accreted onto the sink, we form one massive star whose initial mass is randomly sampled in the interval 9--120~M$_\odot$ from a Salpeter-like IMF \citep[][]{salpeter1955}. The remaining mass is assumed to form low-mass stars inside the cluster, which we consider only for their gravitational effects and for their far-ultraviolet (FUV) radiation. We do not form isolated massive stars unless sink accretion is really inefficient. We follow the evolution of individual massive stars employing the BoOST massive star model tracks \citep[][]{brott2011, szecsi2020} or the Geneva models (\citealt{ekstrom2012}), as described in the following section. 

During their lifetime massive stars shape the surrounding ISM via stellar feedback in the form of stellar winds, their radiation, supernova explosions, and cosmic rays. At the end of their lifetime all massive stars explode as Type II supernovae, injecting either energy or momentum into the ISM depending on the ambient density of the supernova. If the radius at the end of the Sedov-Taylor phase can be resolved with at least three grid cells (corresponding to around 11.7~pc), thermal energy of 0.9~$\times$~10$^{51}$~erg is injected in a volume centred on the supernova and with a radius of three cells. Otherwise, the momentum at the end of the Sedov-Taylor phase is injected \citep[see][]{kim2015, gatto2015, Naab2017}{}{}. The remaining mass of the progenitor is added to the mass already present in the injection region.  

\subsection{Stellar models}
\label{sec:stellar_models}
We use Geneva stellar tracks \citep{ekstrom2012} for our solar-metallicity model as done in previous works of the SILCC collaboration \citep[e.g][]{silcc3, silcc6, silcc7, silcc8}. For all our subsolar-metallicity runs we employ the stellar models for single, slowly rotating massive stars from the BoOST project version 1.3 (\citealt{brott2011}, \citealt{szecsi2020}).

For FLASH, we provide input data consisting of the time evolution of the effective temperature, bolometric luminosity, mass loss rate, and wind terminal velocity for 112 models for each metallicity. These models cover an initial mass range from 9 to 120~M$_\odot$ with increments of 1~M$_\odot$.
 
The BoOST models describe the evolution of massive stars from the Zero-Age Main Sequence (ZAMS) until the core-helium depletion, whereas the Geneva models also include the core-carbon-burning phase, which does not affect considerably the duration of the lifetimes of the models. In Appendix~\ref{sec:app_stellarmodels} we analyse the parameters that mainly influence the stellar feedback in our simulations, such as the lifetime of stellar models, their wind and bolometric luminosities, and the fraction of ionising radiation for both Geneva and BoOST stellar models.

\subsection{Shielding and ionising radiation with \textsc{TreeRay}}

To treat the (self-)shielding of the gas by dust and molecules, as well as the radiation transport of extreme ultraviolet (EUV) radiation from massive stars, we employ the \textsc{TreeRay} module \citep{wunsch2018, wunsch2021}, which makes use of the octal tree already used by the gravity solver. \textsc{TreeRay} creates a \textsc{HEALPix} sphere \citep{gorski2005} with (in this work) 48 pixels around the center of each cell and for all cells in the computational domain. Rays are cast through the centre of each \textsc{HEALPix} pixel, and the radiation transport, as well as the shielding column densities, are computed along these directions.  

Using the \textsc{TreeRay/OpticalDepth} module, we compute the shielding column densities. \textsc{TreeRay/OpticalDepth} is based on the original \textsc{TreeCol} algorithm from \citet{Clark2012}. The local visual extinction $A_\mathrm{V, 3D}$ is computed as \citep{Bohlin1978}

\begin{equation}
    A_\mathrm{V, 3D} = \frac{N_\mathrm{H, tot}}{1.87 \times 10^{21} \ \mathrm{cm}^{-2}} \times Z,
\end{equation}
with $N_\mathrm{H, tot}$ the local 3D-averaged column density of the gas computed by \textsc{TreeRay/OpticalDepth}, and $Z$ is the metallicity in units of solar metallicity Z$_\odot$. 

To solve the radiation transport equation, all Lyman continuum photons ($E_\gamma$~$>$~13.6~eV) are treated with the module \textsc{TreeRay/OnTheSpot}, a backward-propagating radiation transport scheme that makes use of the On-The-Spot approximation \citep{Osterbrock_1988}. In this method, the propagation of radiation is almost independent of the number of sources and each cell can be a source of radiation. We compute the fraction of ionising photons assuming the spectrum of each massive star to be a black body with a given effective temperature and integrating the black body spectrum for energies higher than 13.6~eV. For each timestep, the energy density of ionising radiation emitted by all stars within a given sink particle is injected into the cell where the sink is located. Hence, star cluster sink particles are sources of radiation. The FUV radiation emitted from massive and low-mass stars is approximated with a simpler method than \textsc{TreeRay} (see below, Sec.~\ref{sec:meth_chemistry}).

\subsection{Chemistry, heating and cooling mechanisms}
\label{sec:meth_chemistry} 
Comprehending how metallicity influences the evolution of the ISM is heavily dependent on our treatment of the heating and cooling processes. Many of them depend on the chemical abundances of metals, which in turn are given by the metallicity. We compute these processes on-the-fly using a chemical network based on \citet{Nelson_1997} and \citet{glover2007a, Glover_2007b}, which calculates the non-equilibrium abundances of seven species: H, H$^{+}$, H$_2$, CO, C$^{+}$, O, and free electrons. We assume fixed, metallicity-dependent elemental abundances for C, O, and Si, for which we adopt the values of $x_\mathrm{C}$~=~1.4~$\times$~10$^{-4}$, $x_\mathrm{O}$~=~3.2~$\times$~10$^{-4}$, $x_\mathrm{Si}$~=~1.5~$\times$~10$^{-5}$ \citep{sembach2000} at solar metallicity. For sub-solar metallicities, we linearly scale these values with $Z$. This implies that the ratio of the gas-phase metals to the total amount of metals is metallicity-invariant, as we do not consider depletion onto dust.

We assume that H$_2$ can only form on the surface of dust grains \citep{Hollenbach1989}, neglecting gas phase formation through the H$^{-}$ and H$_2^{+}$ ions and the three-body channel \citep{Glover2003}, which becomes important at high gas densities \citep{Palla1983} and lower metallicities than those treated in this work \citep{Omukai_2005}. The dominant process that destroys H$_2$ is photodissociation by the interstellar radiation field (ISRF), although the model also accounts for collisional dissociation of H$_{2}$ in hot gas.

Regarding cooling processes, if the temperature of the gas is higher than 10$^4$~K we assume collisional ionisation equilibrium for helium and metals, and use the tabulated cooling rates from \citealt{Gnat2012}. Cooling from atomic hydrogen is always calculated using the non-equilibrium H and e$^{-}$ abundances provided by the chemical model. At lower temperatures, we also account for cooling from a number of other processes, including the fine structure lines of C$^{+}$ and O, and the rotational and vibrational lines of H$_2$, as described in more detail in \citet{Glover2012b}. Concerning heating processes, we include, among others, photoelectric heating (PE) by dust grains and heating due to low-energy cosmic rays.  

PE heating is variable in space and time and is modelled using the new \textsc{AdaptiveG0} module from \citet{silcc8}. We computed the intensity of FUV radiation from each star cluster by integrating the black body spectrum of each massive star and at each timestep in the range of 6 -- 13.6~eV, plus employing \textsc{Starburst99} single-stellar population synthesis models for all low-mass stars in the cluster sink. The total FUV radiation emitted by each cluster sink $i$ is called $G_\mathrm{cluster, i}$ and given in units of the Habing field \citep{habing1968}.
The local, unattenuated FUV field in each grid cell, $G_0$, is then obtained by summing up the contributions from all sources within a maximum distance of 50~pc from the cell
\begin{equation}
    G_0 = \sum_i G_\mathrm{cluster} \times R^{-2}_i,
\end{equation}
with $R_i$ the distance of the star cluster $i$ from the cell of interest and the inverse square law is applied. We then apply a background ISFR $G_\mathrm{bg}$~=~0.0948 to the computed $G_0$ as done in \citet{silcc8}.
The $G_0$ field is also attenuated by dust to obtain the effective ISRF $G_\mathrm{eff}$ for each cell:
\begin{equation}
    G_\mathrm{eff} = G_0 \times \mathrm{exp}(-2.5 A_\mathrm{V, 3D}).
\end{equation}
The PE heating rate is computed as \citep{Bakes1994, Bergin2004}, 
\begin{equation}
    \Gamma_\mathrm{pe} = 1.3 \times 10^{-24} \epsilon G_\mathrm{eff} n d \ [\mathrm{erg \ s}^{-1} \mathrm{cm}^{-3} ],
    \label{eq:gamma_pe}
\end{equation}
where
\begin{equation}
    \epsilon = \frac{0.049}{1 + (\psi/963)^{0.73}} + \frac{0.037 (T/10^4)^{0.7}}{1 + (\psi/2500)},
\end{equation}
with
\begin{equation}
    \psi = \frac{G_\mathrm{eff} T^{0.5}}{n_e},
\end{equation}
with $n$ the number density of hydrogen nuclei, $d$ the dust-to-gas mass ratio in per cent, with $d$~=~1 (1 per cent) in solar-neighbourhood conditions, $T$ the temperature of the gas, and $n_e$ the electron number density. This prescription of the PE heating is valid in solar-neighbourhood conditions, however, we extend it to metal-poor environments for simplicity. This approximation may break at the lowest metallicities that we treat in our work, as the abundance of polycyclic aromatic hydrocarbons (PAHs) seems to experience a superlinear drop with decreasing metallicity \citep{Draine2007, Sandstrom2012, Whitcomb2024}. Since PAHs contribute substantially to the total PE heating rate, the real PE heating rate could be smaller than the value obtained in Eq.~\ref{eq:gamma_pe}.   

 Regarding cosmic ray heating, we employ the new method developed by \citet{Brugaletta2025}, where the CR ionisation rate, $\zeta$, is computed scaling the local energy density of CRs, which is already computed by our CR-MHD solver. If the column density of the gas, $N_\mathrm{H, tot}$, is lower than a threshold value $N_\mathrm{H, thresh}$~=~10$^{20}$~cm$^{-2}$, we scale $\zeta$ linearly with the energy density of CRs, $e_\mathrm{cr}$, following
\begin{equation}
    \zeta = 3 \ \times 10^{-17} \ \Big( \frac{e_\mathrm{cr}}{1 \ \text{eV cm}^{-3}} \Big) \ \text{s}^{-1}.
    \label{eq:cr_linear}
\end{equation}
If $N_\mathrm{H, tot} > N_\mathrm{H, thresh}$, we multiply the linear scaling of Eq.~\ref{eq:cr_linear} by the attenuation factor 
\begin{equation}
    c_\mathrm{att} = (N_\mathrm{H, tot}/N_\mathrm{H, thresh})^{-0.423},
\end{equation}
where the exponent is given by \citet{Padovani2009}, who obtained it for the spectrum of protons and heavy nuclei, and considered the H$_2$ column density rather than the total column density. Since no equivalent is provided for the atomic gas column density, we apply the prescription from \citet{Padovani2009, Padovani2022} obtained for the H$_2$ column density using the H column density instead. Given the small amount of H$_2$ formed at the low metallicities of this work, we would underestimate the CR attenuation when only using the H$_2$ column density. With this choice, we suppose that our estimate of the CR attenuation is around a factor of 2 \citep{Glassgold1974} different compared to the attenuation shown in Fig. C1 in \citet{Padovani2022}.

The rate for CR heating is computed assuming that each ionisation deposits 20~eV as heat, following the prescription from \citet{goldsmith1978},
\begin{align}
   \nonumber \Gamma_\mathrm{cr} & =  \ 20 \ \zeta \ (n_\mathrm{H_2} + n_\mathrm{H}) \ \ \ \ \ [\text{eV s}^{-1}\text{cm}^{-3}]   \\ & = 3.2 \times 10^{-11} \zeta (n_\mathrm{H_2} + n_\mathrm{H}) \ \ \ \ \ [ \text{erg s}^{-1} \text{cm}^{-3}],
\label{eq:CRheating}
\end{align}
where $n_\mathrm{H_2}$ is the number density of H$_2$ and $n_\mathrm{H}$ is the number density of H.

\subsection{The low-metallicity ISM}
\label{sec:eq_curves}
\begin{figure}
	\includegraphics[width=\columnwidth]{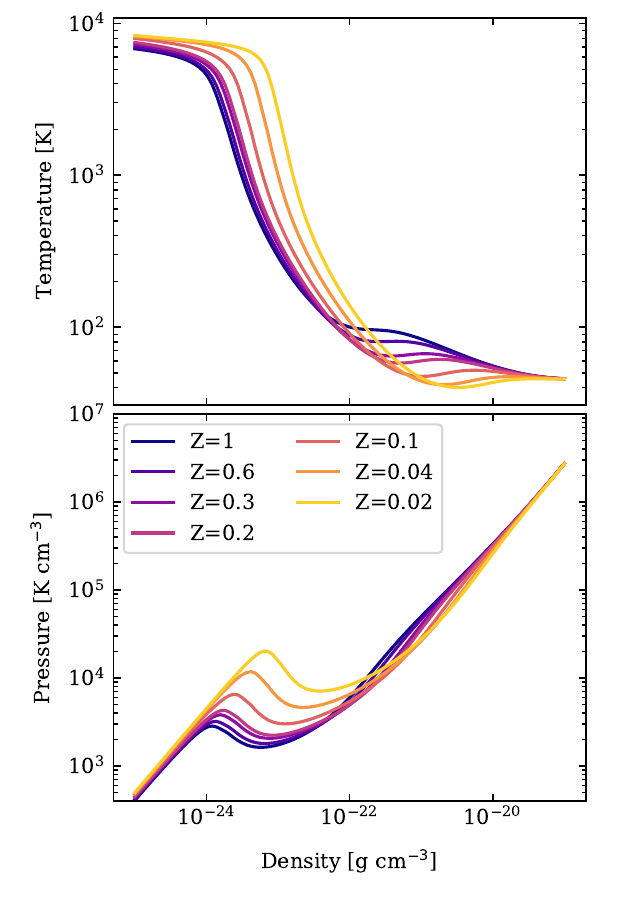}
    \caption{Equilibrium temperature (upper panel) and pressure (bottom panel) as a function of the gas density, for every metallicity analysed in this work. These equilibrium curves are computed using our chemical network without considering the self-shielding of H$_2$ and CO, assuming a constant external hydrogen column density $N_\text{H, tot}$~=~10$^{20}$~cm$^{-2}$, a constant $G_0$~=~1.7 and a constant $\zeta$~=~3~$\times$~10$^{-17}$~s$^{-1}$.}
    \label{fig:eq_curves}
\end{figure}

In this section, we briefly describe the thermal structure of the ISM at solar metallicity and discuss how it changes in metal-poor environments. For a more detailed treatment of all relevant heating and cooling processes, the reader can refer to \citet{Wolfire1995, glover2007a, Glover_2007b, Glover2014, BialySternberg2019}, and Sec.~\ref{sec:meth_chemistry}. Here we compute the equilibrium curves (see Fig.~\ref{fig:eq_curves}) corresponding to the conditions of the gas simulated in our runs using our chemical network for a single cell evolving for 1~Gyr, after which the gas has surely reached thermal and chemical equilibrium. For this test, we assume an optically thin gas with $A_\mathrm{V, 3D}$~=~0, G$_0$~=~1.7 and $\zeta$~=~3~$\times$~10$^{-17}$~s$^{-1}$ and neglect any self-shielding due to surrounding H$_2$ and CO. Note that all these parameters are locally variable in the 3D simulations. 

If we consider for now only the equilibrium curve at solar metallicity (dark violet line in Fig.~\ref{fig:eq_curves}), we notice that for a density lower than 10$^{-24}$~g~cm$^{-3}$ the temperature remains almost constant. In fact, for densities below 10$^{-24}$~g~cm$^{-3}$, the PE heating is balanced by the Ly$\alpha$ cooling, which is a strong function of the temperature. In this regime, the Ly$\alpha$ cooling rate slowly increases with increasing density and, correspondingly, the gas temperature slowly decreases to balance the almost constant heating rate. This is why the equilibrium temperature for densities below 10$^{-24}$~g~cm$^{-3}$ does not vary significantly. Note that this is different in the 3D simulations, where shock heating dominates this density regime \citep[see e.g.][]{Hu2016}.

For a threshold density of around 10$^{-24}$~g~cm$^{-3}$ the Ly$\alpha$ cooling becomes comparable to the metal (C$^+$, O) fine-structure line cooling, which dominates for higher densities \citep{Dalgarno1972}. In this regime, we observe a steep drop in temperature, which is due to the weak dependency of the C$^+$ and O fine structure cooling rates on temperature. In fact, when the density increases, the metal cooling rates increase as well; as a result,  heating and cooling balance at a significantly lower temperature \citep{Field1969}. The temperature drop occurs just after the local pressure peak shown in the bottom panel of Fig.~\ref{fig:eq_curves}. 

At a density of around 10$^{-22}$~g~cm$^{-3}$, corresponding to an equilibrium temperature of $\sim$~10$^2$~K, the curve slightly flattens again because metal line cooling becomes exponentially dependent on temperature. For higher densities, H$_2$ formation heating becomes important, increasing the equilibrium temperature. This effect is strongly seen here because we compute the equilibrium curves without H$_2$ self-shielding. In a real cloud, the gas would be almost fully molecular at these densities; therefore the H$_2$ formation heating could be less important than what we find here. For even higher densities, the conversion of atomic to molecular hydrogen suppresses the H$_2$ heating, as this is efficient in atomic gas \citep{BialySternberg2019}, causing the drop for densities higher than 10$^{-21}$~g~cm$^{-3}$. 

With decreasing Z, the equilibrium curves are shifted to higher temperatures at the same density, or analogously they shift towards higher pressures. In our approach, the dust-to-gas ratio scales linearly with the metallicity, meaning that the PE heating rate and the metal cooling rate scale in the same way. However, as the importance of PE heating decreases for metal-poor gas, CR heating becomes increasingly important because it does not depend on metallicity (\citealt{KimJG2023}, \citealt{Brugaletta2025}). Therefore, the cooling and heating rates are balanced at a higher temperature, for constant $\zeta$, than in solar-neighbourhood conditions. 

\begin{figure*}
	\includegraphics[width=0.9\textwidth]{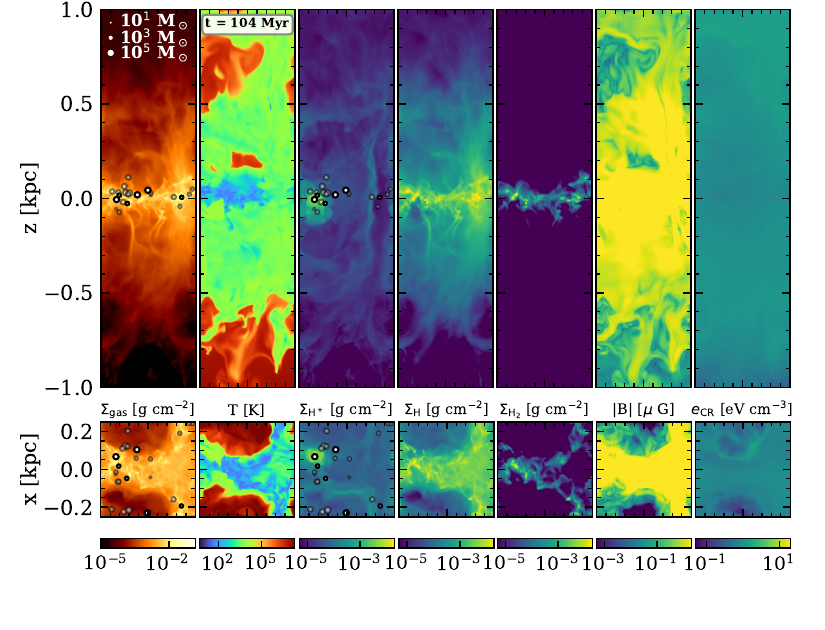}
    \caption{Snapshot of the $\Sigma$010-Z0.2 run at $t$~=~104~Myr. The top row represents an edge–on view of the simulation box, the bottom squares show the corresponding face–on views. From left to right we depict the total gas column density $\Sigma_{\text{gas}}$ (projection), the temperature $T$ (slice), the column densities of H$^{+}$, H, and H$_2$ (projections), the magnetic field strength $B$ (slice), and the energy density of CRs $e_{\text{CR}}$ (slice). The top elongated panels show only a part of the box (1~kpc around the midplane instead of 4~kpc). All slices are taken at $y$~=~0 (upper panels), and $z$~=~0 (bottom panels). The projections are computed along the y-axis (upper panels) or along the z-axis (bottom panels), respectively. The white circles in the first and third panel represent the active star clusters, whereas the transparent circles represent star clusters that are no longer active. The stellar feedback due to the presence of star clusters shapes the structure and governs the evolution of the multiphase ISM.} 
    \label{fig:snapshot}
\end{figure*}

\subsection{Simulation parameters}
\label{sec:sim_param}

\begin{table}
	\centering
	\caption{List of the performed simulations. The main parameter is the initial gas metallicity, $Z_\mathrm{gas}$, according to which we rescale the abundances of C, O, and Si relative to hydrogen and the dust-to-gas ratio as seen in Sec.~\ref{sec:meth_chemistry}. The metallicity of the stars in the stellar tracks is $Z_\mathrm{stars}$, and it has been obtained from Table~1 in \citet{szecsi2020} and dividing by 0.014 \citep{Asplund2009}. Note that we presented the $\Sigma$010-Z0.02 run in \citealt{Brugaletta2025} in greater detail.}
	\label{tab:sim_parameters}
	\begin{tabular}{lcccc} 
		\hline
		Run name & $Z_\mathrm{gas}$ & $Z_\mathrm{stars}$ &  Object with  & Stellar \\
		& [Z$_\odot$] & [Z$_\odot$] &  comparable $Z$ &  models \\
		\hline
		$\Sigma$010-Z1 & 1 & 1 & Solar neighbourhood & Geneva\\
		$\Sigma$010-Z0.6 & 0.63 & 0.63 &  Milky Way & BoOST\\
		$\Sigma$010-Z0.3 & 0.34 & 0.34 & Large Magellanic Cloud & BoOST\\
		$\Sigma$010-Z0.2 & 0.24 & 0.15 & Small Magellanic Cloud & BoOST\\
		$\Sigma$010-Z0.1 & 0.10 & 0.075 &  Sextans A or NGC 362 & BoOST\\
        $\Sigma$010-Z0.04 & 0.04 & 0.03 &  NGC 1904 & BoOST\\
        $\Sigma$010-Z0.02 & 0.02 & 0.015 & I Zwicky 18 & BoOST\\
		\hline
	\end{tabular}
\end{table}

 We run seven different simulations with $Z$ between 0.02~--~1~Z$_\odot$ (see Table~\ref{tab:sim_parameters}). The chosen values for the gas metallicity in our runs follow the availability of stellar models with the same initial metallicity (see Sec.~\ref{sec:stellar_models}). Since no stellar models with solar metallicity are provided within BoOST, we run the solar-metallicity run, i.e. run $\Sigma$010-Z1, employing the Geneva tracks from \citet{ekstrom2012}. We use almost identical initial conditions, such as a constant gas surface density of 10~M$_\odot$~pc$^{-2}$, except for those parameters affected by metallicity. We change the initial metallicity of the gas by linearly scaling the carbon, oxygen and silicon abundances as well as the dust-to-gas mass ratio (see Sec.~\ref{sec:meth_chemistry}). 
 
 Overall, the names of the runs listed in the first column of Table~\ref{tab:sim_parameters}, indicate the simulated metallicity in units of solar metallicity. We note that run $\Sigma$010-Z0.02 has been presented in \citealt{Brugaletta2025}, where it was called run Z0.02-vG$_0$-v$\zeta$. For simplicity and because we use a variable FUV and cosmic ray ionisation rate in all runs presented here, we shorten the name of this simulation.

\section{Results}
\label{sec:results}

\subsection{Qualitative evolution}
\label{sec:global_evolution}
 In the beginning of our simulations, we stir turbulence in the gas to avoid a fast collapse towards the midplane that would generate an unphysical starburst. As described in Sec.~\ref{sec:numerical_methods}, the initial stirring is applied for the first 20~Myr and influences the local conditions for the formation of the first stars. As overdense regions are formed, stars form in accreting star cluster sink particles. 

As soon as massive stars are born, they start shaping the ISM via their stellar winds and ionising radiation that leads to the formation of expanding H~II regions. At the end of their life, all massive stars are assumed to explode as Type II supernovae, injecting energy or momentum and CRs into the ISM. The combined explosion of several supernovae distributed in different star clusters residing near the midplane dramatically shapes the ISM creating superbubbles and launching outflows from the disc. Galactic fountains and outflows remove gas mass available for new star formation near the disc midplane, slowing down the star formation rate. When the majority of the formed stars have exploded and new star formation proceeds at a lower rate, the outward force exerted on the gas by stellar feedback weakens, until it stops pushing the gas in the vertical direction. Subsequently, the gas that is not unbound can fall back onto the midplane because of the overall gravity of the disc and reforms a reservoir for future star formation. This cycle continues until one eventually runs out of gas (far beyond the simulated time). A snapshot of run $\Sigma$010-Z0.2 at a time of 104~Myr can be seen in Fig.~\ref{fig:snapshot}.

\subsection{Gas phases}
\label{sec:gas_phase}
\begin{figure}
	\includegraphics[width=\columnwidth]{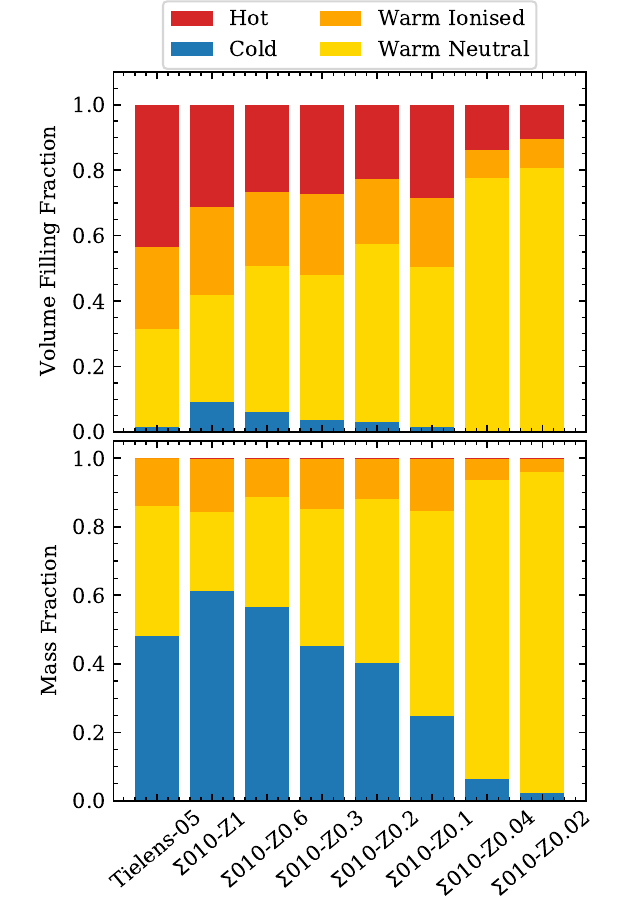}
    \caption{Volume-weighted average filling fraction (top) and average mass fraction (bottom) for the different gas phases defined in the text. The averages are computed in the 200~Myr period after the beginning of star formation, in the region where $|z|<$~250~pc. We note that the warm gas occupies around 50\% or more of the volume, and constitutes a large fraction of the gas mass, especially at low metallicity. The hot gas volume filling fraction shows a correlation with metallicity, whereas the cold gas volume and mass fractions decrease for lower metallicity. We add the values measured by \citet{Tielens2005} in the Solar neighbourhood as a comparison.}
    \label{fig:vff}
\end{figure}

\begin{table*}
	\centering
	\caption{Average volume filling fractions ($\overline{\text{VFF}}$) and average mass fractions ($\overline{\text{MF}}$) for all gas phases. The values are computed in the time interval [$t_\text{SF}$, $t_\text{SF}$~+~200~Myr]. We add the values reported from \citealt{Drainebook2011} and \citealt{Tielens2005} for solar-neighbourhood conditions.}
	\label{tab:mean_vff}
	\begin{tabular}{llllllllr} 
		\hline
		Run & $\overline{\text{VFF}}_\text{cold}$ & $\overline{\text{VFF}}_\text{WNM}$ & $\overline{\text{VFF}}_\text{WIM}$ & $\overline{\text{VFF}}_\text{hot}$ & $\overline{\text{MF}}_\text{cold}$ & $\overline{\text{MF}}_\text{WNM}$ & $\overline{\text{MF}}_\text{WIM}$ & $\overline{\text{MF}}_\text{hot}$ \\
         & [\%] & [\%] & [\%] & [\%] & [\%] & [\%] & [\%] & $[~10^{-2}$~\%] \\
		\hline
        Draine & 1 & 40 & 10 & 50 & - & - & - & - \\
        Tielens-05 & 1.05 & 30 & 25 & $\sim$50 & 48 & 38 & 14 & - \\
		$\Sigma$010-Z1 & 9.3 $\pm$ 0.1   & 32.5 $\pm$ 0.4& 27.1 $\pm$ 0.4& 31.2 $\pm$ 0.4& 61.2 $\pm$ 0.3& 23.2 $\pm$ 0.2 & 15.6 $\pm$ 0.3 & 4.3 $\pm$ 0.1\\
		$\Sigma$010-Z0.6 & 6.1 $\pm$ 0.1 & 44.6 $\pm$ 0.5& 22.5 $\pm$ 0.3& 26.8 $\pm$ 0.4 & 56.6 $\pm$ 0.3& 32.1 $\pm$ 0.3 & 11.2 $\pm$ 0.2 & 5.4 $\pm$ 0.1\\
        $\Sigma$010-Z0.3 & 3.6 $\pm$ 0.1 & 44.4 $\pm$ 0.5& 24.7 $\pm$ 0.4& 27.3 $\pm$ 0.4& 45.1 $\pm$ 0.3& 40.0 $\pm$ 0.3 & 14.8 $\pm$ 0.3& 5.6 $\pm$ 0.1\\
		$\Sigma$010-Z0.2 & 3.2 $\pm$ 0.1 & 54.4 $\pm$ 0.7& 19.7 $\pm$ 0.3& 22.8 $\pm$ 0.5& 40.4 $\pm$ 0.3& 47.7 $\pm$ 0.3 & 11.9 $\pm$ 0.3 & 6.1 $\pm$ 0.2\\
        $\Sigma$010-Z0.1 & 1.4 $\pm$ 0.1 & 48.9 $\pm$ 0.7& 21.2 $\pm$ 0.5& 28.5 $\pm$ 0.6& 24.8 $\pm$ 0.4& 59.7 $\pm$ 0.5 &  15.5 $\pm$ 0.5 & 5.2 $\pm$ 0.1 \\
        $\Sigma$010-Z0.04& 0.3 $\pm$ 0.1 & 77.2 $\pm$ 0.8& 8.5 $\pm$ 0.4 & 13.9 $\pm$ 0.6& 6.5 $\pm$ 0.3& 87.1 $\pm$ 0.5  & 6.4 $\pm$ 0.4 &  2.2 $\pm$ 0.1\\
        $\Sigma$010-Z0.02 & 0.05$\pm$ 0.01  & 80.7 $\pm$ 0.4& 8.7 $\pm$ 0.2 & 10.6 $\pm$ 0.3& 2.3 $\pm$ 0.1 & 93.6 $\pm$ 0.2& 4.0 $\pm$ 0.2 & 0.87 $\pm$ 0.04\\
		\hline
	\end{tabular}
\end{table*}

We distinguish four different gas phases, following the definitions already adopted in \citealt{silcc1, silcc3, silcc6, silcc7, silcc8, Brugaletta2025} (here we omit the definition of molecular gas):
\begin{itemize}
\item T~$\le$~300~K : cold gas (CM);
\item  300~$<$~T~$<$~3~$\times$~10$^5$~K and H mass fraction above 50\%: warm neutral medium (WNM);
\item 300~$<$~T~$<$~3~$\times$~10$^5$~K and H$^+$ mass fraction above 50\%: warm ionised medium WIM;
\item T~$\ge$~3~$\times$~10$^5$ K: hot gas.
\end{itemize}
The presence of hot gas is mainly due to supernovae that (overlap to) heat the gas, whereas stellar winds, radiation, and CRs influence the amount of warm gas \citep{Naab2017, silcc6}. 

We compute the volume-weighted average volume filling fractions (VFF) and mass fractions (MF) of the three phases for each run, as shown in Fig.~\ref{fig:vff} and reported in Table~\ref{tab:mean_vff}. To reduce the impact of our initial conditions, we compute these averages in a time interval [$t_\text{SF}$, $t_\text{SF}$~+~200~Myr], with $t_\text{SF}$ being the time at which the first star is born (see Table~\ref{tab:sfr_feedback_features}), and inside the region $|z|<$~250~pc. For a comparison, we provide the VFFs and/or MFs of the gas phases as reported by \citet{Drainebook2011} and \citet{Tielens2005} for solar-neighbourhood conditions in Table~\ref{tab:mean_vff}, and we add the values from \citet{Tielens2005} in Fig.~\ref{fig:vff}.

The warm gas phase (neutral plus ionised) overall dominates the VFF, contributing more than 50\% of the total volume (top panel of Fig.~\ref{fig:vff}). The VFF of the WNM is similar for runs with $Z$~$\ge$~0.1~Z$_\odot$, but it reaches up to 70 -- 80\% for runs $\Sigma$010-Z0.04 and $\Sigma$010-Z0.02. Its MF increases with decreasing metallicity, going from 23\% at solar metallicity to 97\% at 0.02~Z$_\odot$. On the other hand, the VFF of the WIM is around 20 -- 27\% for the runs with $Z$~$\ge$~0.1~Z$_\odot$, but it is around 8\% in the $\Sigma$010-Z0.04 and $\Sigma$010-Z0.02 runs. A similar trend is seen for its MF, which is around 11 -- 15\% for $Z$~$\ge$~0.1~Z$_\odot$ and below 7\% for the two most metal-poor runs. This trend can be explained in relation the results of Sec.~\ref{sec:SF}. Metal-poor runs form fewer stars; therefore, the total amount of radiation emitted by star clusters is lower than in metal-rich runs, despite metal-poor stars having a higher surface temperature, leading to a lower VFF of the WIM. Therefore, even though the fraction of energy emitted in the EUV increases with decreasing metallicity, as shown in Fig.~\ref{fig:ion_en_frac}, the effect of having a reduced star formation rate dominates.

The hot phase occupies around 10 -- 30\% of the volume around the midplane, depending on the metallicity. Since the presence of the hot phase is connected to the presence of supernovae, the decrease of the hot phase VFF with decreasing metallicity is partly linked to the lower number of supernovae present in the low-metallicity runs (see Sec.~\ref{sec:stellar_feedback}).  The hot gas MF is negligible (less than 1\%), since the hot gas is very diffuse, with densities typically around 10$^{-28}$ -- 10$^{-27}$~g~cm$^{-3}$. 

The CM occupies only a few per cent of the total volume, about 0.05 -- 9\%, scaling with the metallicity. Its MF dominates at metallicities higher than 0.3~Z$_\odot$, with an average mass fraction in the range 56 -- 61.2\%, showing a positive correlation with metallicity. This is due to more efficient cooling at higher $Z$. Here we notice that in the $\Sigma$010-Z1 run the MF of the CM is higher than the MF of the CM reported by \citet{Tielens2005} by around 13\%. Moreover, the VFF of the CM in the $\Sigma$010-Z1 run is around 9 times higher than the value measured by \citet{Tielens2005}. The reason for these discrepancies could be attributed to our new treatment of the CR heating, since having a variable $\zeta$ allows the gas to reach very low values of around $\zeta \sim 10^{-20}$-- 10$^{-19}$~s$^{-1}$ (see Sec.~\ref{sec:CR_ion_rate}), which triggers the formation of cold gas, instead of the typical value of 3~$\times$~10$^{-17}$~s$^{-1}$ (see Sec.~\ref{sec:eq_curves}). Moreover, the presence of diffuse cold gas at solar metallicity (see Fig.~\ref{sec:mol_hydrogen}, \citealt{silcc8}) can explain the higher VFF of the cold gas. In the $\Sigma$010-Z0.02 run we get cold gas only due to our variable CR ionization rate \citep[see][]{Brugaletta2025}.

\subsection{Temperature-density and pressure-density phase diagrams}
\label{sec:phase_plots}

\begin{figure*}
	\includegraphics[width=0.87\textwidth]{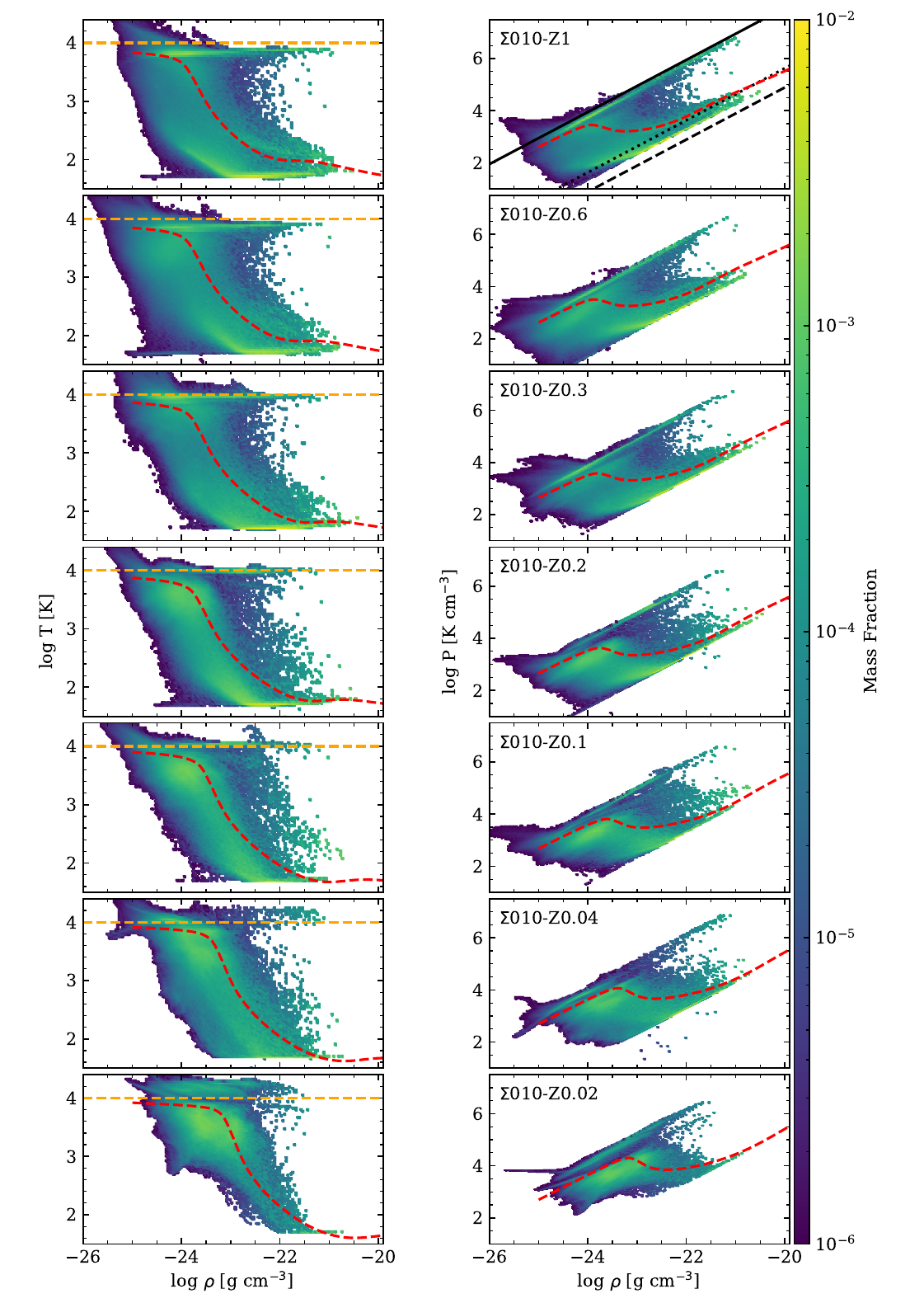}
    \caption{Temperature-density (left column) and pressure-density (right column) phase diagrams of all our simulations, at a time in which the H~II region branch is most visible in every run. From top to bottom, the metallicity goes from 1 to 0.02~Z$_\odot$, and the selected snapshots are at simulation times $t$~=~45.3, 68.4, 43.1, 104.5, 72.4, 52.2, 45.3~Myr, respectively. We take into consideration the region $|z|<$~250~pc. We overplot the unshielded equilibrium curves (red dashed lines) from Fig.~\ref{fig:eq_curves}, computed assuming $G_0$~=~1.7 and $\zeta$~=~3~$\times$~10$^{-17}$~s$^{-1}$, and a (orange dashed) line corresponding to T~=~10$^4$~K in the left column. In the pressure panel for $\Sigma$010 we add black isothermal lines corresponding to 10$^4$~K (solid), 100~K (dotted), 30~K (dashed). The presence of the two-phase medium is less evident at low Z due to the lack of cold gas.}
    \label{fig:phaseplot}
\end{figure*}

In this section, we analyse the temperature and pressure distributions as a function of density. In Fig.~\ref{fig:phaseplot} we present the temperature-density (left column) and pressure-density (right column) phase diagrams, computed in a region $|z|<$~250~pc, for all our simulations at a time at which the presence of the H~II regions (at around 10$^4$~K) is mostly visible by eye, and we overplot the unshielded equilibrium curves from Fig.~\ref{fig:eq_curves}. We also add a line indicating a temperature $T$~=~10$^4$~K. Regarding the temperature-density plots, we choose a temperature range in the y-axis that highlights the warm and cold gas phases already defined in Sec.~\ref{sec:gas_phase} for all our runs. It can be observed that, with decreasing metallicity, the presence of the two-phase medium (warm and cold gas) becomes less evident, since less cold gas is formed (compare with Fig.~\ref{fig:vff}). This can be attributed to the reduced cooling in metal-poor environments. However, at solar metallicity, the mass fraction of the cold gas is of a few per cent in the density range 10$^{-24}$--10$^{-22}$~g~cm$^{-3}$. This gas is located in cells that are far away from star clusters and hence that are characterized by a small value of $G_0$.

We observe a broad distribution of temperature values associated with each density value.  As already seen in \citealt{silcc1}, the reason for that is the different conditions that characterise every cell. In fact, shielded cells can cool to temperatures lower than those expected from unshielded equilibrium curves. For example, it can be seen in Fig.~\ref{fig:phaseplot} that the temperature distribution in the higher-metallicity runs is offset below the equilibrium curves, which is due to shielding. Furthermore, local turbulence and/or the presence of shocks can quickly heat cells to temperatures higher than those predicted by the equilibrium curves. Moreover, the broader density distribution can be attributed to a variable value of the G$_0$ parameter (as seen already in \citealt{silcc1, silcc8}) and of the $\zeta$ parameter \citep{Brugaletta2025}, which results in different equilibrium conditions in every cell. We note that the "branch" in the temperature-density plots at around 10$^4$~K corresponding to H~II regions becomes slightly hotter with decreasing metallicity, starting from around $\sim$~7000~K at solar metallicity and reaching $\sim$~12000 K in the most metal-poor run \citep{Haid2018}. Depending on the snapshot considered, some of this cold gas is found in a more diffuse phase (as already discussed in \citealt{silcc8}, and further analysed in a follow-up paper). Moreover, the relative differences in the mass distribution appear to be smaller in the pressure-density plots for varying metallicities.

\subsection{Molecular hydrogen}
\label{sec:mol_hydrogen}

\begin{figure}
	\includegraphics[width=\columnwidth]{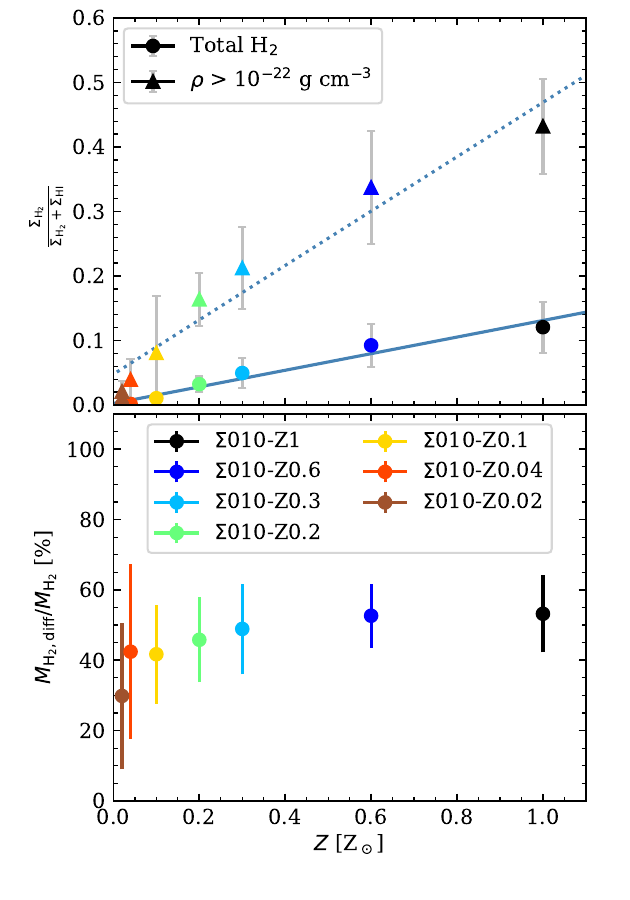}
    \caption{Top panel: average ratio of the molecular hydrogen surface density over the sum of molecular and atomic surface densities, as a function of the metallicity. The average has been computed considering a time interval of 200~Myr. In the region $|z|$~$<$~250~pc we consider either the total amount of H$_2$ and H (round markers) or the amount of H$_2$ and H found for a gas denser than 10$^{-22}$~g~cm$^{-3}$ (triangles). The solid and dotted lines represent the best fit for the total amount and that for dense gas, respectively. Bottom panel: average mass fraction of diffuse H$_2$, as a function of metallicity. We find this fraction to be around 50\% for the highest Z, and around 30\% for the lowest.}
    \label{fig:H2_mass}
\end{figure}

\begin{figure}
	\includegraphics[width=\columnwidth]{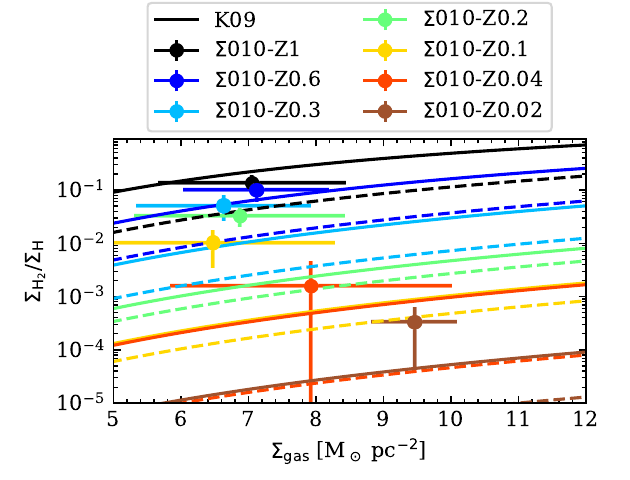}
    \caption{Time-averaged ratio of molecular and atomic hydrogen surface density as a function of the gas surface density. The coloured lines represent the predicted values from the \citet{Krumholz2009} model, where the colour matches that of the marker they refer to. The dashed lines have been computed using the fiducial values from \citet{Krumholz2009}, whereas the solid lines have been computed using the average density of the CM, and the average $G_0$, taken from the simulations.}
    \label{fig:H2_ratio}
\end{figure}

\begin{figure}
	\includegraphics[width=0.9\columnwidth]{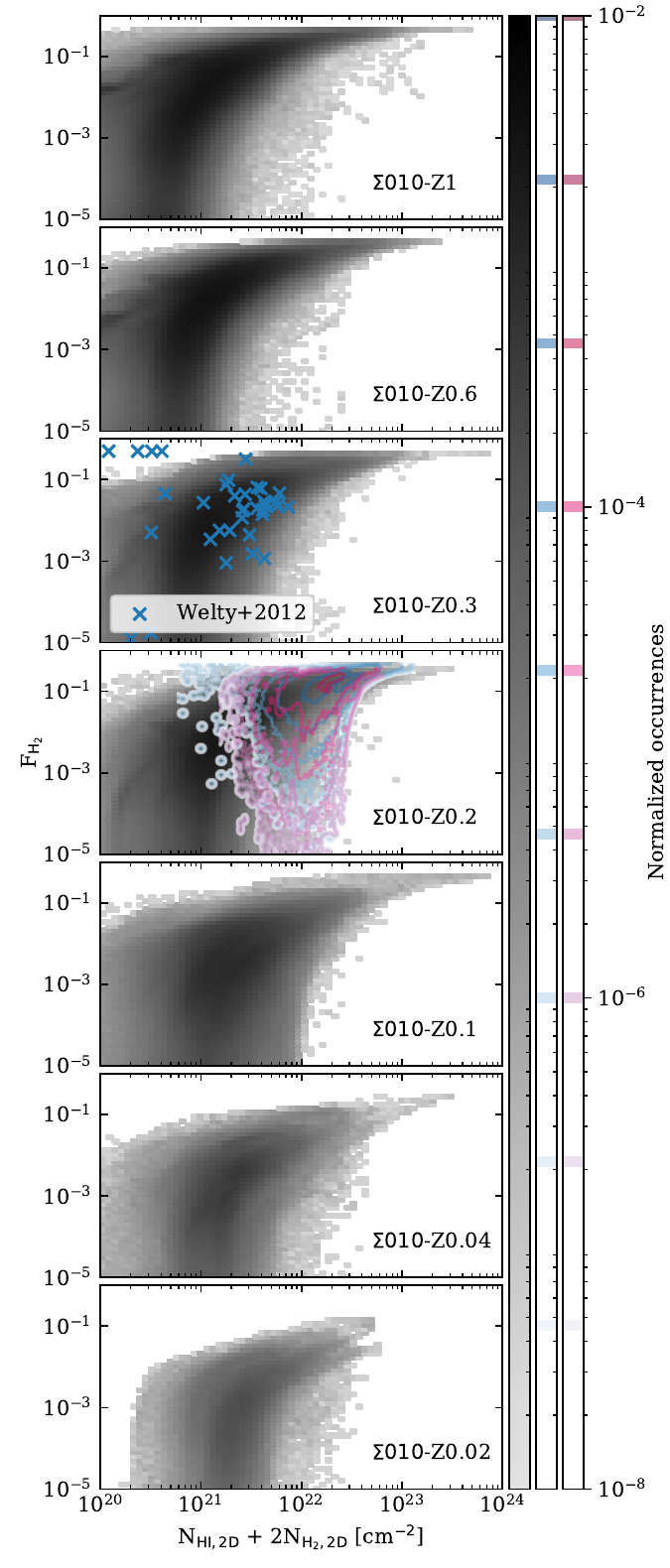}
    \caption{2D histogram of the projected molecular hydrogen fraction $F_\mathrm{H_2}$ as a function of column density of atomic and molecular gas. The column densities of the atomic and molecular hydrogen have been computed by projecting the 3D HI and H$_2$ densities along the z-direction. The histogram has been computed using all snapshots after the onset of star formation for all runs, and the occurrences have been normalized by the total number of occurrences in each run. We over plot the observational data from \citet{Welty2012} for the Large Magellanic Cloud, and from \citet{bolatto2011, Jameson2016}, in blue and pink contours, respectively, for the Small Magellanic Cloud.}
    \label{fig:H2_frac}
\end{figure}

\begin{figure}
	\includegraphics[width=\columnwidth]{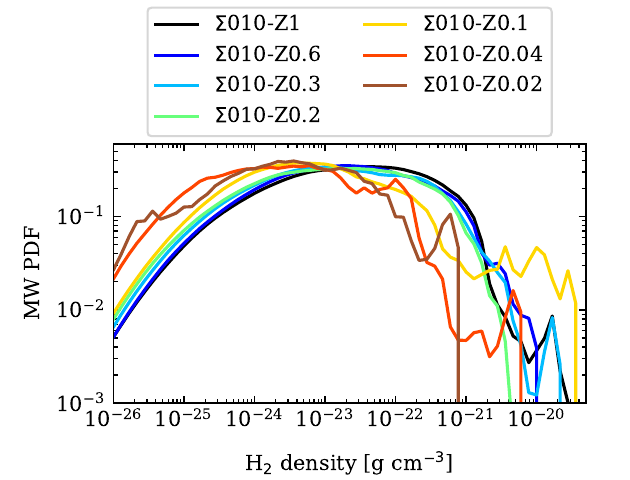}
    \caption{Mass-weighted PDF of the H$_2$ density, for all our runs, computed using their entire evolution after the onset of star formation. We note that H$_2$ spans a wide range of densities, from 10$^{-26}$ to slightly above 10$^{-20}$~g~cm$^{-3}$.}
    \label{fig:H2_dens_pdf}
\end{figure}

Several studies have seen a correlation between star formation and the presence of molecular gas \citep[see e.g.][]{Wong2002, Kennicutt2007, Bigiel2008, Leroy2008, Krumholz2011}. Therefore, in this section we aim to investigate how the amount of molecular hydrogen present in our simulations varies with metallicity. The formation of molecular hydrogen mainly occurs via reaction of two hydrogen atoms on the surface of dust grains \citep{Omukai_2010}. However, molecular hydrogen gets photodissociated by FUV photons. Dust grains can shield H$_2$ molecules from FUV radiation, preventing them from being photodissociated. The formation and destruction of molecular hydrogen is therefore dependent on the availability of dust grains and hence the value of the dust-to-gas mass ratio, which is a function of metallicity. However, we should point out that the amount of molecular gas that forms in our simulations also depends on our numerical resolution \citep{Seifried2017, Joshi2019} and the sink particle density threshold $\rho_\mathrm{sink}$, especially for the low-metallicity runs. In fact, given the chosen value of $\rho_\mathrm{sink}$, for the low-metallicity runs the H$_2$ formation time is usually longer than the typical molecular cloud lifetime; therefore, we find smaller H$_2$ fractions. At higher resolution, we would allow for the formation of denser regions in the ISM that would help to form more molecular gas.

\subsubsection{Molecular hydrogen mass fraction}

The mass fraction of H$_2$ is expected to decrease for lower metallicity \citep{Krumholz2009, Polzin2024}, and at very low metallicities star formation can possibly occur in atomic gas \citep{Krumholz2012, Glover2012a, Hu2016, Hu2017}. Therefore, in the top panel of Fig.~\ref{fig:H2_mass}, we show how the time-averaged ratio of $\Sigma_{\rm H_2}$ over the sum of $\Sigma_{\rm H_2}$~+~$\Sigma_{\rm HI}$ varies as a function of $Z$ for all gas (circles) and dense gas with $\rho > 10^{-22}$g~cm$^{-3}$ (triangles). We note that in both cases this ratio decreases for lower metallicity, as expected from the lower dust-to-gas mass ratio that characterises metal-poor environments. The different behaviour, considering the total gas or only the dense gas, can be attributed to the lower amount of atomic gas present in the denser gas, which increases the ratio. We fit a linear function to the dependence of the $\Sigma_\mathrm{H_2}$/($\Sigma_\mathrm{HI} + \Sigma_\mathrm{H_2}$) ratio with metallicity, and we find a slope of 0.13~$\pm$~0.01 when considering all of the H$_2$, and a slope of 0.42~$\pm$~0.04 when considering only the dense gas. We adopt a linear function for our best fit, as the dust-to-gas mass ratio, which is a key ingredient for the formation of H$_2$, scales linearly with the metallicity in our simulations. The two different slopes that we find indicate that with increasing metallicity, there is a higher fraction of molecular gas in denser environments. 
\begin{figure}
	\includegraphics[width=\columnwidth]{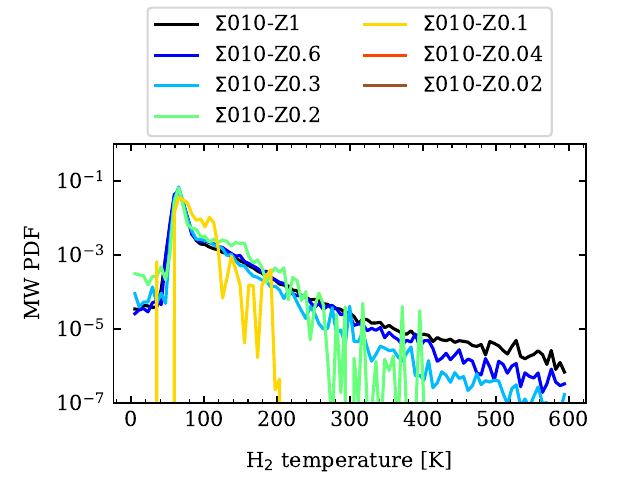}
    \caption{Mass-weighted PDF of the H$_2$ temperature. We consider the temperature of all the cells with an H$_2$ mass fraction above 50\%. In the $\Sigma$010-Z0.04 and $\Sigma$010-Z0.02 runs, no cell fulfils this criterion, therefore no H$_2$ temperature PDF is shown. We note that the PDF peaks at around 60~K, however temperatures up to 600~K are present. }
    \label{fig:H2_temp}
\end{figure}

\begin{figure*}
	\includegraphics[width=\textwidth]{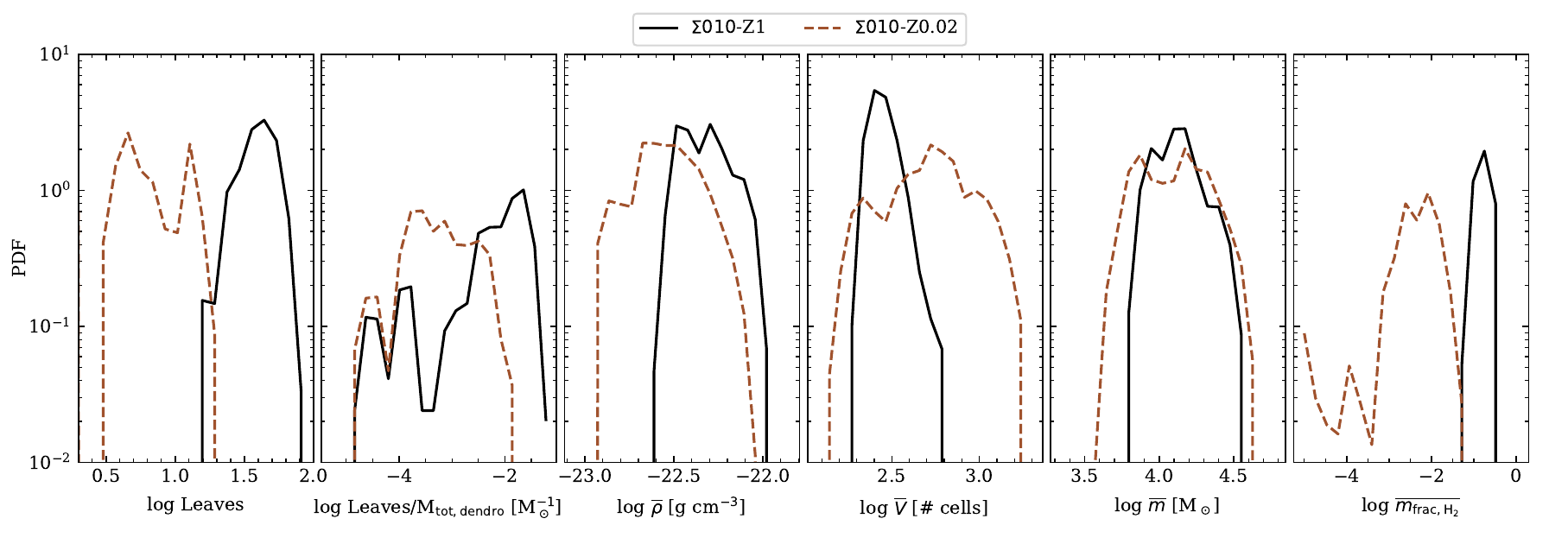}
    \caption{Probability density functions computed for the dendrogram sub-structures found. From left to right, we plot the distribution of the number of leaves, the ratio of the number of leaves and the total gas mass $M_\mathrm{tot,dendro}$ that enters the dendrogram calculation, the distribution of the snapshot-averaged density of the leaves, as well as their snapshot-averaged volume, mass, and H$_2$ mass fraction. The distributions for all simulation runs can be found in Fig.~\ref{fig:dendro_allruns}.}
    \label{fig:dendro_2runs}
\end{figure*}

Moreover, we investigate how much H$_2$ can be found in the diffuse gas ($\rho <$~10$^{-22}$~g~cm$^{-3}$) as a function of metallicity. In the bottom panel of Fig.~\ref{fig:H2_mass} we show the mass fraction of diffuse H$_2$, defined as $M_\mathrm{H_2, diff}/M_\mathrm{H_2}$, with $M_\mathrm{H_2, diff}$~=~$M_\mathrm{H_2, tot}$ - $M_\mathrm{H_2, dense}$. The latter term is the mass of H$_2$ found at densities higher than 10$^{-22}$~g~cm$^{-3}$. We note that for a metallicity higher than 0.2~Z$_\odot$ the mass fraction of the diffuse H$_2$ is almost constant and around 50\%. At lower metallicity, this fraction drops down to around 30\% for the most metal-poor run. This is due to the fact that in the metal-rich ISM the colder gas and higher shielding allow the H$_2$ to survive in diffuser environments. 

In Fig.~\ref{fig:H2_ratio} we show the time-averaged $\Sigma_\mathrm{H_2}$/$\Sigma_\mathrm{H}$ ratio, $\Sigma_\mathrm{H}$ being the surface density of atomic hydrogen, as a function of the gas surface density. We overplot the expected values from the \citet{Krumholz2009} model as coloured lines, where the colour of each line is the same as that of the point to which it refers. The \citet{Krumholz2009} model takes as input the metallicity of the gas, the gas surface density, the average density of the cold neutral medium, and the average value of the $G_0$ parameter. We compute the dashed lines using the fiducial values they adopt in \citet{Krumholz2009}. The solid lines are computed using the values of the temporal and spatial average number density of the cold gas and of $G_0$ provided by our data. We note that, in certain cases, the solid lines are one order of magnitude higher than the dashed ones. However, except for the two most metal-rich runs, we tend to overestimate the molecular hydrogen fraction compared to that predicted by \citet{Krumholz2009} by about one or two orders of magnitude. This difference can be attributed to the uniform density adopted in the model from \citet{Krumholz2009}, which hinders the formation of cold gas at very low metallicity, since the conversion from warm to cold gas occurs at higher densities and pressures (see Fig.~\ref{fig:eq_curves}). In our simulations, turbulence and shocks can compress the gas, and facilitate the formation of cold gas, and consequently, of H$_2$ gas. Nevertheless, the $\Sigma_\mathrm{H_2}/\Sigma_\mathrm{H}$ ratio for the two most-metal rich runs is comparable to the values shown in Fig.~13 in \citet{Schruba2011}, which consider massive spirals with metallicities close to solar. 

In Fig.~\ref{fig:H2_frac} we show a 2D-histogram of the projected molecular hydrogen fraction $F_\mathrm{H_2} \equiv N_\mathrm{H_2, 2D}$/$(N_\mathrm{HI, 2D} + 2N_\mathrm{H_2, 2D})$ as a function of the total column density of atomic and molecular hydrogen $N_\mathrm{HI, 2D} + 2N_\mathrm{H_2, 2D}$. The column densities $N_\mathrm{H_2, 2D}$ and $N_\mathrm{HI, 2D}$ are the projection of the molecular and atomic gas 3D density along the z-axis, respectively. For every run, we calculate the corresponding histogram using the entire evolution after the onset of star formation and normalize by the total number of occurrences. The value of $F_\mathrm{H_2}$ increases for higher densities, until it flattens to a value of around 0.5 at column densities higher than 10$^{23}$~cm$^{-2}$. This saturation at 0.5 means that the gas becomes fully molecular. However, in the $\Sigma$010-Z0.04 run the high-density distribution of $F_\mathrm{H_2}$ flattens to a value of around 0.3, and for run $\Sigma$010-Z0.02 there is no high column density distribution present, and the maximum value of $F_\mathrm{H_2}$ is around 0.1. This is consistent with the lower H$_2$ mass fractions shown in Fig.~\ref{fig:H2_mass}-\ref{fig:H2_ratio}. Moreover, this is well in accordance with the idea that, at the column densities in which star formation takes place, the metal-poor gas is still dominated by the atomic rather than the molecular gas, in accordance with \citet{Glover2012a} and \citet{Krumholz2012}. Comparing our results for the $\Sigma$010-Z0.6 and $\Sigma$010-Z0.2 simulations with the values of $F_\mathrm{H_2}$ in Fig.~2 in \citet{Polzin2024}, we note that our simulations reach column densities that are between one and two orders of magnitude higher. This can be explained because our midplane resolution of 3.9~pc is higher than the 10~pc resolution in the simulations by \citet{Polzin2024}. We also over-plot observational data from \citet{Welty2012} for the Large Magellanic Cloud, and from \citet{bolatto2011}, as blue contours, and \citet{Jameson2016}, as pink contours, for the Small Magellanic Cloud. We note a very good agreement between the observational data and the values from our simulations. 

\subsubsection{H$_2$ density and temperature}
In this section we want to give an overview of the conditions in which the H$_2$ present in our simulations is found. Fig.~\ref{fig:H2_dens_pdf} shows the mass-weighted probability density function (PDF) of the H$_2$ density, computed accounting for the entire evolution of the runs after the onset of star formation. We see that the H$_2$ density PDF spans a large density range, from 10$^{-26}$ to slightly above 10$^{-20}$~g~cm$^{-3}$. There is no clear peak, rather a wide density range (10$^{-24}$--10$^{-21}$~g~cm$^{-3}$) where the PDF is almost constant and at its maximum. This demonstrates that the large amount of diffuse H$_{2}$ that we find in our simulations is not an artifact of our choice to adopt a density threshold of 10$^{-22}$~g~cm$^{-3}$ to distinguish between diffuse and dense gas; we could decrease this value by an order of magnitude or more and would still find very similar results.
For the two most metal-poor runs, the broad peak in the PDF is slightly shifted (10$^{-25}$~g~cm$^{-3}$--10$^{-23}$~g~cm$^{-3}$), 
which possibly just reflects the fact that these runs form less dense gas overall.
In the high-density regime, we see a power-law tail for all runs. We also note that the amount of H$_2$ above the density threshold for star formation, $\rho_\mathrm{sink}$~=~2~$\times$~10$^{-21}$~g~cm$^{-3}$, is less than the amount of diffuse H$_2$. 

In Fig.~\ref{fig:H2_temp} we show the mass-weighted PDF of the H$_2$ temperature. In this computation, we take into account all cells in our computational domain whose H$_2$ mass fraction is higher than 50\%. In the case of the two most metal-poor runs, $\Sigma$010-Z0.02 and $\Sigma$010-Z0.04, no cell fulfils this criterion, therefore no temperature PDF is computed. Lowering this threshold implies that the majority of the gas in the cell is atomic, which biases the cell's temperature toward that of the atomic phase. The H$_2$ temperature PDF peaks at around 60~K for all runs, but it extends to much higher temperatures. In particular, the maximum temperature depends on the metallicity. For a metallicity equal or higher than 0.3~Z$_\odot$, we find the H$_2$ temperature to reach up to 600~K. On the other hand, the $\Sigma$010-Z0.2 run reaches an H$_2$ temperature of around 400~K, whereas  the $\Sigma$010-Z0.1 run reaches 200~K. This anticorrelation with metallicity can be explained by the reduced dust shielding in metal-poor gas, which prevents H$_2$ from surviving at higher temperatures. Our results are in agreement with the idea of an H$_2$ temperature power law presented by \citet{Togi2016}, necessary to recover the distribution of mid-infrared rotational H$_2$ emission observed by Spitzer at solar metallicity. They also find warm H$_2$ with a temperature of above 100~K.

\subsection{Gas fragmentation}
\label{sec:gas_frag}

\begin{table*}
	\centering
	\caption{Mean values of the quantities shown in Fig.~\ref{fig:dendro_2runs} and Fig.~\ref{fig:dendro_allruns}. We also provide the volume in units of pc$^3$, obtained multiplying the number of cells by the volume of one cell, (3.9 pc)$^3$.}
	\label{tab:mean_dendro}
	\begin{tabular}{lccccccr} 
		\hline
		Run & Leaves & Leaves/$M_\mathrm{tot, dendro}$ & Density & Volume & Volume & Mass & H$_2$ Mass Fraction\\
         &  & [$10^{-3}$ M$^{-1}_\odot$] & [10$^{-23}$~g cm$^{-3}$] & [$\#$ cells] & [10$^3$ pc$^3$] & [10$^{3}~$M$_\odot$] & $[\%]$ \\
		\hline
        $\Sigma$010-Z1 & 38.7 $\pm$ 10.2  & 10.0 $\pm$ 8.7 & 4.8 $\pm$ 1.5 & 292.4 $\pm$ 54.6 & 17.3 $\pm$ 3.2 & 14.0 $\pm$ 5.1 &  0.14 $\pm$ 0.05 \\
		$\Sigma$010-Z0.6 &35.9 $\pm$ 8.5 & 7.4 $\pm$ 7.7 & 5.1 $\pm$ 1.8 & 305.7 $\pm$ 68.0 & 18.1 $\pm$ 4.0 & 15.0 $\pm$ 6.2 & 0.12 $\pm$ 0.04 \\
        $\Sigma$010-Z0.3 & 31.3 $\pm$ 8.9 & 7.2 $\pm$ 7.1 & 5.1 $\pm$ 2.1 & 317.9 $\pm$ 56.2 & 18.9 $\pm$ 3.3 & 15.5 $\pm$ 6.7 & 0.07 $\pm$ 0.03 \\
		$\Sigma$010-Z0.2 & 30.0 $\pm$ 6.5 & 7.1 $\pm$ 6.9 & 5.0 $\pm$ 1.5 & 312.1 $\pm$ 67.8 & 18.5 $\pm$ 4.0 & 14.8 $\pm$ 5.1 & 0.05 $\pm$ 0.02 \\
        $\Sigma$010-Z0.1 & 22.3 $\pm$ 5.0 & 4.0 $\pm$ 5.0 & 4.7 $\pm$ 2.1& 354.1 $\pm$ 94.0 & 21.0 $\pm$ 5.6 & 15.1 $\pm$ 7.1 & 0.02 $\pm$ 0.01 \\
        $\Sigma$010-Z0.04& 6.9 $\pm$ 6.5 & 1.4 $\pm$ 2.3 & 4.1 $\pm$ 2.6 & 328.2 $\pm$ 101.2 & 19.5 $\pm$ 6.0 & 11.4 $\pm$ 7.7 & 0.01 $\pm$ 0.01  \\
        $\Sigma$010-Z0.02 & 7.5 $\pm$ 4.1& 1.2 $\pm$ 1.7 & 3.1 $\pm$ 1.2 & 583.9 $\pm$ 299.3 & 34.6 $\pm$ 17.8 & 13.9 $\pm$ 7.0 & 0.007 $\pm$ 0.006 \\
		\hline
	\end{tabular}
\end{table*}

As seen in Sec.~\ref{sec:gas_phase}, the amount of cold gas found in the ISM decreases with metallicity. The degree of fragmentation of the gas and the subsequent formation of dense clumps can influence the formation of stars. In this section, we investigate the fragmentation of the cold and dense gas in our simulations using the open-source dendrogram algorithm \textsc{Astrodendro}\footnote{\url{https://dendrograms.readthedocs.io/en/stable/}}. 

We provide \textsc{Astrodendro} with the logarithm of the 3D total gas density. As a minimum density to be considered for the calculation, we assume the threshold of $\rho_\mathrm{min, dendro}$~=~10$^{-23}$~g~cm$^{-3}$. We choose this threshold value because the low-metallicity ISM tends to be more diffuse than a solar metallicity ISM, and choosing a higher value for the density threshold would underestimate the number of substructures detected with the dendrogram method. Further, we assume a step width (min\_delta value) in log scale of 0.1 dex and a minimum number of 100 cells required per substructure in order to reduce the noise due to the presence of very small, fluctuating substructures that do not form stars anyway. The outcome of our dendrogram analysis is shown in Figs.~\ref{fig:dendro_2runs} and ~\ref{fig:dendro_allruns}, and reported in Table~\ref{tab:mean_dendro}. Our analysis takes into account the fragmentation of the gas in the 200~Myr interval after the onset of star formation.

For clarity reasons, in Fig.~\ref{fig:dendro_2runs} we show the distributions of some key dendrogram quantities only for the $\Sigma$010-Z1 and the $\Sigma$010-Z0.02 runs. An important quantity to describe the degree of fragmentation of the gas is the number of leaves of the dendrogram, since they are the smallest independent structures found in the gas. As we can see in the left panel of Fig.~\ref{fig:dendro_2runs}, the PDF of the distribution of the number of leaves in the $\Sigma$010-Z0.02 run is shifted to the left compared to that of the $\Sigma$010-Z1 run, meaning that at low metallicity the number of leaves formed is much smaller than at solar metallicity. 

Moreover, in the second panel we show the PDF of the distribution of the ratio of the number of leaves and the respective mass that enters the dendrogram calculation, $M_\mathrm{tot,dendro}$, meaning the total mass above the density threshold of 10$^{-23}$~g~cm$^{-3}$. Note that $M_\mathrm{tot,dendro}$ changes for each run and every snapshot, since the structure of the ISM continuously changes. The value of this ratio can be understood as the amount of leaves that can form in the ISM per M$_\odot$ of gas denser than $\rho_\mathrm{min, dendro}$. We see a similar behaviour as for the PDF of the number of leaves, showing that the solar-metallicity ISM is able to form more sub-structures per unit mass. Since we use the entire evolution after the onset of star formation, we compute for each snapshot the average volume and mass of all leaves, and the average of the leaf-averaged density. We represent these quantities in the third to fifth panels. The leaves in the $\Sigma$010-Z1 run are on average denser and smaller in volume than those of the $\Sigma$010-Z0.02 run. However, both exhibit almost the same average mass distribution. 
Finally, we plot the average mass fraction of molecular hydrogen found in the leaves (right panel). As already seen in Fig.~\ref{fig:H2_mass}, the amount of H$_2$ found in our simulations depends on metallicity, which explains why metal-poor leaves exhibit a much smaller amount of H$_2$. 

We represent the same quantities computed for all runs in Fig.~\ref{fig:dendro_allruns} in the Appendix and report the mean values of the analysed quantities in Table~\ref{tab:mean_dendro}. Regarding the number of leaves and its ratio with $M_\mathrm{tot, dendro}$, we see that the ability of the gas to fragment and form smaller, denser cores scales with metallicity. However, for a metallicity larger than 0.1~Z$_\odot$, there seems to be only a weak trend. Within the uncertainty, all results are more or less comparable. 
On the other hand, we see a much stronger trend for metallicities lower than 0.1~Z$_\odot$. Regarding the average density of the leaves, we note no substantial difference for metallicities larger than 0.1~Z$_\odot$, whereas at the lowest metallicities the leaves tend to be more diffuse. Furthermore, we see larger average volumes for lower metallicities, but similar average masses. The average H$_2$ mass fraction in the leaves clearly increases with metallicity.

\subsection{Star formation}
\label{sec:SF}
\subsubsection{Local conditions for star formation}
\label{sec:local_cond_sf}

\begin{figure*}
	\includegraphics[width=\textwidth]{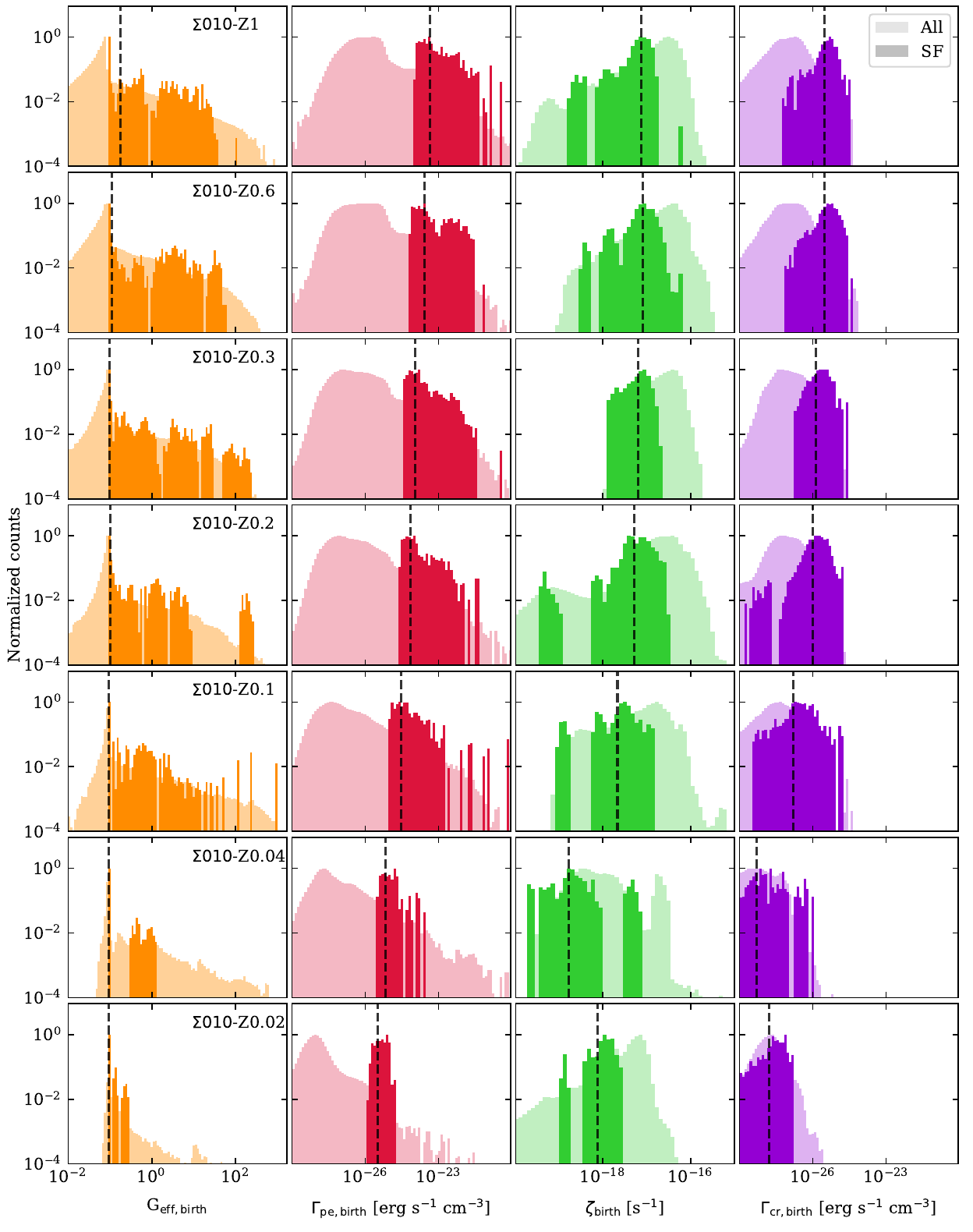}
    \caption{Distributions of $G_\mathrm{eff}$, $\Gamma_\mathrm{pe}$, $\zeta$ and $\Gamma_\mathrm{cr}$ for all gas and for star-forming (SF) gas for all runs. The black vertical lines are the medians of the SF gas distributions as listed in Table \ref{tab:birth_cond}.}
    \label{fig:birthconditions}
\end{figure*}

\begin{figure}
	\includegraphics[width=0.95\columnwidth]{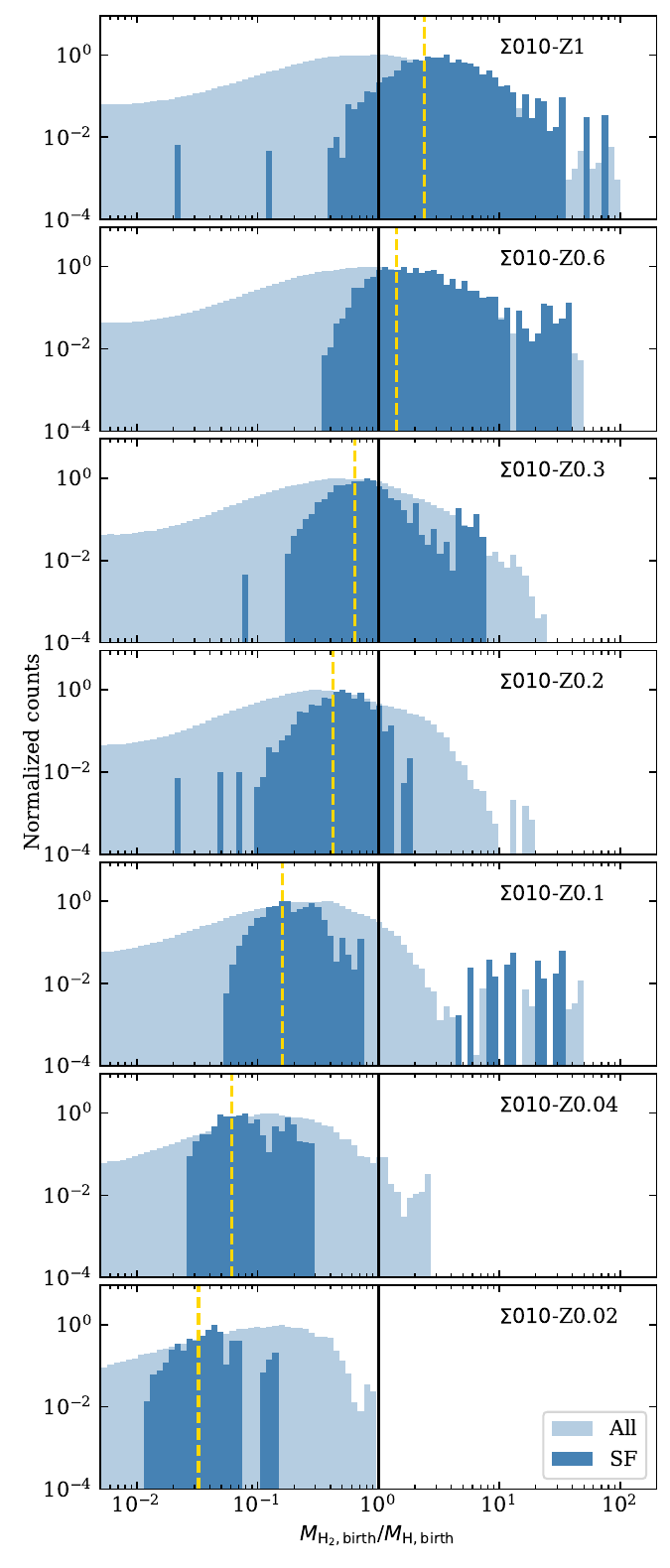}
    \caption{Distribution of the H$_2$ to HI mass ratio for the total gas (transparent distribution), and the star-forming gas (opaque). The dashed yellow line is the median of the distribution for the star-forming gas, whereas the black line indicates $M_\mathrm{H_2}$/$M_\mathrm{HI}$~=~1.}
    \label{fig:birthconditions_H2}
\end{figure}

\begin{table*}
	\centering
	\caption{Median values of $G_\mathrm{eff, birth}$, $\Gamma_\mathrm{pe, birth}$, $\zeta_\mathrm{birth}$, $\Gamma_\mathrm{cr, birth}$ and $M_\mathrm{H_2, birth}/M_\mathrm{H, birth}$ distributions computed for the star-forming gas in Fig.~\ref{fig:birthconditions}. The subscripts and superscripts are the 25$^{\mathrm{th}}$ and 75$^{\mathrm{th}}$ percentiles, respectively.}
	\label{tab:birth_cond}
	\begin{tabular}{lccccr} 
		\hline
		Run & $G_\mathrm{eff, birth}$ & $\Gamma_\mathrm{pe, birth}$ & $\zeta_\mathrm{birth}$ & $\Gamma_\mathrm{cr, birth}$ & $M_\mathrm{H_2, birth}/M_\mathrm{H, birth}$\\
         & & [10$^{-25}$~erg s$^{-1}$ cm$^{-3}$] & $[$10$^{-18}$~s$^{-1}$] & $[$10$^{-27}$~erg s$^{-1}$ cm$^{-3}$]  & \\
		\hline
       $\Sigma$010-Z1 & 0.18~$_{0.09}^{1.76}$ & 44~$_{20}^{220}$ & 7.6~$_{4.7}^{10}$ & 31~$_{17}^{48}$ & 2.4$_{1.6}^{3.8}$ \\[1.2mm]
       $\Sigma$010-Z0.6 & 0.11~$_{0.09}^{3.2}$  & 27~$_{12}^{220}$ & 8.2~$_{5.2}^{12}$  & 31~$_{16}^{53}$ & 1.4$_{1.0}^{2.3}$\\[1.2mm]
       $\Sigma$010-Z0.3 & 0.10~$_{0.09}^{0.8}$ & 11~$_{6.2}^{52}$ & 6.5~$_{2.6}^{9.5}$  & 13~$_{5.7}^{25}$ & 0.6$_{0.5}^{0.9}$\\[1.2mm]
       $\Sigma$010-Z0.2 & 0.10~$_{0.09}^{0.84}$ & 7.1~$_{4.0}^{26}$ & 5.3~$_{3.4}^{9.2}$ & 10~$_{4.9}^{19}$ & 0.4$_{0.3}^{0.6}$\\[1.2mm]
       $\Sigma$010-Z0.1 & 0.09~$_{0.09}^{0.54}$ & 2.9~$_{1.6}^{8.1}$ & 2.2~$_{0.8}^{3.7}$ & 1.6~$_{0.6}^{3.3}$ & 0.2$_{0.1}^{0.2}$\\[1.2mm]
       $\Sigma$010-Z0.04 & 0.09~$_{0.09}^{0.10}$ & 0.7~$_{0.5}^{1.3}$& 0.2~$_{0.02}^{3.1}$ & 0.1~$_{0.02}^{0.2}$ & 0.06$_{0.05}^{0.1}$\\[1.2mm]
       $\Sigma$010-Z0.02 & 0.09~$_{0.09}^{0.09}$ & 0.3~$_{0.2}^{0.5}$ & 0.8~$_{0.5}^{1.5}$ & 0.2~$_{0.05}^{0.3}$ & 0.03$_{0.02}^{0.04}$\\
       
		\hline
	\end{tabular}
\end{table*}

As seen in the previous paragraph, the degree of fragmentation of the dense gas is similar for metallicities higher than 0.1~Z$_\odot$, but it is clearly lower for the metal-poor gas. This suggests that the conditions under which stars form vary with metallicity. To explore this aspect, we investigate the distributions of $G_\mathrm{eff}$, the PE heating rate, $\Gamma_\mathrm{pe}$, $\zeta$, and the CR heating rate, $\Gamma_\mathrm{cr}$ for all gas (transparent) and for star-forming gas (opaque) in Fig.~\ref{fig:birthconditions}. 
All distributions are mass-weighted and normalized such that they fall between 0 and 1.
For all gas, we consider the gas near the midplane within $|z|$~$<$~250~pc. 
To trace the star-forming gas, we find the coordinates of each star cluster at the time of its creation. We then calculate the coordinates that the star-forming gas would have had one snapshot before its formation given its velocity at birth. Then we compute above quantities in a region centred on these coordinates and with a radius $r_\mathrm{accr}$~$\sim$~11.7~pc. We consider only cells whose density is larger than 5~$\times$~10$^{-22}$~g~cm$^{-3}$, which is a bit lower than the density threshold for sink particle formation. 
We also show the median values of the distributions for the star-forming gas as black vertical lines, and we report them with their 25$^{\mathrm{th}}$ and 75$^{\mathrm{th}}$ percentiles in Table~\ref{tab:birth_cond}. 

From Fig.~\ref{fig:birthconditions} we see that the median value of $G_\mathrm{eff}$ is comparable to the $G_\mathrm{bg}$ value. Since $G_\mathrm{eff}$ is computed within a spherical radius of 50~pc from each star cluster, a large volume fraction has $G_\mathrm{eff} \lesssim G_\mathrm{bg}$. Smaller values are possible because $G_\mathrm{bg}$ is locally attenuated (see Sec.~\ref{sec:numerical_methods}). In run $\Sigma$010-Z1 the median of $G_\mathrm{eff}$ is slightly higher than the background because more stars are formed. The median $\Gamma_\mathrm{pe}$ is in the range of 10$^{-26}$ -- 10$^{-24}$ erg s$^{-1}$ cm$^{-3}$ and scales with metallicity. This is expected as $\Gamma_\mathrm{pe}$ depends on the dust-to-gas ratio, which scales linearly with the metallicity. The median $\zeta$ is around 2 -- 8~$\times$~10$^{-18}$~s$^{-1}$ for the most metal-rich runs, and around 2 -- 8~$\times$~10$^{-19}$~s$^{-1}$ for the most metal-poor runs. The median $\Gamma_\mathrm{cr}$ scales with metallicity and is in the range of 10$^{-28}$ -- 10$^{-26}$ erg s$^{-1}$ cm$^{-3}$. The only exception is the $\Sigma$010-Z0.04 run, for which the median of $\Gamma_\mathrm{cr}$ is around 5~$\times$~10$^{-29}$ erg s$^{-1}$ cm$^{-3}$. This discrepancy from the scaling with metallicity is due to the lower number of formed stars, and will be addressed in Sec.~\ref{sec:sfr}. 

Moreover, we notice that in every run the PE heating rate dominates over the CR heating rate, both in all gas and in the star-forming gas. Furthermore, the PE and CR heating rates above which star formation is suppressed drop with decreasing metallicity, owing to the less efficient cooling, and this behaviour is particularly accentuated in the two most metal-poor runs. Comparing the star-forming gas distributions with their respective total distributions, we notice a few features. For example, we see selection effects in the value of $G_\mathrm{eff}$ in the star-forming gas mostly at $Z <$~0.1~Z$_\odot$, whereas at higher metallicity larger values of $G_\mathrm{eff}$ are apparent. We also note that the distribution of $\Gamma_\mathrm{pe}$ for the star-forming gas matches the higher-end of the corresponding total gas distribution. The reason for this is the dependence of $\Gamma_\mathrm{pe}$ on the density, which in the star-forming gas is around the value of $\rho_\mathrm{sink}$. A similar behaviour is found for the $\Gamma_\mathrm{cr}$ distribution in the star-forming gas. Moreover, we observe selection effects in the $\zeta$ distribution for the star-forming gas with respect to all gas. In fact, we notice that the highest values of $\zeta$, beyond which star formation is suppressed, are around two orders of magnitude lower than the highest values found in all gas.

To summarise, the star-forming gas at low metallicity experiences lower PE and CR heating than in solar-neighbourhood conditions. This is due to the interplay of two different, yet connected, effects. First, in the lower metallicity runs the star formation rate surface density is lower than for solar-neighbourhood conditions (see Sec.~\ref{sec:sfr}), therefore there is a lower production of UV radiation and cosmic rays. As a consequence, the total distributions of $G_\mathrm{eff}$ and $\zeta$ show reduced values. Second, the inefficient cooling at low metallicity implies that lower PE and CR heating rates are needed to prevent the gas from cooling and collapsing. 

Using the same method as in Fig.~\ref{fig:birthconditions}, we show in Fig.~\ref{fig:birthconditions_H2} the distribution of the molecular to atomic hydrogen mass ratio for both the total (transparent distribution) and star-forming gas (opaque). The median values of this ratio for the star-forming gas are shown as vertical dashed yellow lines, and their values are reported in Table~\ref{tab:birth_cond} together with their 25$^\mathrm{th}$ and 75$^\mathrm{th}$ percentile. Moreover, we add a vertical solid black line that indicates $M_\mathrm{H_2, birth}/M_\mathrm{H, birth}$~=~1, meaning when the mass of molecular and atomic hydrogen are the same. We note that the median for the star-forming gas decreases as a function of the metallicity, and becomes lower than 1 at a metallicity of 0.3~Z$_\odot$. However, when considering the entire distribution for the star-forming gas, we note that the gas is almost fully atomic at a metallicity of 0.04~Z$_\odot$ and lower. We also note, that the highest values of the molecular to atomic hydrogen mass ratios, computed for the total gas, do not correspond to the values for the star-forming gas, apart for the two most metal-rich runs. This means that, for metallicities of 0.3~Z$_\odot$ or lower, there are cells in our domain which present a high mass fraction of H$_2$, without forming stars. This can happen even if the total density of the gas in the cell is higher than $\rho_\mathrm{sink}$, but the other criteria for sink formation are not fulfilled \citep[see e.g.][for more details]{silcc3}. All in all, we show here that star formation occurs in atomic gas in the low-metallicity ISM.

\subsubsection{Star formation rates}
\label{sec:sfr}

\begin{figure}
	\includegraphics[width=0.97\columnwidth]{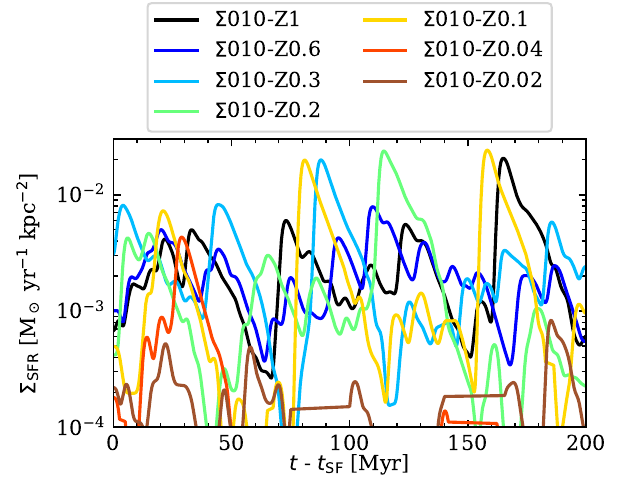}
    \caption{Observable star formation rate surface density $\Sigma_{\text{SFR}}$ as a function of time. $\Sigma_{\text{SFR}}$ oscillates in time, therefore a star-forming phase is followed by a more quiescent phase during which star formation diminishes or halts completely. The $\Sigma_{\text{SFR}}$ for the $\Sigma$010-Z0.1 and $\Sigma$010-Z0.04 are significantly lower than those of the more metal-rich runs.}
    \label{fig:sfr}
\end{figure}

Star formation plays a central role in determining the dynamical and chemical evolution of the gas.  We measure the star formation rate in our simulations by means of the "observable" star formation rate surface density, $\Sigma_{\text{SFR}}$, described in \citealt{silcc3} as
\begin{equation}
\Sigma_{\text{SFR}} (t) = \frac{\mathrm{SFR} (t)}{A} = \frac{1}{A} \sum_{i = 1}^{N_*(t)} \frac{120~\text{M}_\odot}{t_{\text{OB}, i }},
\label{eq:SFR}
\end{equation}
where $A$ = (0.5~kpc)$^2$ is the area of the computational domain in the midplane, $N_*(t)$ is the number of active massive stars at time $t$, and $t_{\text{OB}, i}$ is the lifetime of the i$^{\mathrm{th}}$ formed massive star. Eq.~\ref{eq:SFR} takes into account that for each formed massive star, a total gas mass of 120~M$_\odot$ is converted into stars. 
On the x-axis, we subtract the time at which the first star is born in every simulation, $t_\text{SF}$ (see Table~\ref{tab:sfr_feedback_features}).
We show the time evolution of $\Sigma_{\text{SFR}}$ in Fig.~\ref{fig:sfr}, and we list the median values in Table~\ref{tab:sfr_feedback_features}, where we also report the 25$^{\mathrm{th}}$ and 75$^{\mathrm{th}}$ percentiles. 
 
In the following paragraphs, we will discuss physical quantities evolving in the time interval [$t_\text{SF}$, $t_\text{SF}$~+~200~Myr].  

We find that $\Sigma_{\text{SFR}}$ oscillates in time with a period of a few tens of Myr, depending on the particular simulation. As already seen in Sec.~\ref{sec:global_evolution}, the stellar feedback originating from all existing clusters pushes the gas away from the midplane in the vertical direction, depleting the gas reservoir of the midplane. At the same time, the gas that has been pushed in the vertical direction is too hot and diffuse to give birth to new star clusters; therefore, the total star formation activity in the simulation box decreases. This drop is also seen in the $\Sigma_{\text{SFR}}$, which reaches a local minimum in this case. However, after all clusters have become inactive, meaning that all the massive stars have exploded as supernovae, the expelled gas is no longer exposed to the high pressure driven by stellar feedback. It therefore falls back towards the midplane due to gravity. As the gas accumulates further near the midplane, it again reaches the conditions for another starburst. 

From Fig.~\ref{fig:sfr} we note that the more metal-rich runs show only small differences in $\Sigma_{\text{SFR}}$, whereas the $\Sigma$010-Z0.04 and $\Sigma$010-Z0.02 runs clearly show a much lower $\Sigma_{\text{SFR}}$. These differences are quantified with the median $\Sigma_{\text{SFR}}$ (see Table~\ref{tab:sfr_feedback_features}), which decreases roughly with metallicity. In particular, in run $\Sigma$010-Z0.02 the number of formed stars is only around 4\% of those formed in run $\Sigma$010-Z1. Although $\Sigma_\mathrm{SFR}$ depends on the availability of cold gas, which we have seen decreasing as a function of metallicity in Fig.~\ref{fig:vff}, it also depends on the degree of fragmentation of the gas. As seen in Table~\ref{tab:mean_dendro}, there are significant differences in the fragmentation of the gas for the lowest metallicity runs.

\subsection{Stellar feedback}
\label{sec:stellar_feedback}
\subsubsection{Stellar winds, supernovae, radiation}

\begin{table}
	\centering
	\caption{A few key quantities regarding star formation and feedback in every simulation. $t_\text{SF}$ is the time at which the first star cluster forms, $N_\text{stars}$ is the number of massive stars formed in the time interval [$t_\text{SF}$, $t_\text{SF}$ + 200 Myr]. We add the median of $\Sigma_\mathrm{SFR}$ computed in the same interval. The value of 0 for the $\Sigma$010-Z0.04 is justified since for the majority of its evolution, the $\Sigma_\mathrm{SFR}$ is zero, and in the median calculation repetitions are counted. The mean final CR ionisation rate $\overline{\zeta}_\mathrm{fin}$ is computed at $t_\mathrm{SF}$+~200~Myr for every run (see Sec.~\ref{sec:CR_ion_rate}).}
	\label{tab:sfr_feedback_features}
	\begin{tabular}{lcccc} 
		\hline
		Simulation & $t_\text{SF}$  &$N_\text{stars}$& $\Sigma_\mathrm{SFR}$ & $\overline{\zeta}_\mathrm{fin}$\\
         & [Myr] & & [10$^{-3}$~M$_\odot$~yr$^{-1}$~kpc$^{-2}$] & [~10$^{-17}$~s$^{-1}$]\\
		\hline
        $\Sigma$010-Z1 & 23.6 & 1231 & 1.8~$_{1.1}^{3.5}$  & 5.7\\[1.2mm]
		$\Sigma$010-Z0.6 & 23.9 & 1013 & 1.9~$_{1.0}^{2.7}$ &  7.3\\[1.2mm]
		$\Sigma$010-Z0.3 & 26.1 & 1118 & 1.8~$_{0.8}^{3.3}$ & 4.9\\[1.2mm]
		$\Sigma$010-Z0.2 & 21.1 & 975 & 0.9~$_{0.3}^{2.0}$ & 9.9\\[1.2mm]
		$\Sigma$010-Z0.1 & 28.7 & 1121 & 0.7~$_{0.2}^{2.5}$ & 2.6\\[1.2mm]
        $\Sigma$010-Z0.04 & 33.3 & 307 & 0~$_{0}^{0.07}$ & 1.8\\[1.2mm]
        $\Sigma$010-Z0.02 & 44.2 & 54 & 0.07~$_{0}^{0.2}$ & 0.6\\
		\hline
	\end{tabular}
\end{table}

\begin{figure*}
	\includegraphics[width=0.9\textwidth]{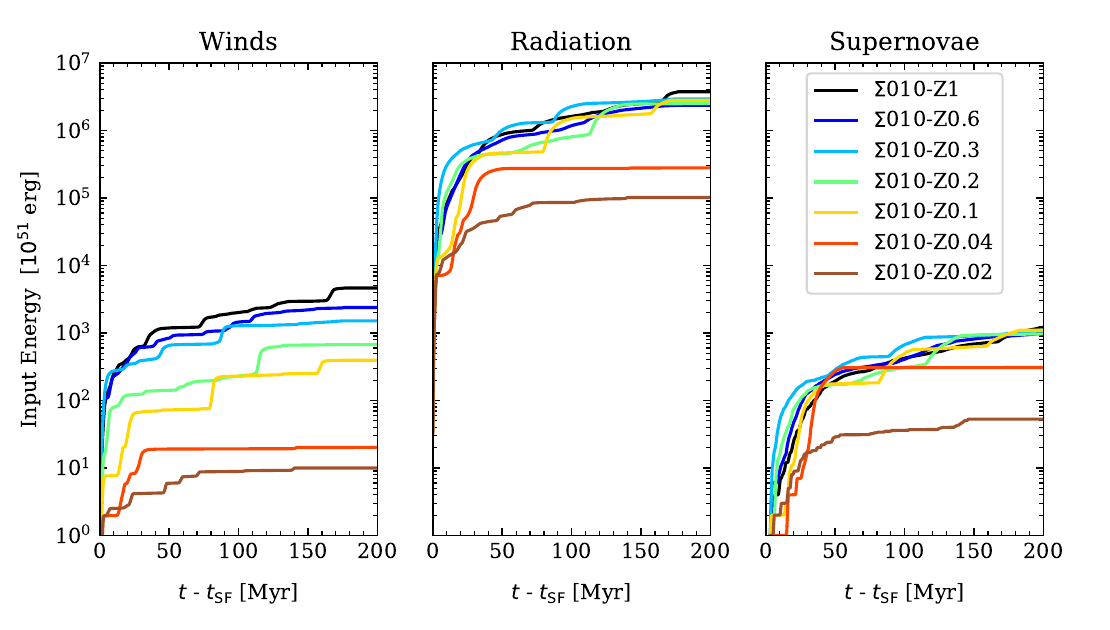}
    \caption{Cumulative energy input per feedback channel as a function of time. From left to right we show stellar winds, radiation, and supernovae. The contribution of stellar winds scales with metallicity, and overall it is comparable to the contribution by supernovae. The contribution of the emitted radiation is around four orders of magnitude higher than winds and supernovae, and, as for supernovae, it mostly depends on the number of formed stars, rather than metallicity. Note however, that not all of the emitted radiative energy is converted to thermal or kinetic energy of the gas.}
    \label{fig:energy_input}
\end{figure*}

During their entire lifetime, stars inject energy in the ISM in the form of stellar winds, radiation, and supernova explosions. However, these three different energy inputs depend on several factors, e.g. the metallicity, the number of formed stars, and for how long stars live. In turn, these energy injections profoundly affect the shape and evolution of the ISM. Therefore, it is worth analysing the total amount of energy injected due to these three different feedback channels. 

In Fig.~\ref{fig:energy_input}, we show the cumulative distribution of the energy injected because of winds, radiation, and supernovae, which sums up the contribution of all stars formed in our simulations. In the case of stellar winds, the injected energy has been computed by integrating the wind luminosity of every model in time, as defined in Eq.~\ref{eq:wind_lum}. We calculate the cumulative contribution of radiation by integrating the bolometric luminosity of the stellar models in time. The energy injection due to supernovae is derived by multiplying the number of supernovae, which corresponds to the number of formed massive stars, by 10$^{51}$~erg (as seen in Sec.~\ref{sec:numerical_methods}). Therefore, the cumulative energy input from supernovae is directly dependent on the number of massive stars formed (as reported in Table~\ref{tab:sfr_feedback_features}).

We note that the total contribution of stellar winds scales as a function of metallicity, as expected from the bottom panel of Fig.~\ref{fig:lbol_time}, and because low-metallicity runs form fewer stars. For the most metal-rich runs, the energy injected by winds is higher than the energy injected by SNe in the same runs. Moreover, we do not see a clear trend of the cumulative amount of emitted radiation with metallicity for $Z>0.1$~Z$_\odot$. We also note that the energy input from the emitted radiation is three orders of magnitude higher than that due to winds and SNe. However, e.g. \citet{walch2012, Haid2018} report that less than 0.1 per cent of the ionising radiation emitted by a star couples with the surrounding gas as the conversion of radiative energy to thermal and kinetic energy is very inefficient \citep{silcc4}. This implies that all feedback channels are of comparable importance in shaping ISM dynamics.

\subsubsection{Cosmic rays}
\label{sec:CR_ion_rate}
\begin{figure}
	\includegraphics[width=0.95\columnwidth]{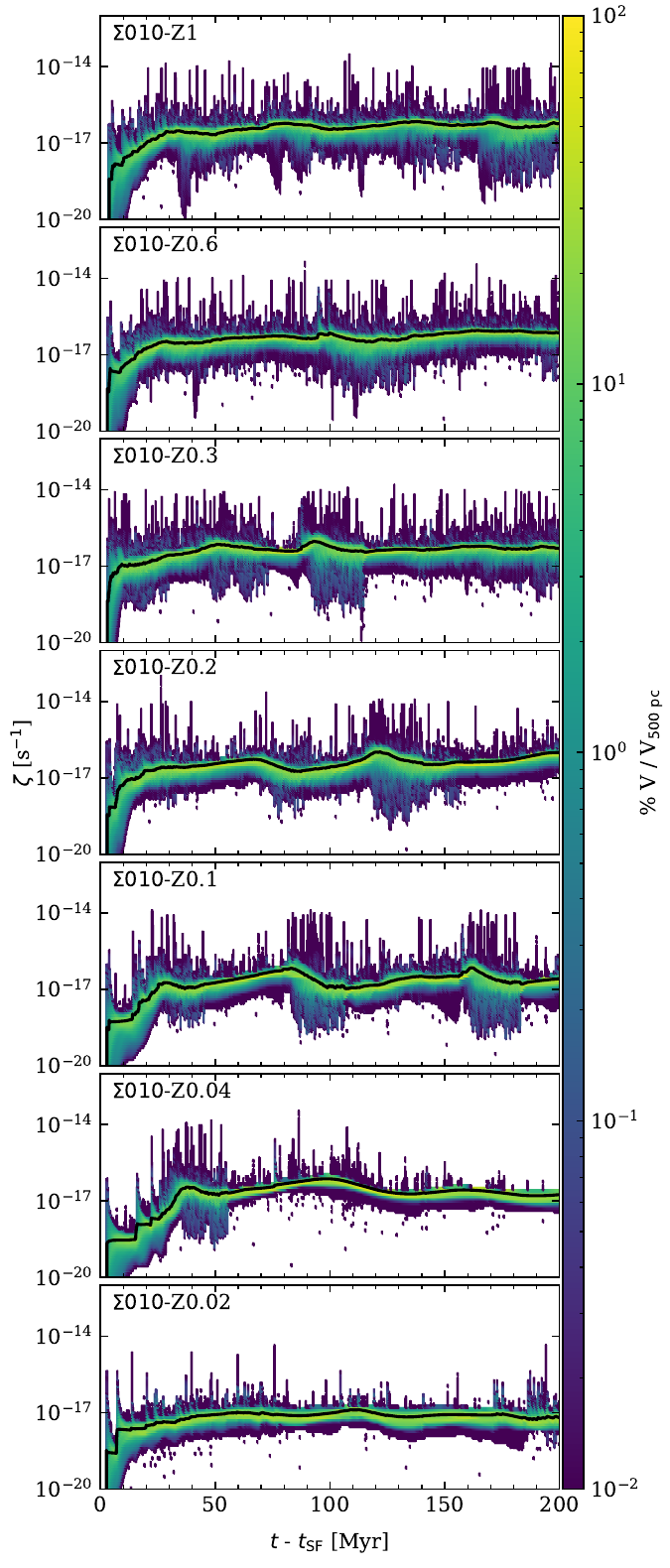}
    \caption{Variations of the CR ionisation rate in space and time in all our runs, computed within $|z|<$~250~pc around the midplane. The black line is the average of $\zeta$ computed in the same region. }
    \label{fig:crir}
\end{figure}

In this section, we analyse the temporal and spatial variability of $\zeta$. We compute $\zeta$ for every cell in a region $|z|<$~250~pc. We then bin the values of $\zeta$ in 100 logarithmically spaced bins and compute the relative frequency of each bin. We do this for every snapshot in all our runs, obtaining the time evolution shown in Fig.~\ref{fig:crir}.

We note that at fixed time, $\zeta$ varies by several orders of magnitude, ranging from around 10$^{-19}$ to 10$^{-14}$~s$^{-1}$. The highest spikes (10$^{-15}$ -- 10$^{-14}$~s$^{-1}$) correspond to regions in the vicinity of SNe, where the injection of CR energy takes place. The lower values (10$^{-18}$ -- 10$^{-17}$~s$^{-1}$) correspond, as already seen in Fig.~\ref{fig:birthconditions}, to the star-forming gas. We do not observe dramatic changes in the values of $\zeta$ for different metallicities, and only for the two lowest-metallicity runs $\zeta$ is overall smaller. This is due to the number of SNe exploding in our simulations, dependent on the SFR (see Sec.~\ref{sec:sfr}), which determines the amount of CR energy injection. We report in Table~\ref{tab:sfr_feedback_features} the average value of $\zeta$ at $t_\mathrm{SF}+200$~Myr, $\overline{\zeta}_\mathrm{fin}$, when the ISM is the most evolved in our runs. However, we do not see a clear correlation with metallicity, except for the two most metal-poor runs.

\subsection{Outflows}
\subsubsection{Mass outflows}

\begin{figure}
	\includegraphics[width=\columnwidth]{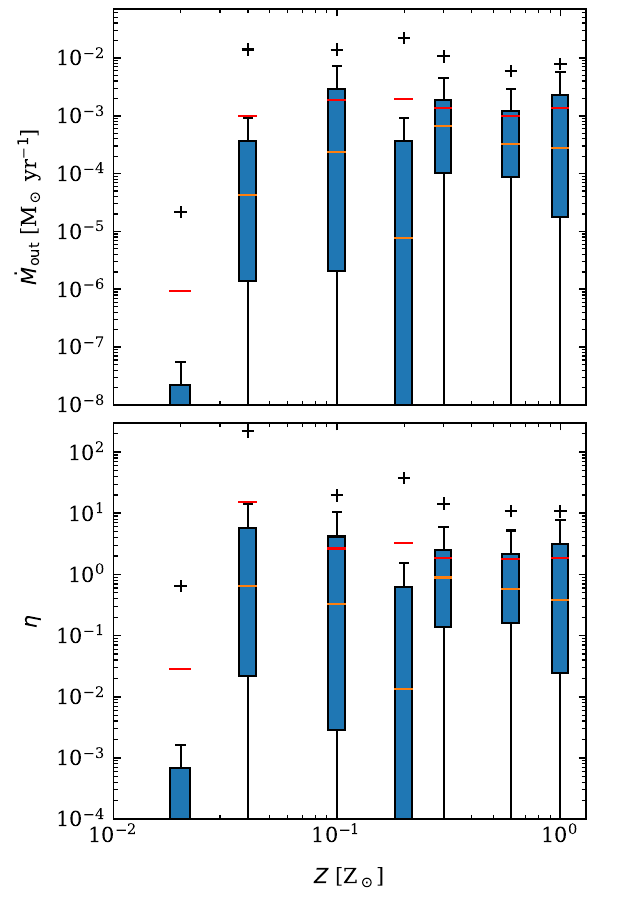}
    \caption{Box plot of the distributions of $\dot{M}_\mathrm{out}$ (top panel), and of the mass loading $\eta$ (bottom panel), as a function of metallicity. The orange lines are the median of each distribution, the red lines the mean. The boxes extend from the 75$^{\mathrm{th}}$ percentile down to the 25$^{\mathrm{th}}$ percentile. The plus markers represent the maximum of each distribution. We observe no strong correlation of the $\dot{M}_\mathrm{out}$ with $Z$ for $Z >$~0.1~$Z_\odot$. The most metal-poor run does not have enough SNe to experience a strong outflow. A similar behaviour is observed for $\eta$. 
    }
    \label{fig:mdot}
\end{figure}

As mentioned above, stellar feedback leads to the outflow of gas from the midplane. We compute the amount of mass instantaneously traversing the surface $|z|$~=~1~kpc, meaning the mass outflow rate $\dot{M}_\text{out}$, and we show its distribution as a function of metallicity in the top panel of Fig.~\ref{fig:mdot}. We also show the distribution of the mass loading $\eta$~=~$\dot{M}_\mathrm{out}$/$\langle$SFR$\rangle$ in the bottom panel. Red lines indicate the mean for every run, and orange lines the median. The boxes extend from the 75$^{\mathrm{th}}$ percentile down to the 25$^{\mathrm{th}}$ percentile. The plus markers represent the maximum of each distribution.  

We do not see major differences in the mean $\dot{M}_\mathrm{out}$ and $\eta$ values for all runs, except for the most metal-poor one, which has a much lower outflow rate and mass loading, as expected from the fact that it does not form many stars. This behaviour suggests that, if the star formation rate is high enough, the outflows do not seem to strongly depend on metallicity if the medium has the same surface density. Since outflows are mainly driven by supernova feedback, this result can also be observed in the values of the VFF of the hot gas in Fig.~\ref{fig:vff}, which do not change substantially down to a metallicity of 0.1~Z$_\odot$.

\subsubsection{Vertical acceleration profiles}

\begin{figure}
	\includegraphics[width=\columnwidth]{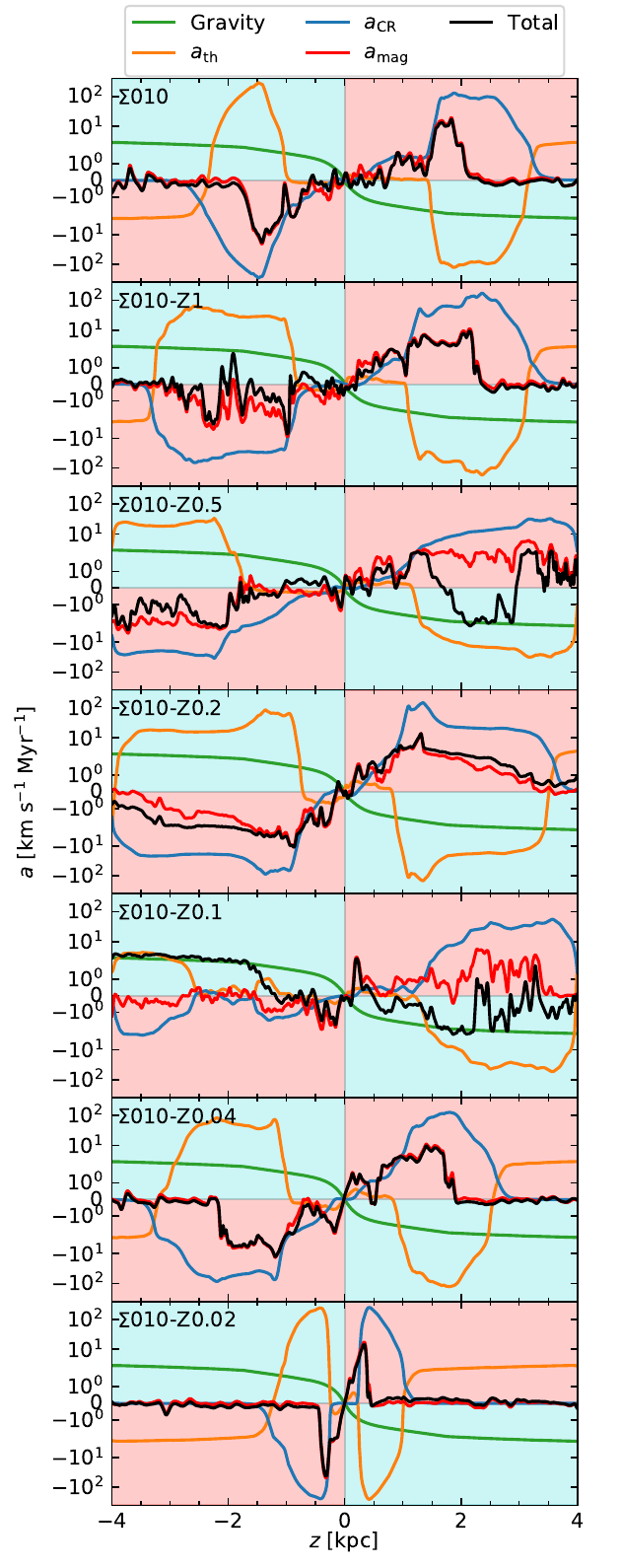}
    \caption{Vertical acceleration profiles for all runs. We consider the acceleration due to CR (blue), thermal (orange), turbulent (purple), magnetic (red) pressure gradients, as well as the acceleration due to total gravity (green). The black line is the sum of these five accelerations. The red-shaded regions indicate the accelerations that push the gas in the outward direction from the midplane, the turquoise-shaded ones those in the inward direction.} 
    \label{fig:acc_prof}
\end{figure}

We analyse the outflows taking place in our runs by calculating vertical acceleration profiles, as shown in Fig.~\ref{fig:acc_prof}. For this calculation, we consider the entire box and average the profiles in the interval [$t_\mathrm{SF}$~+~100~Myr, $t_\mathrm{SF}$~+~150~Myr] to avoid the influence of the initial conditions. We decide to average the profiles in a time interval of 50~Myr to eliminate fluctuations occurring on very short timescales, without, at the same time, losing the contributions of the outflow/inflow motions, which would average out to some degree when considering the entire time span of the simulations. Positive accelerations are oriented in the direction of positive $z$-coordinates. We define the accelerations in the vertical direction due to pressure forces, as
\begin{equation}
    a_i = -\frac{1}{\rho} \frac{dP_i}{dz}, 
\end{equation}
where the subscript $i$ denotes the different pressure components, meaning due to the thermal,  magnetic and CR pressure gradients. The total pressure is defined as
\begin{align}
    P_\mathrm{tot} &= P_\mathrm{th} + P_\mathrm{mag} + P_\mathrm{CR} \\
    & = (\gamma -1)e_\mathrm{th} +  \frac{B^2}{8\pi} + (\gamma_\mathrm{CR} - 1)e_\mathrm{CR},
    \label{eq:pressures}
\end{align}
where $\gamma$~=~5/3 and $\gamma_\mathrm{CR}$~=~4/3. In the calculation of $P_\mathrm{mag}$ we do not account for the magnetic tension, which is likely negligible since we are dominated by small-scale fields. The relative importance of the magnetic tension with respect to the magnetic pressure in SILCC has been shown in Fig.~C1 of \citet{Girichidis2021}. The acceleration due to gravity is defined as 
\begin{equation}
    a_\mathrm{grav} = - \frac{d\Phi_\mathrm{tot}}{dz},
\end{equation}
where $\Phi_\mathrm{tot}$ is the total gravitational potential, which takes into account the contributions of gas, sink particles, the external potential due to the presence of an old stellar population, and dark matter. We calculate all the accelerations in our 3D box, and then we compute the vertical acceleration profiles taking into account all the cells $j$ present in a slice $k$, with the thickness of one cell, and averaging in mass as
\begin{equation}
    a_k = \frac{\sum_j m_\mathrm{\textit{j, k}} a_\mathrm{\textit{j, k}}}{\sum_j m_\mathrm{\textit{j, k}}}
\end{equation}
where $m_\mathrm{\textit{j, k}}$ is the mass of the $j^{\mathrm{th}}$ cell in the $k^{\mathrm{th}}$ slice. 

The vertical profiles of the total acceleration, computed as $a_i$~+~$a_\mathrm{grav}$, are shown as black lines in Fig.~\ref{fig:acc_prof}.  For better comprehension, the red-shaded regions in Fig.~\ref{fig:acc_prof} indicate accelerations that push the gas out of the midplane in the outward direction - both for positive and negative $z$ - and the turquoise-shaded regions indicate accelerations that push the gas back to the midplane.

We notice that in all runs $a_\mathrm{CR}$ is almost zero near the midplane and up to 100~km~s$^{-1}$~Myr$^{-1}$ in the outward direction for higher heights, helping to lift the gas off the midplane, which is in agreement with the results from Fig. 8 in \citet{GirichidisEtAl2018a}. On the other hand, $a_\mathrm{th}$ is either almost zero or it weakly pushes the gas out of the midplane for the first few hundred parsec, and inward for larger heights. The profiles for $a_\mathrm{CR}$ and $a_\mathrm{th}$ are similar in shape but opposite in sign, for many heights, therefore, they tend to compensate each other. This occurs because thermal pressure dominates in high-temperature regions, where the gas is heated by supernovae, while CR pressure is more significant in colder regions. At the interfaces between these regions, the pressure gradients have opposite signs, resulting in their partial cancellation. If there were regions with both high (or both low) thermal and CR pressure, they would rapidly expand or collapse, causing one form of pressure to quickly dominate over the other. Since $a_\mathrm{th}$ and $a_\mathrm{CR}$ tend to compensate each other, the total acceleration is dominated by the acceleration due to the magnetic pressure gradient. However, for the $\Sigma$010-Z0.02 run, the total acceleration is significant only within 500 pc from the midplane, whereas the gas at higher heights experiences almost no acceleration. This explains why the values of $\dot{M}_\mathrm{out}$ measured at 1~kpc shown in Fig.~\ref{fig:mdot} for $\Sigma$010-Z0.02 are much lower compared to the other runs. A more detailed analysis of how these different accelerations impact the dynamics of the ISM will be presented in a following paper.

\section{Discussion}
\label{sec:discussion}

\subsection{Validity of the model}
In Table~\ref{tab:sim_parameters} we have compared the metallicity of our runs with that of known real galaxies. In this regard, it is good to remember that such galaxies might have different conditions with respect to those of our runs, e.g. a different gas surface density, strength of the magnetic field, stellar surface density, etc., so a direct comparison is not trivial. In our simulations many parameters can be fine-tuned in order to get more realistic results to be compared to observations; therefore, the parameter space that can be analysed with our simulations is quite large. However, given that our computational resources are limited, we need to restrict ourselves to a set of parameters for which it is also possible to compare with our previous work. Since in this study we are interested in understanding the effects of metallicity on the ISM, we choose the same setup as in \citealt{Brugaletta2025}, which adds the implementation of a variable CR ionisation rate to the setup used in \citealt{silcc8}, and we change only the metallicity, the metal abundances and the dust-to-gas ratio, as already described in Sec.~\ref{sec:sim_param}. SILCC setups have been originally implemented to study environments in solar-neighbourhood conditions, therefore the only change in metallicity and dust-to-gas ratio might not be enough to accurately describe the metal-poor ISM. In our treatment of low-metallicity environments we are neglect that nearby dwarf galaxies with these metallicities might have a much higher gas surface density, for which they could cool more efficiently and form more stars. For example, a gas surface density in the range 10 -- 60~~M$_\odot$~pc$^{-2}$ has been observed in the LMC \citep[see Fig.~7 in][]{Hughes2010}, 10 -- 200~M$_\odot$~pc$^{-2}$ in the SMC \citep[see e.g. Fig.~3 in][]{bolatto2011}, and up to 100~M$_\odot$~pc$^{-2}$ in IZw18 \citep{Lelli2012}. Therefore, if a comparison with a specific galaxy is intended, our parameters would have to be fine-tuned to match those of the target environment.  

\subsection{Comparison to previous works}
\citet{Hu2016, Hu2017} simulate an isolated dwarf galaxy employing a similar chemical network based on \citet{Nelson_1997, glover2007a, Glover2012b} to treat non-equilibrium cooling and chemistry. They also model a variable ISFR (in \citealt{Hu2017}), star formation and stellar feedback. They assume a constant CR ionization rate $\zeta$~=~10$^{-18}$~s$^{-1}$, and a fixed metallicity of 0.1~Z$_\odot$. The G1D01 run in \citet{Hu2016} is the most similar to our $\Sigma$010-Z0.1 run, however with a fixed G$_0$~=~1.7. The MF of formed cold gas in this run is around 2-3\%, an order of magnitude lower than what we find for the $\Sigma$010-Z0.1 run (see Table~\ref{tab:mean_vff}). This difference can be traced back to the assumption of fixed G$_0$ and $\zeta$ parameters. The PE-PI-SN run in \citet{Hu2017} is the most similar to our $\Sigma$010-Z0.1 run. This run has a $\Sigma_\mathrm{SFR}$ of around 2~$\times$~10$^{-5}$~M$_\odot$~yr$^{-1}$~kpc$^{-2}$ at a gas surface density of around 1~M$_\odot$~pc$^{-2}$, whereas in our $\Sigma$010-Z0.1 we have a mean $\Sigma_\mathrm{SFR}$~=~7~$\times$~10$^{-4}$~M$_\odot$~yr$^{-1}$~kpc$^{-2}$ at a gas surface density of 10~M$_\odot$~pc$^{-2}$. Scaling the $\Sigma_\mathrm{SFR}$ value of the PE-PI-SN run by the gas surface density of $\Sigma$010-Z0.1, we obtain a value similar to the $\Sigma_\mathrm{SFR}$ for the $\Sigma$010-Z0.1 run.

\citet{Lahen2020} simulate a gas-rich dwarf galaxy merger, adopting a metallicity of 0.1~Z$_\odot$, and the same chemical network as in \citet{Hu2016}. Between 0 and 100~Myr, the $\Sigma_\mathrm{SFR}$ they find is around 10$^{-3}$~M$_\odot$~yr$^{-1}$~kpc$^{-2}$, which is in accordance with our mean value for the $\Sigma$010-Z0.1 run. With a similar setup, \citet{Lahen2023} simulate a dwarf galaxy with an initial metallicity of around 0.02~Z$_\odot$, including stellar feedback as stellar winds, supernovae and radiation, by means of the BoOST stellar tracks. They find a $\Sigma_\mathrm{SFR}$ in range 10$^{-5}$--10$^{-3}$~M$_\odot$~yr$^{-1}$~kpc$^{-2}$, which is in accordance with the median value reported in Table~\ref{tab:sfr_feedback_features} for the $\Sigma$010-Z0.02 run.

\citet{Whitworth2022} simulate an isolated dwarf galaxy with varying metallicity and UV strength, non-equilibrium chemistry, shielding, and supernova feedback. The chosen values of metallicity are 0.1~Z$_\odot$ and 0.01~Z$_\odot$, that can be compared with our $\Sigma$010-Z0.1 and $\Sigma$010-Z0.02. They assume a fixed value of $G_0$ in the range 0.017~--~0.17 and a fixed $\zeta$ in the range 3~$\times$~10$^{-19}$ -- 3~$\times$~10$^{-18}$~s$^{-1}$. They find that the MF of H$_2$ is always lower than 1\%, it scales with metallicity and decreases for higher values of the $G_0$ parameter. The MF of cold gas scales with metallicity when $G_0$ is fixed, and it anticorrelates with $G_0$ at fixed metallicity. The MF of cold gas they find at 0.1~Z$_\odot$ and 0.01~Z$_\odot$ is around one to two orders of magnitude lower than the values reported in Table~\ref{tab:mean_vff} for the $\Sigma$010-Z0.1 and $\Sigma$010-Z0.02 runs. This difference can be partially explained as we define the cold gas phase for $T<$~300~K, whereas \citet{Whitworth2022} use a temperature threshold of 100~K. Moreover, they did not observe a significant effect on the SFR when reducing both the metallicity and UV field strength by a factor of 10.

Our runs have a similar setup to the TIGRESS-NCR simulations presented in \citealt{Kim2024}, where the influence of metallicity is investigated with respect to the pressure-regulated feedback-modulated star formation theory. They consider metallicities in a range 0.1~--~3~Z$_\odot$ and gas surface densities in the range 5~--~150~M$_\odot$~pc$^{-2}$. They scale the dust abundance linearly with metallicity, except for the runs with $Z$~=~0.1~Z$_\odot$ where they assume a dust abundance of 0.025. The simulations include supernova feedback, metal cooling, PE heating, heating and cooling processes due to H$_2$ formation and dissociation processes, among others. They also add CR heating assuming that the CR ionisation rate $\zeta$ (in our nomenclature) scales as $\zeta$~$\propto$~$\Sigma_\mathrm{SFR}$/$\Sigma_\mathrm{gas}$. Here the $\Sigma_\mathrm{SFR}$ is computed from the stars formed in the past 40~Myr of evolution. This scaling is such that a value $\zeta_0$~=~2~$\times$~10$^{-16}$~s$^{-1}$ is assumed in solar-neighbourhood conditions when $\Sigma_\mathrm{SFR}$~=~2.5~$\times$~10$^{-3}$~M$_\odot$~yr$^{-1}$~kpc$^{-2}$ and $\Sigma_\mathrm{gas}$~=~10~M$_\odot$. In addition to this linear scaling, they include an attenuation $\zeta$~$\propto$~$N_\mathrm{eff}^{-1}$, with $N_\mathrm{eff}$~=~1.5~$\times$~10$^{21}$~cm$^{-2}$($n_\mathrm{H}$/100~cm$^{-3}$)$^{0.3}$, when the local column density exceeds $N_0$~=~9.35~$\times$~10$^{20}$~cm$^{-2}$ \citep{Neufeld2017}. Given our gas surface density of 10~M$_\odot$~pc$^{-2}$, we can only compare our runs with their R8-Z simulations, which adopt an initial gas surface density of 12~M$_\odot$~pc$^{-2}$. In the case of solar metallicity and 0.1~Z$_\odot$, they find $\Sigma_\mathrm{SFR}$ to be about a factor of two higher. However, they also observe that the $\Sigma_\mathrm{SFR}$ anticorrelates with metallicity.

\subsection{Caveats}

In our simulations, we assume for simplicity that the dust-to-gas ratio always scales linearly with respect to metallicity. However, some studies have pointed out that this assumption might not hold in very metal-poor environments. 
\citet{Remy2014} analyse the variation of the dust-to-gas ratio with metallicity for a set of 126 galaxies, and find that the observed dust-to-gas ratio vs metallicity relation cannot be described with a power law with a single exponent of -1, as for metallicities below that of the SMC this relation becomes steeper. This break in the power law could affect several of our low-metallicity runs. \citet{Feldmann2015} identifies a critical value of the metallicity for which the linear dependence breaks down, which depends on the competition of dust growth in the ISM and the dilution of the gas by infall of dust-poor gas onto the galactic plane. \citet{BialySternberg2019} study the thermal properties and the multiphase behaviour of the gas for varying metallicity, down to very metal-poor conditions. For the scaling of the dust-to-gas ratio they assume a linear dependence for $Z$~$\ge$~0.2~Z$_\odot$, and for lower values a broken power law of the kind 0.2~$\times$~($Z$/0.2~Z$_\odot$)$^3$. \citet{KimJG2023} compare a linear scaling of the dust-to-gas ratio as a function of metallicity with the prescription of \citet{BialySternberg2019}. They note that a superlinear dependence of the dust abundance with respect to the gas metallicity decreases the strength of the PE heating and recombination cooling with respect to the CR heating, which becomes dominant in metal-poor environments. As the CR heating does not depend on the metallicity, this means that a lower overall heating rate is present, and the gas can cool down more efficiently. In a previous related work \citep{Brugaletta2025}, we have investigated how a variable CR ionisation rate affects the cooling of the gas at very low metallicity ($Z$~=~0.02~Z$_\odot$), as well as the impact of a different scaling (linear, or power law) of the dust-to-gas ratio with metallicity at 0.02~Z$_\odot$. We have observed that using the scaling proposed by \citet{BialySternberg2019} instead of a linear relation results in a slightly enhanced average star formation rate at $Z$~=~0.02~Z$_\odot$, since the photoelectric heating is lower because of the reduced dust-to-gas ratio. However, at higher metallicities up to 0.1~Z$_\odot$ this difference might become more and more important, since photoelectric heating is more efficient at higher metallicity.   

Moreover, for simplicity, we have assumed that the abundances of C, O, and Si scale linearly with the metallicity. However, these elements, especially carbon and oxygen, are enriched differently due to stellar nucleosynthesis. Furthermore, since at low metallicity a smaller fraction of metals is locked up in dust grains, this changes the relative abundances of C, O and Si. \citet{Bisbas2024} study the impact of the $\alpha$-enhanced gas on the abundances of C, C$^+$ and CO, concluding that the relative carbon-to-oxygen abundance is of high importance at low metallicity. In this regard, we do not model chemical enrichment from stellar evolution, which would change our local abundances of metals, mainly affecting our metal-line cooling rates.  

Finally, we assume that all the massive stars formed in our simulations will end their life as supernovae. However, at sub-solar metallicities this assumption might not be correct, as pointed out by \citet{Heger_2003}. In fact, metal-poor single massive stars above 40~M$_\odot$ are predicted to collapse directly into a black hole, whereas those in range 25 $< M < 40$~M$_\odot$ produce a black hole by fallback, without exploding as supernovae. Therefore, in our prescription, we tend to overestimate of around 25\% the impact of supernova feedback in low-metallicity environments.

\section{Summary and Conclusions}
\label{sec:conclusions}

In this work, we analyse the structure and evolution of the ISM at low metallicity, by means of a set of SILCC simulations. We include stellar feedback in the form of radiation, stellar winds, cosmic rays, and supernovae. Considering a gas surface density of 10~M$_\odot$~pc$^{-2}$, a variable strength of the interstellar radiation field in Habing units $G_0$, and a variable CR ionisation rate $\zeta$. We vary the initial metallicity of the gas as well as the metallicity of the stellar evolutionary tracks (BoOST) from a solar value down to 0.02~Z$_\odot$. We employ a non-equilibrium chemical network to follow the time evolution of the hydrogen and carbon abundances, which is particularly important to understand the heating and cooling rates for metal-poor environments. We add an additional simulation at solar metallicity using the Geneva stellar models, employed in previous SILCC works. We highlight our most important results in the following.

Metal-poor environments tend to be warmer than media at solar metallicity. This is due to the fact that metals are the main coolants for the atomic gas up to a few 10$^6$~K, therefore a lack in metals means a higher temperature of the medium on average. We find that the mass fraction of cold gas decreases from 61\% at solar metallicity to 2.3\% at 0.02~Z$_\odot$. Moreover, we find that the warm medium (both neutral and ionized) is the dominant phase in mass and volume filling fractions at low metallicity. The amount of hot gas is dependent on the number of supernovae, and this decreases for lower metallicity.

We find that the molecular hydrogen mass fraction scales linearly with metallicity, with a slope of 0.13~$\pm$~0.01 when considering the total H$_2$, and a slope of 0.42~$\pm$~0.04 if we take into account only the dense gas with $\rho >$~10$^{-22}$~g~cm$^{-3}$. We also find that the mass of H$_2$ found in the diffuse phase ($\rho < 10^{-22}$~g~cm$^{-3}$) is around 50\% for the most metal-rich runs and it drops to around 30\% for the most metal-poor run. Moreover, we find that the density of H$_2$ spans a wide range, from 10$^{-26}$ to slightly above 10$^{-20}$~g~cm$^{-3}$ for all runs. We measure H$_2$ temperatures up to 600~K for the most metal-rich runs, and the maximum of the temperature anticorrelates with metallicity. 
    
Using a dendrogram analysis, we find that metallicity is not the main physical parameter affecting the fragmentation of the gas in the range 0.1--1~Z$_\odot$. However, at lower metallicity, the gas forms a smaller number of fragments. We see that low-metallicity fragments are more diffuse and bigger than at solar metallicity, however with comparable masses. The mass fraction of H$_2$ present in the fragments also decreases with metallicity. The degree of fragmentation of the gas directly affects the star formation rate.
    
We observe that stars in low-metallicity runs form in a medium that experiences lower PE and CR heating rates, compared to the solar-neighbourhood values. This is due to two effects, namely the smaller number of formed stars, that decrease both the PE and CR heating rates; and less efficient cooling of the gas implies that lower PE and CR heating rates are needed to allow the gas to collapse and form stars. We also see that in low-metallicity runs, stars are formed in atomic gas, in contrast to solar-metallicity environments where star formation occurs in molecular gas. 

 Regarding stellar feedback, we note that the contribution of winds increases for higher metallicity, and is higher than the contribution from supernovae in the same runs, if the metallicity is higher than 0.1~Z$_\odot$. The distribution of $\zeta$ varies between 10$^{-18}$ and 10$^{-14}$~s$^{-1}$ irrespective of the metallicity, except for the two most metal-poor runs, which have less supernovae. 

 The distribution of mass outflow rates does not show dramatic changes with metallicity, except for the run with 0.02~Z$_\odot$ metallicity. This implies that, if the star formation rate is high enough, the mass outflow rate does not strongly depend on the metallicity at fixed gas surface density.

 We investigate the role of the thermal, magnetic and CR pressures in determining the acceleration of the gas in the vertical direction. Accelerations due to thermal and CR pressures tend to balance each other, therefore the total acceleration is dominated by the magnetic pressure gradient, and it is directed in the outward direction in most cases.

\section*{Acknowledgements}
VB thanks Masato Kobayashi for the fruitful discussions regarding gas fragmentation at low metallicity. VB, SW, TER, DS, and PCN thank the Deutsche Forschungsgemeinschaft (DFG) for funding through the SFB 1601 “Habitats of massive stars across cosmic time” (sub-
projects B1, B4 and B6). SW, TER, and DS further acknowledge support by the project ”NRW-Cluster for data-intensive radio astronomy: Big Bang to Big Data (B3D)” funded through the programme ”Profilbildung 2020”, an initiative of the Ministry of Culture and Science of the State of North Rhine-Westphalia. VB and SW thank the Bonn-Cologne Graduate School. TN acknowledges support from the DFG under Germany’s Excellence Strategy - EXC-2094 - 390783311 from the DFG Cluster of Excellence "ORIGINS". PG and SCOG acknowledge funding by the European Research Council via the ERC Synergy Grant “ECOGAL” (project ID 855130). RW acknowledges support by the institutional project RVO:67985815 and by the INTER-COST LUC24023 project
of the Czech Ministry of Education, Youth and Sports. SCOG also acknowledges support from the
Heidelberg Cluster of Excellence EXC 2181 (Project-ID 390900948)
‘STRUCTURES: A unifying approach to emergent phenomena in
the physical world, mathematics, and complex data’ supported by
the German Excellence Strategy. The software used in this work was in part developed
by the DOE NNSA-ASC OASCR Flash Centre at the University of Rochester \citep{Fryxell_2000, dubey2009}. Part of the data visualisation has been done with the Python package \textsc{YT} \citep{Turk2011} and the \textsc{flash\_amr\_tools} Python package\footnote{\url{https://pypi.org/project/flash-amr-tools/}} developed by PCN. The data analysis has been performed using the following Python packages: \textsc{numpy} \citep{vanderWalt2011}, \textsc{matplotlib} \citep{Hunter2007}, \textsc{h5py} \citep{Collette2020}, \textsc{IPython} \citep{Perez2007}, \textsc{SciPy} \citep{scipy}, \textsc{astrodendro}\footnote{\url{http://www.dendrograms.org/}}.

\section*{Data Availability}
The derived data underlying this article will be shared on reasonable request to the corresponding author. The simulation data will be made available on the SILCC data web page: \url{http://silcc.mpa-garching.mpg.de}.



\bibliographystyle{mnras}
\bibliography{main} 

\begin{thebibliography}{}
\makeatletter
\relax
\def\mn@urlcharsother{\let\do\@makeother \do\$\do\&\do\#\do\^\do\_\do\%\do\~}
\def\mn@doi{\begingroup\mn@urlcharsother \@ifnextchar [ {\mn@doi@}
  {\mn@doi@[]}}
\def\mn@doi@[#1]#2{\def\@tempa{#1}\ifx\@tempa\@empty \href
  {http://dx.doi.org/#2} {doi:#2}\else \href {http://dx.doi.org/#2} {#1}\fi
  \endgroup}
\def\mn@eprint#1#2{\mn@eprint@#1:#2::\@nil}
\def\mn@eprint@arXiv#1{\href {http://arxiv.org/abs/#1} {{\tt arXiv:#1}}}
\def\mn@eprint@dblp#1{\href {http://dblp.uni-trier.de/rec/bibtex/#1.xml}
  {dblp:#1}}
\def\mn@eprint@#1:#2:#3:#4\@nil{\def\@tempa {#1}\def\@tempb {#2}\def\@tempc
  {#3}\ifx \@tempc \@empty \let \@tempc \@tempb \let \@tempb \@tempa \fi \ifx
  \@tempb \@empty \def\@tempb {arXiv}\fi \@ifundefined
  {mn@eprint@\@tempb}{\@tempb:\@tempc}{\expandafter \expandafter \csname
  mn@eprint@\@tempb\endcsname \expandafter{\@tempc}}}

\bibitem[\protect\citeauthoryear{{Ackermann}, {Ajello}, {Allafort}
  et~al.}{{Ackermann} et~al.}{2013}]{Ackermann2013}
{Ackermann} M.,  {Ajello} M.,  {Allafort} A.,   et~al., 2013, \mn@doi [Science]
  {10.1126/science.1231160}, \href
  {https://ui.adsabs.harvard.edu/abs/2013Sci...339..807A/abstract} {339, 807}

\bibitem[\protect\citeauthoryear{{Aller}}{{Aller}}{1942}]{Aller1942}
{Aller} L.~H.,  1942, \mn@doi [\apj] {10.1086/144372}, \href
  {https://ui.adsabs.harvard.edu/abs/1942ApJ....95...52A} {95, 52}

\bibitem[\protect\citeauthoryear{{Aloisi}, {Tosi}  \& {Greggio}}{{Aloisi}
  et~al.}{1999}]{Aloisi1999}
{Aloisi} A.,  {Tosi} M.,   {Greggio} L.,  1999, \mn@doi [\aj] {10.1086/300924},
  \href {https://ui.adsabs.harvard.edu/abs/1999AJ....118..302A} {118, 302}

\bibitem[\protect\citeauthoryear{{Aloisi}, {Clementini}, {Tosi}
  et~al.}{{Aloisi} et~al.}{2007}]{Aloisi2007}
{Aloisi} A.,  {Clementini} G.,  {Tosi} M.,   et~al., 2007, \mn@doi [\apjl]
  {10.1086/522368}, \href
  {https://ui.adsabs.harvard.edu/abs/2007ApJ...667L.151A} {667, L151}

\bibitem[\protect\citeauthoryear{{Asplund}, {Grevesse}, {Sauval}  \&
  {Scott}}{{Asplund} et~al.}{2009}]{Asplund2009}
{Asplund} M.,  {Grevesse} N.,  {Sauval} A.~J.,   {Scott} P.,  2009, \mn@doi
  [\araa] {10.1146/annurev.astro.46.060407.145222}, \href
  {https://ui.adsabs.harvard.edu/abs/2009ARA&A..47..481A} {47, 481}

\bibitem[\protect\citeauthoryear{{Bakes} \& {Tielens}}{{Bakes} \&
  {Tielens}}{1994}]{Bakes1994}
{Bakes} E.~L.~O.,  {Tielens} A.~G.~G.~M.,  1994, \mn@doi [\apj]
  {10.1086/174188}, \href
  {https://ui.adsabs.harvard.edu/abs/1994ApJ...427..822B} {427, 822}

\bibitem[\protect\citeauthoryear{{Balser}, {Rood}, {Bania}  et~al.}{{Balser}
  et~al.}{2011}]{balser2011}
{Balser} D.~S.,  {Rood} R.~T.,  {Bania} T.~M.,   et~al., 2011, \mn@doi [\apj]
  {10.1088/0004-637X/738/1/27}, \href
  {https://ui.adsabs.harvard.edu/abs/2011ApJ...738...27B/abstract} {738, 27}

\bibitem[\protect\citeauthoryear{{Bate}}{{Bate}}{2019}]{Bate2019}
{Bate} M.~R.,  2019, \mn@doi [\mnras] {10.1093/mnras/stz103}, \href
  {https://ui.adsabs.harvard.edu/abs/2019MNRAS.484.2341B} {484, 2341}

\bibitem[\protect\citeauthoryear{{Bate}}{{Bate}}{2025}]{Bate2025}
{Bate} M.~R.,  2025, \mn@doi [\mnras] {10.1093/mnras/staf059}, \href
  {https://ui.adsabs.harvard.edu/abs/2025MNRAS.537..752B} {537, 752}

\bibitem[\protect\citeauthoryear{Bate, Bonnell  \& Price}{Bate
  et~al.}{1995}]{bate1995}
Bate M.~R.,  Bonnell I.~A.,   Price N.~M.,  1995, \mn@doi [MNRAS]
  {10.1093/mnras/277.2.362}, \href
  {https://ui.adsabs.harvard.edu/abs/1995MNRAS.277..362B/abstract} {277, 362}

\bibitem[\protect\citeauthoryear{{Baumgardt}, {Parmentier}  et~al.}{{Baumgardt}
  et~al.}{2013}]{Baumgardt2012}
{Baumgardt} H.,  {Parmentier} G.,   et~al., 2013, \mn@doi [\mnras]
  {10.1093/mnras/sts667}, \href
  {https://ui.adsabs.harvard.edu/abs/2013MNRAS.430..676B} {430, 676}

\bibitem[\protect\citeauthoryear{{Bekki}, {Beasley}, {Forbes}  \&
  {Couch}}{{Bekki} et~al.}{2004}]{Bekki2003}
{Bekki} K.,  {Beasley} M.~A.,  {Forbes} D.~A.,   {Couch} W.~J.,  2004, \mn@doi
  [\apj] {10.1086/381171}, \href
  {https://ui.adsabs.harvard.edu/abs/2004ApJ...602..730B} {602, 730}

\bibitem[\protect\citeauthoryear{{Belfiore} et~al.,}{{Belfiore}
  et~al.}{2017}]{Belfiore2017}
{Belfiore} F.,  et~al., 2017, \mn@doi [\mnras] {10.1093/mnras/stx789}, \href
  {https://ui.adsabs.harvard.edu/abs/2017MNRAS.469..151B} {469, 151}

\bibitem[\protect\citeauthoryear{{Bergin}, {Hartmann}  \& {et al.,}}{{Bergin}
  et~al.}{2004}]{Bergin2004}
{Bergin} E.~A.,  {Hartmann} L.~W.,   {et al.,} 2004, \mn@doi [\apj]
  {10.1086/422578}, \href
  {https://ui.adsabs.harvard.edu/abs/2004ApJ...612..921B} {612, 921}

\bibitem[\protect\citeauthoryear{{Bialy} \& {Sternberg}}{{Bialy} \&
  {Sternberg}}{2019}]{BialySternberg2019}
{Bialy} S.,  {Sternberg} A.,  2019, \mn@doi [\apj] {10.3847/1538-4357/ab2fd1},
  \href {https://ui.adsabs.harvard.edu/abs/2019ApJ...881..160B} {881, 160}

\bibitem[\protect\citeauthoryear{{Bigiel}, {Leroy}, {Walter}  et~al.}{{Bigiel}
  et~al.}{2008}]{Bigiel2008}
{Bigiel} F.,  {Leroy} A.,  {Walter} F.,   et~al., 2008, \mn@doi [\aj]
  {10.1088/0004-6256/136/6/2846}, \href
  {https://ui.adsabs.harvard.edu/abs/2008AJ....136.2846B} {136, 2846}

\bibitem[\protect\citeauthoryear{{Bisbas}, {Zhang}, {Gjergo}  et~al.}{{Bisbas}
  et~al.}{2024}]{Bisbas2024}
{Bisbas} T.~G.,  {Zhang} Z.-Y.,  {Gjergo} E.,   et~al., 2024, \mn@doi [\mnras]
  {10.1093/mnras/stad3782}, \href
  {https://ui.adsabs.harvard.edu/abs/2024MNRAS.527.8886B} {527, 8886}

\bibitem[\protect\citeauthoryear{{Bohlin}, {Savage}  \& {Drake}}{{Bohlin}
  et~al.}{1978}]{Bohlin1978}
{Bohlin} R.~C.,  {Savage} B.~D.,   {Drake} J.~F.,  1978, \mn@doi [\apj]
  {10.1086/156357}, \href
  {https://ui.adsabs.harvard.edu/abs/1978ApJ...224..132B} {224, 132}

\bibitem[\protect\citeauthoryear{{Bolatto}, {Leroy}, {Rosolowsky}
  et~al.}{{Bolatto} et~al.}{2008}]{Bolatto2008}
{Bolatto} A.~D.,  {Leroy} A.~K.,  {Rosolowsky} E.,   et~al., 2008, \mn@doi
  [\apj] {10.1086/591513}, \href
  {https://ui.adsabs.harvard.edu/abs/2008ApJ...686..948B} {686, 948}

\bibitem[\protect\citeauthoryear{{Bolatto}, {Leroy}, {Jameson}
  et~al.}{{Bolatto} et~al.}{2011}]{bolatto2011}
{Bolatto} A.~D.,  {Leroy} A.~K.,  {Jameson} K.,   et~al., 2011, \mn@doi [\apj]
  {10.1088/0004-637X/741/1/12}, \href
  {https://ui.adsabs.harvard.edu/abs/2011ApJ...741...12B/abstract} {741, 12}

\bibitem[\protect\citeauthoryear{{Bortolini}, {{\"O}stlin}, {Habel}
  et~al.}{{Bortolini} et~al.}{2024}]{Bortolini2024}
{Bortolini} G.,  {{\"O}stlin} G.,  {Habel} N.,   et~al., 2024, \mn@doi [\aap]
  {10.1051/0004-6361/202450632}, \href
  {https://ui.adsabs.harvard.edu/abs/2024A&A...689A.146B} {689, A146}

\bibitem[\protect\citeauthoryear{{Brott}, {de Mink}, {Cantiello}
  et~al.}{{Brott} et~al.}{2011}]{brott2011}
{Brott} I.,  {de Mink} S.~E.,  {Cantiello} M.,   et~al., 2011, \mn@doi [\aap]
  {10.1051/0004-6361/201016113}, \href
  {https://ui.adsabs.harvard.edu/abs/2011A%26A...530A.115B/abstract} {530,
  A115}

\bibitem[\protect\citeauthoryear{{Brugaletta}, {Walch}, {Naab}
  et~al.}{{Brugaletta} et~al.}{2025}]{Brugaletta2025}
{Brugaletta} V.,  {Walch} S.,  {Naab} T.,   et~al., 2025, \mn@doi [\mnras]
  {10.1093/mnras/staf039}, \href
  {https://ui.adsabs.harvard.edu/abs/2025MNRAS.537..482B} {537, 482}

\bibitem[\protect\citeauthoryear{{Cameron}, {Saxena}, {Bunker}
  et~al.}{{Cameron} et~al.}{2023}]{Cameron2023}
{Cameron} A.~J.,  {Saxena} A.,  {Bunker} A.~J.,   et~al., 2023, \mn@doi [\aap]
  {10.1051/0004-6361/202346107}, \href
  {https://ui.adsabs.harvard.edu/abs/2023A%26A...677A.115C/abstract} {677,
  A115}

\bibitem[\protect\citeauthoryear{Chevance, Madden, Lebouteiller
  et~al.}{Chevance et~al.}{2016}]{Chevance2016}
Chevance M.,  Madden S.~C.,  Lebouteiller V.,   et~al., 2016, \aap, \href
  {https://ui.adsabs.harvard.edu/abs/2016A%26A...590A..36C/abstract} {590, A36}

\bibitem[\protect\citeauthoryear{{Clark}, {Glover}  \& {Klessen}}{{Clark}
  et~al.}{2012}]{Clark2012}
{Clark} P.~C.,  {Glover} S. C.~O.,   {Klessen} R.~S.,  2012, \mn@doi [\mnras]
  {10.1111/j.1365-2966.2011.20087.x}, \href
  {https://ui.adsabs.harvard.edu/abs/2012MNRAS.420..745C} {420, 745}

\bibitem[\protect\citeauthoryear{{Collette}, {Kluyver}, {Caswell}
  et~al.}{{Collette} et~al.}{2020}]{Collette2020}
{Collette} A.,  {Kluyver} T.,  {Caswell} T.~A.,   et~al., 2020, h5py/h5py:
  3.1.0, \mn@doi{10.5281/zenodo.4250762}

\bibitem[\protect\citeauthoryear{{Contreras Ramos}, {Annibali}, {Fiorentino}
  et~al.}{{Contreras Ramos} et~al.}{2011}]{ContrerasRamos2011}
{Contreras Ramos} R.,  {Annibali} F.,  {Fiorentino} G.,   et~al., 2011, \mn@doi
  [\apj] {10.1088/0004-637X/739/2/74}, \href
  {https://ui.adsabs.harvard.edu/abs/2011ApJ...739...74C/abstract} {739, 74}

\bibitem[\protect\citeauthoryear{{Corbett Moran}, {Grudi{\'c}}  \&
  {Hopkins}}{{Corbett Moran} et~al.}{2018}]{Moran2018}
{Corbett Moran} C.,  {Grudi{\'c}} M.~Y.,   {Hopkins} P.~F.,  2018, \mn@doi
  [arXiv e-prints] {10.48550/arXiv.1803.06430}, \href
  {https://ui.adsabs.harvard.edu/abs/2018arXiv180306430C} {p. arXiv:1803.06430}

\bibitem[\protect\citeauthoryear{{Curti}, {D'Eugenio}, {Carniani}
  et~al.}{{Curti} et~al.}{2023}]{Curti2023}
{Curti} M.,  {D'Eugenio} F.,  {Carniani} S.,   et~al., 2023, \mn@doi [\mnras]
  {10.1093/mnras/stac2737}, \href
  {https://ui.adsabs.harvard.edu/abs/2023MNRAS.518..425C} {518, 425}

\bibitem[\protect\citeauthoryear{{Dalgarno} \& {McCray}}{{Dalgarno} \&
  {McCray}}{1972}]{Dalgarno1972}
{Dalgarno} A.,  {McCray} R.~A.,  1972, \mn@doi [\araa]
  {10.1146/annurev.aa.10.090172.002111}, \href
  {https://ui.adsabs.harvard.edu/abs/1972ARA&A..10..375D} {10, 375}

\bibitem[\protect\citeauthoryear{{Dinnbier} \& {Walch}}{{Dinnbier} \&
  {Walch}}{2020}]{Dinnbier2020}
{Dinnbier} F.,  {Walch} S.,  2020, \mn@doi [\mnras] {10.1093/mnras/staa2560},
  \href {https://ui.adsabs.harvard.edu/abs/2020MNRAS.499..748D/abstract} {499,
  748}

\bibitem[\protect\citeauthoryear{{Doan} et~al.}{{Doan} et~al.}{2024}]{Doan2024}
{Doan} S.,  et~al., 2024, \mn@doi [arXiv e-prints] {10.48550/arXiv.2408.04774},
  \href {https://ui.adsabs.harvard.edu/abs/2024arXiv240804774D} {p.
  arXiv:2408.04774}

\bibitem[\protect\citeauthoryear{{Draine}}{{Draine}}{2011}]{Drainebook2011}
{Draine} B.~T.,  2011, {Physics of the Interstellar and Intergalactic Medium}

\bibitem[\protect\citeauthoryear{{Draine}, {Dale}, {Bendo}  et~al.}{{Draine}
  et~al.}{2007}]{Draine2007}
{Draine} B.~T.,  {Dale} D.~A.,  {Bendo} G.,   et~al., 2007, \mn@doi [\apj]
  {10.1086/518306}, \href
  {https://ui.adsabs.harvard.edu/abs/2007ApJ...663..866D} {663, 866}

\bibitem[\protect\citeauthoryear{Dubey, Reid  \& Fisher}{Dubey
  et~al.}{2008}]{Dubey_2008}
Dubey A.,  Reid L.~B.,   Fisher R.,  2008, \mn@doi [Physica Scripta]
  {10.1088/0031-8949/2008/t132/014046}, T132, 014046

\bibitem[\protect\citeauthoryear{{Dubey}, {Reid}  \& {et al., }}{{Dubey}
  et~al.}{2009}]{dubey2009}
{Dubey} A.,  {Reid} L.~B.,   {et al., } 2009, arXiv e-prints, p.
  arXiv:0903.4875

\bibitem[\protect\citeauthoryear{{Dufour}, {Esteban}  \& {Castaneda}}{{Dufour}
  et~al.}{1996}]{Dufour1996}
{Dufour} R.~J.,  {Esteban} C.,   {Castaneda} H.~O.,  1996, \mn@doi [\apjl]
  {10.1086/310341}, \href
  {https://ui.adsabs.harvard.edu/abs/1996ApJ...471L..87D/abstract} {471, L87}

\bibitem[\protect\citeauthoryear{{Ekstr{\"o}m} et~al.}{{Ekstr{\"o}m}
  et~al.}{2011}]{ekstrom2011}
{Ekstr{\"o}m} S.,  et~al., 2011, in {Neiner} C.,  {Wade} G.,  {Meynet} G.,
  {Peters} G.,  eds,  IAU Symposium Vol. 272, Active OB Stars: Structure,
  Evolution, Mass Loss, and Critical Limits. pp 62--72 (\mn@eprint {arXiv}
  {1010.3838}), \mn@doi{10.1017/S1743921311009987}

\bibitem[\protect\citeauthoryear{{Ekstr{\"o}m}, {Georgy}  \& {et al.,
  }}{{Ekstr{\"o}m} et~al.}{2012}]{ekstrom2012}
{Ekstr{\"o}m} S.,  {Georgy} C.,   {et al., } 2012, \mn@doi [\aap]
  {10.1051/0004-6361/201117751}, \href
  {https://ui.adsabs.harvard.edu/abs/2012A%26A...537A.146E/abstract} {537,
  A146}

\bibitem[\protect\citeauthoryear{{Elmegreen}}{{Elmegreen}}{1989}]{Elmegreen1989}
{Elmegreen} B.~G.,  1989, \mn@doi [\apj] {10.1086/167192}, \href
  {https://ui.adsabs.harvard.edu/abs/1989ApJ...338..178E} {338, 178}

\bibitem[\protect\citeauthoryear{Federrath, Banerjee, Clark  et~al.}{Federrath
  et~al.}{2010}]{Federrath_2010}
Federrath C.,  Banerjee R.,  Clark P.~C.,   et~al., 2010, \mn@doi [ApJ]
  {10.1088/0004-637x/713/1/269}, \href
  {https://ui.adsabs.harvard.edu/abs/2010ApJ...713..269F/abstract} {713, 269}

\bibitem[\protect\citeauthoryear{{Feldmann}}{{Feldmann}}{2015}]{Feldmann2015}
{Feldmann} R.,  2015, \mn@doi [\mnras] {10.1093/mnras/stv552}, \href
  {https://ui.adsabs.harvard.edu/abs/2015MNRAS.449.3274F/abstract} {449, 3274}

\bibitem[\protect\citeauthoryear{{Field}, {Goldsmith}  \& {Habing}}{{Field}
  et~al.}{1969}]{Field1969}
{Field} G.~B.,  {Goldsmith} D.~W.,   {Habing} H.~J.,  1969, \mn@doi [\apjl]
  {10.1086/180324}, \href
  {https://ui.adsabs.harvard.edu/abs/1969ApJ...155L.149F} {155, L149}

\bibitem[\protect\citeauthoryear{Fryxell, Olson, Ricker  et~al.}{Fryxell
  et~al.}{2000}]{Fryxell_2000}
Fryxell B.,  Olson K.,  Ricker P.,   et~al., 2000, \mn@doi [ApJS]
  {10.1086/317361}, \href
  {https://ui.adsabs.harvard.edu/abs/2000ApJS..131..273F/abstract} {131, 273}

\bibitem[\protect\citeauthoryear{Fujii, Minamidani, Mizuno  et~al.}{Fujii
  et~al.}{2014}]{Fujii2014}
Fujii K.,  Minamidani T.,  Mizuno N.,   et~al., 2014, \apj, \href
  {https://ui.adsabs.harvard.edu/abs/2014ApJ...796..123F/abstract} {796, 123}

\bibitem[\protect\citeauthoryear{{Fukushima}, {Yajima}, {Sugimura}
  et~al.}{{Fukushima} et~al.}{2020}]{Fukushima2020}
{Fukushima} H.,  {Yajima} H.,  {Sugimura} K.,   et~al., 2020, \mn@doi [\mnras]
  {10.1093/mnras/staa2062}, \href
  {https://ui.adsabs.harvard.edu/abs/2020MNRAS.497.3830F} {497, 3830}

\bibitem[\protect\citeauthoryear{{Garnett}, {Skillman}, {Dufour}
  et~al.}{{Garnett} et~al.}{1997}]{Garnett1997}
{Garnett} D.~R.,  {Skillman} E.~D.,  {Dufour} R.~J.,   et~al., 1997, \mn@doi
  [\apj] {10.1086/304058}, \href
  {https://ui.adsabs.harvard.edu/abs/1997ApJ...481..174G} {481, 174}

\bibitem[\protect\citeauthoryear{{Gatto}, {Walch}  \& {et al., }}{{Gatto}
  et~al.}{2015}]{gatto2015}
{Gatto} A.,  {Walch} S.,   {et al., } 2015, \mn@doi [\mnras]
  {10.1093/mnras/stv324}, \href
  {https://ui.adsabs.harvard.edu/abs/2015MNRAS.449.1057G/abstract} {449, 1057}

\bibitem[\protect\citeauthoryear{{Gatto}, {Walch}, {Naab}  et~al.}{{Gatto}
  et~al.}{2017}]{silcc3}
{Gatto} A.,  {Walch} S.,  {Naab} T.,   et~al., 2017, \mn@doi [\mnras]
  {10.1093/mnras/stw3209}, \href
  {https://ui.adsabs.harvard.edu/abs/2017MNRAS.466.1903G/abstract} {466, 1903}

\bibitem[\protect\citeauthoryear{{Girichidis}}{{Girichidis}}{2021}]{Girichidis2021}
{Girichidis} P.,  2021, \mn@doi [\mnras] {10.1093/mnras/stab2157}, \href
  {https://ui.adsabs.harvard.edu/abs/2021MNRAS.507.5641G} {507, 5641}

\bibitem[\protect\citeauthoryear{{Girichidis}, {Walch}  \& {et al.,
  }}{{Girichidis} et~al.}{2016a}]{silcc2}
{Girichidis} P.,  {Walch} S.,   {et al., } 2016a, \mn@doi [\mnras]
  {10.1093/mnras/stv2742}, \href
  {https://ui.adsabs.harvard.edu/abs/2016MNRAS.456.3432G/abstract} {456, 3432}

\bibitem[\protect\citeauthoryear{{Girichidis}, {Naab}, {Walch}
  et~al.}{{Girichidis} et~al.}{2016b}]{GirichidisEtAl2016a}
{Girichidis} P.,  {Naab} T.,  {Walch} S.,   et~al., 2016b, \mn@doi [\apjl]
  {10.3847/2041-8205/816/2/L19}, \href
  {http://adsabs.harvard.edu/abs/2016ApJ...816L..19G} {816, L19}

\bibitem[\protect\citeauthoryear{{Girichidis}, {Naab}, {Hanasz}
  et~al.}{{Girichidis} et~al.}{2018a}]{GirichidisEtAl2018a}
{Girichidis} P.,  {Naab} T.,  {Hanasz} M.,   et~al., 2018a, \mn@doi [\mnras]
  {10.1093/mnras/sty1653}, \href
  {http://adsabs.harvard.edu/abs/2018MNRAS.479.3042G} {479, 3042}

\bibitem[\protect\citeauthoryear{{Girichidis}, {Seifried}  \& {et al.,
  }}{{Girichidis} et~al.}{2018b}]{silcc5}
{Girichidis} P.,  {Seifried} D.,   {et al., } 2018b, \mn@doi [\mnras]
  {10.1093/mnras/sty2016}, \href
  {https://ui.adsabs.harvard.edu/abs/2018MNRAS.480.3511G/abstract} {480, 3511}

\bibitem[\protect\citeauthoryear{{Girichidis}, {Pfrommer}, {Hanasz}
  et~al.}{{Girichidis} et~al.}{2020}]{Girichidis2020}
{Girichidis} P.,  {Pfrommer} C.,  {Hanasz} M.,   et~al., 2020, \mn@doi [\mnras]
  {10.1093/mnras/stz2961}, \href
  {https://ui.adsabs.harvard.edu/abs/2020MNRAS.491..993G} {491, 993}

\bibitem[\protect\citeauthoryear{{Glassgold} \& {Langer}}{{Glassgold} \&
  {Langer}}{1974}]{Glassgold1974}
{Glassgold} A.~E.,  {Langer} W.~D.,  1974, \mn@doi [\apj] {10.1086/153130},
  \href {https://ui.adsabs.harvard.edu/abs/1974ApJ...193...73G} {193, 73}

\bibitem[\protect\citeauthoryear{{Glover}}{{Glover}}{2003}]{Glover2003}
{Glover} S. C.~O.,  2003, \mn@doi [\apj] {10.1086/345684}, \href
  {https://ui.adsabs.harvard.edu/abs/2003ApJ...584..331G} {584, 331}

\bibitem[\protect\citeauthoryear{{Glover} \& {Clark}}{{Glover} \&
  {Clark}}{2012a}]{Glover2012a}
{Glover} S. C.~O.,  {Clark} P.~C.,  2012a, \mn@doi [\mnras]
  {10.1111/j.1365-2966.2011.19648.x}, \href
  {https://ui.adsabs.harvard.edu/abs/2012MNRAS.421....9G} {421, 9}

\bibitem[\protect\citeauthoryear{{Glover} \& {Clark}}{{Glover} \&
  {Clark}}{2012b}]{Glover2012b}
{Glover} S. C.~O.,  {Clark} P.~C.,  2012b, \mn@doi [\mnras]
  {10.1111/j.1365-2966.2011.20260.x}, \href
  {https://ui.adsabs.harvard.edu/abs/2012MNRAS.421..116G} {421, 116}

\bibitem[\protect\citeauthoryear{{Glover} \& {Clark}}{{Glover} \&
  {Clark}}{2012c}]{Glover2012c}
{Glover} S. C.~O.,  {Clark} P.~C.,  2012c, \mn@doi [\mnras]
  {10.1111/j.1365-2966.2012.21737.x}, \href
  {https://ui.adsabs.harvard.edu/abs/2012MNRAS.426..377G} {426, 377}

\bibitem[\protect\citeauthoryear{{Glover} \& {Clark}}{{Glover} \&
  {Clark}}{2014}]{Glover2014}
{Glover} S. C.~O.,  {Clark} P.~C.,  2014, \mn@doi [\mnras]
  {10.1093/mnras/stt1809}, \href
  {https://ui.adsabs.harvard.edu/abs/2014MNRAS.437....9G} {437, 9}

\bibitem[\protect\citeauthoryear{{Glover} \& {Mac Low}}{{Glover} \& {Mac
  Low}}{2007a}]{glover2007a}
{Glover} S. C.~O.,  {Mac Low} M.-M.,  2007a, \mn@doi [\apjs] {10.1086/512238},
  \href {https://ui.adsabs.harvard.edu/abs/2007ApJS..169..239G/abstract} {169,
  239}

\bibitem[\protect\citeauthoryear{{Glover} \& {Mac Low}}{{Glover} \& {Mac
  Low}}{2007b}]{Glover_2007b}
{Glover} S. C.~O.,  {Mac Low} M.-M.,  2007b, \mn@doi [ApJ] {10.1086/512227},
  659, 1317

\bibitem[\protect\citeauthoryear{{Gnat} \& {Ferland}}{{Gnat} \&
  {Ferland}}{2012}]{Gnat2012}
{Gnat} O.,  {Ferland} G.~J.,  2012, \mn@doi [\apjs]
  {10.1088/0067-0049/199/1/20}, \href
  {https://ui.adsabs.harvard.edu/abs/2012ApJS..199...20G/abstract} {199, 20}

\bibitem[\protect\citeauthoryear{{Goldman}}{{Goldman}}{2007}]{Goldman2007}
{Goldman} I.,  2007, \mn@doi [arXiv e-prints]
  {10.48550/arXiv.astro-ph/0703793}, pp astro--ph/0703793

\bibitem[\protect\citeauthoryear{{Goldsmith} \& {Langer}}{{Goldsmith} \&
  {Langer}}{1978}]{goldsmith1978}
{Goldsmith} P.~F.,  {Langer} W.~D.,  1978, \mn@doi [\apj] {10.1086/156206},
  \href {https://ui.adsabs.harvard.edu/abs/1978ApJ...222..881G} {222, 881}

\bibitem[\protect\citeauthoryear{{G{\'o}rski}, {Hivon}  \& {et al.,
  }}{{G{\'o}rski} et~al.}{2005}]{gorski2005}
{G{\'o}rski} K.~M.,  {Hivon} E.,   {et al., } 2005, \mn@doi [\apj]
  {10.1086/427976}, \href
  {https://ui.adsabs.harvard.edu/abs/2005ApJ...622..759G} {622, 759}

\bibitem[\protect\citeauthoryear{{Grishunin}, {Weiss}, {Colombo}
  et~al.}{{Grishunin} et~al.}{2024}]{Grishunin2024}
{Grishunin} K.,  {Weiss} A.,  {Colombo} D.,   et~al., 2024, \mn@doi [\aap]
  {10.1051/0004-6361/202347364}, \href
  {https://ui.adsabs.harvard.edu/abs/2024A&A...682A.137G} {682, A137}

\bibitem[\protect\citeauthoryear{{Guadarrama}, {Vorobyov}, {Rab}
  et~al.}{{Guadarrama} et~al.}{2022}]{Guadarrama2022}
{Guadarrama} R.,  {Vorobyov} E.~I.,  {Rab} C.,   et~al., 2022, \mn@doi [\aap]
  {10.1051/0004-6361/202140995}, \href
  {https://ui.adsabs.harvard.edu/abs/2022A&A...667A..28G} {667, A28}

\bibitem[\protect\citeauthoryear{{Habing}}{{Habing}}{1968}]{habing1968}
{Habing} H.~J.,  1968, \bain, \href
  {https://ui.adsabs.harvard.edu/abs/1968BAN....19..421H/abstract} {19, 421}

\bibitem[\protect\citeauthoryear{{Haid}, {Walch}, {Seifried}  et~al.}{{Haid}
  et~al.}{2018}]{Haid2018}
{Haid} S.,  {Walch} S.,  {Seifried} D.,   et~al., 2018, \mn@doi [\mnras]
  {10.1093/mnras/sty1315}, \href
  {https://ui.adsabs.harvard.edu/abs/2018MNRAS.478.4799H} {478, 4799}

\bibitem[\protect\citeauthoryear{Heger, Fryer  \& {et al., }}{Heger
  et~al.}{2003}]{Heger_2003}
Heger A.,  Fryer C.~L.,   {et al., } 2003, \mn@doi [ApJ] {10.1086/375341},
  \href {https://ui.adsabs.harvard.edu/abs/2003ApJ...591..288H/abstract} {591,
  288}

\bibitem[\protect\citeauthoryear{{Heintz}, {Gim{\'e}nez-Arteaga}, {Fujimoto}
  et~al.}{{Heintz} et~al.}{2023}]{Heintz2023}
{Heintz} K.~E.,  {Gim{\'e}nez-Arteaga} C.,  {Fujimoto} S.,   et~al., 2023,
  \mn@doi [\apjl] {10.3847/2041-8213/acb2cf}, \href
  {https://ui.adsabs.harvard.edu/abs/2023ApJ...944L..30H/abstract} {944, L30}

\bibitem[\protect\citeauthoryear{{Hollenbach} \& {McKee}}{{Hollenbach} \&
  {McKee}}{1989}]{Hollenbach1989}
{Hollenbach} D.,  {McKee} C.~F.,  1989, \mn@doi [\apj] {10.1086/167595}, \href
  {https://ui.adsabs.harvard.edu/abs/1989ApJ...342..306H} {342, 306}

\bibitem[\protect\citeauthoryear{{Hu}, {Naab}, {Walch}  et~al.}{{Hu}
  et~al.}{2016}]{Hu2016}
{Hu} C.-Y.,  {Naab} T.,  {Walch} S.,   et~al., 2016, \mn@doi [\mnras]
  {10.1093/mnras/stw544}, \href
  {https://ui.adsabs.harvard.edu/abs/2016MNRAS.458.3528H} {458, 3528}

\bibitem[\protect\citeauthoryear{{Hu}, {Naab}, {Glover}  et~al.}{{Hu}
  et~al.}{2017}]{Hu2017}
{Hu} C.-Y.,  {Naab} T.,  {Glover} S. C.~O.,   et~al., 2017, \mn@doi [\mnras]
  {10.1093/mnras/stx1773}, \href
  {https://ui.adsabs.harvard.edu/abs/2017MNRAS.471.2151H} {471, 2151}

\bibitem[\protect\citeauthoryear{{Hughes}, {Wong}, {Ott}  et~al.}{{Hughes}
  et~al.}{2010}]{Hughes2010}
{Hughes} A.,  {Wong} T.,  {Ott} J.,   et~al., 2010, \mn@doi [\mnras]
  {10.1111/j.1365-2966.2010.16829.x}, \href
  {https://ui.adsabs.harvard.edu/abs/2010MNRAS.406.2065H} {406, 2065}

\bibitem[\protect\citeauthoryear{Hunter}{Hunter}{2007}]{Hunter2007}
Hunter J.~D.,  2007, \mn@doi [Computing in Science \& Engineering]
  {10.1109/MCSE.2007.55}, \href
  {https://ui.adsabs.harvard.edu/abs/2007CSE.....9...90H/abstract} {9, 90}

\bibitem[\protect\citeauthoryear{{Issa}, {MacLaren}  \& {Wolfendale}}{{Issa}
  et~al.}{1990}]{Issa1990}
{Issa} M.~R.,  {MacLaren} I.,   {Wolfendale} A.~W.,  1990, \aap, \href
  {https://ui.adsabs.harvard.edu/abs/1990A&A...236..237I} {236, 237}

\bibitem[\protect\citeauthoryear{{Jameson} et~al.,}{{Jameson}
  et~al.}{2016}]{Jameson2016}
{Jameson} K.~E.,  et~al., 2016, \mn@doi [\apj] {10.3847/0004-637X/825/1/12},
  \href {https://ui.adsabs.harvard.edu/abs/2016ApJ...825...12J} {825, 12}

\bibitem[\protect\citeauthoryear{Jameson, Bolatto, Wolfire  et~al.}{Jameson
  et~al.}{2018}]{Jameson2018}
Jameson K.~E.,  Bolatto A.~D.,  Wolfire M.,   et~al., 2018, \apj, \href
  {https://ui.adsabs.harvard.edu/abs/2018ApJ...853..111J/abstract} {853, 111}

\bibitem[\protect\citeauthoryear{{Je{\v{r}}{\'a}bkov{\'a}}, {Hasani Zonoozi}
  \& {et al., }}{{Je{\v{r}}{\'a}bkov{\'a}} et~al.}{2018}]{jerabkov2018}
{Je{\v{r}}{\'a}bkov{\'a}} T.,  {Hasani Zonoozi} A.,   {et al., } 2018, \mn@doi
  [\aap] {10.1051/0004-6361/201833055}, \href
  {https://ui.adsabs.harvard.edu/abs/2018A%26A...620A..39J/abstract} {620, A39}

\bibitem[\protect\citeauthoryear{{Joshi}, {Walch}, {Seifried}  et~al.}{{Joshi}
  et~al.}{2019}]{Joshi2019}
{Joshi} P.~R.,  {Walch} S.,  {Seifried} D.,   et~al., 2019, \mn@doi [\mnras]
  {10.1093/mnras/stz052}, \href
  {https://ui.adsabs.harvard.edu/abs/2019MNRAS.484.1735J} {484, 1735}

\bibitem[\protect\citeauthoryear{{Ju}, {Wang}, {Jones}  et~al.}{{Ju}
  et~al.}{2025}]{Ju2025}
{Ju} M.,  {Wang} X.,  {Jones} T.,   et~al., 2025, \mn@doi [\apjl]
  {10.3847/2041-8213/ada150}, \href
  {https://ui.adsabs.harvard.edu/abs/2025ApJ...978L..39J} {978, L39}

\bibitem[\protect\citeauthoryear{{Kennicutt}, {Calzetti}, {Walter}
  et~al.}{{Kennicutt} et~al.}{2007}]{Kennicutt2007}
{Kennicutt} Jr. R.~C.,  {Calzetti} D.,  {Walter} F.,   et~al., 2007, \mn@doi
  [\apj] {10.1086/522300}, \href
  {https://ui.adsabs.harvard.edu/abs/2007ApJ...671..333K} {671, 333}

\bibitem[\protect\citeauthoryear{{Kim} \& {Ostriker}}{{Kim} \&
  {Ostriker}}{2015}]{kim2015}
{Kim} C.-G.,  {Ostriker} E.~C.,  2015, \mn@doi [\apj]
  {10.1088/0004-637X/802/2/99}, \href
  {https://ui.adsabs.harvard.edu/abs/2015ApJ...802...99K} {802, 99}

\bibitem[\protect\citeauthoryear{{Kim}, {Gong}, {Kim}  et~al.}{{Kim}
  et~al.}{2023}]{KimJG2023}
{Kim} J.-G.,  {Gong} M.,  {Kim} C.-G.,   et~al., 2023, \mn@doi [\apjs]
  {10.3847/1538-4365/ac9b1d}, \href
  {https://ui.adsabs.harvard.edu/abs/2023ApJS..264...10K/abstract} {264, 10}

\bibitem[\protect\citeauthoryear{{Kim} et~al.,}{{Kim} et~al.}{2024}]{Kim2024}
{Kim} C.-G.,  et~al., 2024, \mn@doi [\apj] {10.3847/1538-4357/ad59ab}, \href
  {https://ui.adsabs.harvard.edu/abs/2024ApJ...972...67K} {972, 67}

\bibitem[\protect\citeauthoryear{{Kippenhahn}, {Weigert}  \&
  {Weiss}}{{Kippenhahn} et~al.}{2012}]{Kippenhahnbook}
{Kippenhahn} R.,  {Weigert} A.,   {Weiss} A.,  2012, {Stellar Structure and
  Evolution}, \mn@doi{10.1007/978-3-642-30304-3.
}

\bibitem[\protect\citeauthoryear{{Krumholz}}{{Krumholz}}{2012}]{Krumholz2012}
{Krumholz} M.~R.,  2012, \mn@doi [\apj] {10.1088/0004-637X/759/1/9}, \href
  {https://ui.adsabs.harvard.edu/abs/2012ApJ...759....9K} {759, 9}

\bibitem[\protect\citeauthoryear{{Krumholz}, {McKee}  \&
  {Tumlinson}}{{Krumholz} et~al.}{2009}]{Krumholz2009}
{Krumholz} M.~R.,  {McKee} C.~F.,   {Tumlinson} J.,  2009, \mn@doi [\apj]
  {10.1088/0004-637X/693/1/216}, \href
  {https://ui.adsabs.harvard.edu/abs/2009ApJ...693..216K} {693, 216}

\bibitem[\protect\citeauthoryear{{Krumholz}, {Leroy}  \& {McKee}}{{Krumholz}
  et~al.}{2011}]{Krumholz2011}
{Krumholz} M.~R.,  {Leroy} A.~K.,   {McKee} C.~F.,  2011, \mn@doi [\apj]
  {10.1088/0004-637X/731/1/25}, \href
  {https://ui.adsabs.harvard.edu/abs/2011ApJ...731...25K} {731, 25}

\bibitem[\protect\citeauthoryear{{Lah{\'e}n}, {Naab}, {Johansson}
  et~al.}{{Lah{\'e}n} et~al.}{2020}]{Lahen2020}
{Lah{\'e}n} N.,  {Naab} T.,  {Johansson} P.~H.,   et~al., 2020, \mn@doi [\apj]
  {10.3847/1538-4357/ab7190}, \href
  {https://ui.adsabs.harvard.edu/abs/2020ApJ...891....2L} {891, 2}

\bibitem[\protect\citeauthoryear{{Lah{\'e}n}, {Naab}, {Kauffmann}
  et~al.}{{Lah{\'e}n} et~al.}{2023}]{Lahen2023}
{Lah{\'e}n} N.,  {Naab} T.,  {Kauffmann} G.,   et~al., 2023, \mn@doi [\mnras]
  {10.1093/mnras/stad1147}, \href
  {https://ui.adsabs.harvard.edu/abs/2023MNRAS.522.3092L} {522, 3092}

\bibitem[\protect\citeauthoryear{{Langer} \& {Maeder}}{{Langer} \&
  {Maeder}}{1995}]{langer1995}
{Langer} N.,  {Maeder} A.,  1995, \aap, \href
  {https://ui.adsabs.harvard.edu/abs/1995A%26A...295..685L/abstract} {295, 685}

\bibitem[\protect\citeauthoryear{{Legrand}}{{Legrand}}{2000}]{Legrand2000}
{Legrand} F.,  2000, \aap, \href
  {https://ui.adsabs.harvard.edu/abs/2000A%26A...355..891L/abstract} {354, 504}

\bibitem[\protect\citeauthoryear{{Legrand}, {Tenorio-Tagle}, {Silich}
  et~al.}{{Legrand} et~al.}{2001}]{Legrand2001}
{Legrand} F.,  {Tenorio-Tagle} G.,  {Silich} S.,   et~al., 2001, \mn@doi [\apj]
  {10.1086/322960}, \href
  {https://ui.adsabs.harvard.edu/abs/2001ApJ...560..630L} {560, 630}

\bibitem[\protect\citeauthoryear{{Leitherer}, {Robert}  \&
  {Drissen}}{{Leitherer} et~al.}{1992}]{Leitherer1992}
{Leitherer} C.,  {Robert} C.,   {Drissen} L.,  1992, \mn@doi [\apj]
  {10.1086/172089}, \href
  {https://ui.adsabs.harvard.edu/abs/1992ApJ...401..596L/abstract} {401, 596}

\bibitem[\protect\citeauthoryear{{Lelli}, {Verheijen}, {Fraternali}
  et~al.}{{Lelli} et~al.}{2012}]{Lelli2012}
{Lelli} F.,  {Verheijen} M.,  {Fraternali} F.,   et~al., 2012, \mn@doi [\aap]
  {10.1051/0004-6361/201117867}, \href
  {https://ui.adsabs.harvard.edu/abs/2012A%26A...537A..72L/abstract} {537, A72}

\bibitem[\protect\citeauthoryear{{Leroy}, {Walter}, {Brinks}  et~al.}{{Leroy}
  et~al.}{2008}]{Leroy2008}
{Leroy} A.~K.,  {Walter} F.,  {Brinks} E.,   et~al., 2008, \mn@doi [\aj]
  {10.1088/0004-6256/136/6/2782}, 136, 2782

\bibitem[\protect\citeauthoryear{{Leroy}, {Bolatto}, {Bot}  et~al.}{{Leroy}
  et~al.}{2009}]{leroy2009}
{Leroy} A.~K.,  {Bolatto} A.,  {Bot} C.,   et~al., 2009, \mn@doi [\apj]
  {10.1088/0004-637X/702/1/352}, 702, 352

\bibitem[\protect\citeauthoryear{{Lian}, {Bergemann}, {Pillepich}
  et~al.}{{Lian} et~al.}{2023}]{Lian2023}
{Lian} J.,  {Bergemann} M.,  {Pillepich} A.,   et~al., 2023, \mn@doi [Nature
  Astronomy] {10.1038/s41550-023-01977-z}, \href
  {https://ui.adsabs.harvard.edu/abs/2023NatAs...7..951L} {7, 951}

\bibitem[\protect\citeauthoryear{{Maeder} \& {Meynet}}{{Maeder} \&
  {Meynet}}{2000}]{maeder2000}
{Maeder} A.,  {Meynet} G.,  2000, \mn@doi [\araa]
  {10.1146/annurev.astro.38.1.143}, 38, 143

\bibitem[\protect\citeauthoryear{{Maloney} \& {Black}}{{Maloney} \&
  {Black}}{1988}]{Maloney1988}
{Maloney} P.,  {Black} J.~H.,  1988, \mn@doi [\apj] {10.1086/166011}, \href
  {https://ui.adsabs.harvard.edu/abs/1988ApJ...325..389M} {325, 389}

\bibitem[\protect\citeauthoryear{{McKee}}{{McKee}}{1989}]{McKee1989}
{McKee} C.~F.,  1989, \mn@doi [\apj] {10.1086/167950}, \href
  {https://ui.adsabs.harvard.edu/abs/1989ApJ...345..782M} {345, 782}

\bibitem[\protect\citeauthoryear{{M{\'e}ndez-Delgado}, {Amayo}  \& {et al.,
  }}{{M{\'e}ndez-Delgado} et~al.}{2022}]{mendez2022}
{M{\'e}ndez-Delgado} J.~E.,  {Amayo} A.,   {et al., } 2022, \mn@doi [\mnras]
  {10.1093/mnras/stab3782}, 510, 4436

\bibitem[\protect\citeauthoryear{Meschin, Gallart  \& {et al., }}{Meschin
  et~al.}{2013}]{Meschin2013}
Meschin I.,  Gallart C.,   {et al., } 2013, MNRAS, 438, 1067

\bibitem[\protect\citeauthoryear{{Mingozzi}, {Belfiore}, {Cresci}
  et~al.}{{Mingozzi} et~al.}{2020}]{Mingozzi2020}
{Mingozzi} M.,  {Belfiore} F.,  {Cresci} G.,   et~al., 2020, \mn@doi [\aap]
  {10.1051/0004-6361/201937203}, \href
  {https://ui.adsabs.harvard.edu/abs/2020A&A...636A..42M} {636, A42}

\bibitem[\protect\citeauthoryear{{Murray}, {Peek}, {Di Teodoro}
  et~al.}{{Murray} et~al.}{2019}]{Murray2019}
{Murray} C.~E.,  {Peek} J.~E.~G.,  {Di Teodoro} E.~M.,   et~al., 2019, \mn@doi
  [\apj] {10.3847/1538-4357/ab510f}, \href
  {https://ui.adsabs.harvard.edu/abs/2019ApJ...887..267M} {887, 267}

\bibitem[\protect\citeauthoryear{{Musella}, {Marconi}, {Fiorentino}
  et~al.}{{Musella} et~al.}{2012}]{Musella2012}
{Musella} I.,  {Marconi} M.,  {Fiorentino} G.,   et~al., 2012, Memorie della
  Societa Astronomica Italiana Supplementi, 19, 146

\bibitem[\protect\citeauthoryear{{Naab} \& {Ostriker}}{{Naab} \&
  {Ostriker}}{2017}]{Naab2017}
{Naab} T.,  {Ostriker} J.~P.,  2017, \mn@doi [\araa]
  {10.1146/annurev-astro-081913-040019}, \href
  {https://ui.adsabs.harvard.edu/abs/2017ARA&A..55...59N} {55, 59}

\bibitem[\protect\citeauthoryear{{Nava} \& {Gabici}}{{Nava} \&
  {Gabici}}{2013}]{Nava2013}
{Nava} L.,  {Gabici} S.,  2013, \mn@doi [\mnras] {10.1093/mnras/sts450}, \href
  {https://ui.adsabs.harvard.edu/abs/2013MNRAS.429.1643N} {429, 1643}

\bibitem[\protect\citeauthoryear{{Navarro}, {Frenk}  \& {White}}{{Navarro}
  et~al.}{1996}]{NFW1996}
{Navarro} J.~F.,  {Frenk} C.~S.,   {White} S. D.~M.,  1996, \mn@doi [\apj]
  {10.1086/177173}, \href
  {https://ui.adsabs.harvard.edu/abs/1996ApJ...462..563N} {462, 563}

\bibitem[\protect\citeauthoryear{Nelson \& Langer}{Nelson \&
  Langer}{1997}]{Nelson_1997}
Nelson R.~P.,  Langer W.~D.,  1997, \mn@doi [ApJ] {10.1086/304167}, 482, 796

\bibitem[\protect\citeauthoryear{{Neufeld} \& {Wolfire}}{{Neufeld} \&
  {Wolfire}}{2017}]{Neufeld2017}
{Neufeld} D.~A.,  {Wolfire} M.~G.,  2017, \mn@doi [\apj]
  {10.3847/1538-4357/aa6d68}, \href
  {https://ui.adsabs.harvard.edu/abs/2017ApJ...845..163N} {845, 163}

\bibitem[\protect\citeauthoryear{Omukai, Tsuribe, Schneider  et~al.}{Omukai
  et~al.}{2005}]{Omukai_2005}
Omukai K.,  Tsuribe T.,  Schneider R.,   et~al., 2005, \mn@doi [ApJ]
  {10.1086/429955}, 626, 627

\bibitem[\protect\citeauthoryear{Omukai, Hosokawa  \& Yoshida}{Omukai
  et~al.}{2010}]{Omukai_2010}
Omukai K.,  Hosokawa T.,   Yoshida N.,  2010, \mn@doi [ApJ]
  {10.1088/0004-637x/722/2/1793}, 722, 1793

\bibitem[\protect\citeauthoryear{Osterbrock}{Osterbrock}{1988}]{Osterbrock_1988}
Osterbrock D.~E.,  1988, \mn@doi [Publications of the Astronomical Society of
  the Pacific] {10.1086/132188}, 100, 412

\bibitem[\protect\citeauthoryear{{Padovani}, {Galli}  \&
  {Glassgold}}{{Padovani} et~al.}{2009}]{Padovani2009}
{Padovani} M.,  {Galli} D.,   {Glassgold} A.~E.,  2009, \mn@doi [\aap]
  {10.1051/0004-6361/200911794}, \href
  {https://ui.adsabs.harvard.edu/abs/2009A&A...501..619P} {501, 619}

\bibitem[\protect\citeauthoryear{{Padovani}, {Bialy}, {Galli}
  et~al.}{{Padovani} et~al.}{2022}]{Padovani2022}
{Padovani} M.,  {Bialy} S.,  {Galli} D.,   et~al., 2022, \mn@doi [\aap]
  {10.1051/0004-6361/202142560}, \href
  {https://ui.adsabs.harvard.edu/abs/2022A&A...658A.189P} {658, A189}

\bibitem[\protect\citeauthoryear{{Palla}, {Salpeter}  \& {Stahler}}{{Palla}
  et~al.}{1983}]{Palla1983}
{Palla} F.,  {Salpeter} E.~E.,   {Stahler} S.~W.,  1983, \mn@doi [\apj]
  {10.1086/161231}, \href
  {https://ui.adsabs.harvard.edu/abs/1983ApJ...271..632P} {271, 632}

\bibitem[\protect\citeauthoryear{{Papadopoulos}, {Thi}  \&
  {Viti}}{{Papadopoulos} et~al.}{2002}]{Papadopoulos2002}
{Papadopoulos} P.~P.,  {Thi} W.~F.,   {Viti} S.,  2002, \mn@doi [\apj]
  {10.1086/342872}, \href
  {https://ui.adsabs.harvard.edu/abs/2002ApJ...579..270P} {579, 270}

\bibitem[\protect\citeauthoryear{{Pelupessy}, {Papadopoulos}  \& {van der
  Werf}}{{Pelupessy} et~al.}{2006}]{Pelupessy2006}
{Pelupessy} F.~I.,  {Papadopoulos} P.~P.,   {van der Werf} P.,  2006, \mn@doi
  [\apj] {10.1086/504366}, \href
  {https://ui.adsabs.harvard.edu/abs/2006ApJ...645.1024P} {645, 1024}

\bibitem[\protect\citeauthoryear{Perez \& Granger}{Perez \&
  Granger}{2007}]{Perez2007}
Perez F.,  Granger B.~E.,  2007, \mn@doi [Computing in Science \& Engineering]
  {10.1109/MCSE.2007.53}, \href
  {https://ui.adsabs.harvard.edu/abs/2007CSE.....9c..21P/abstract} {9, 21}

\bibitem[\protect\citeauthoryear{{Peters}, {Naab}, {Walch}  et~al.}{{Peters}
  et~al.}{2017}]{silcc4}
{Peters} T.,  {Naab} T.,  {Walch} S.,   et~al., 2017, \mn@doi [\mnras]
  {10.1093/mnras/stw3216}, 466, 3293

\bibitem[\protect\citeauthoryear{{Pfrommer}, {Pakmor}  \& {et al.,
  }}{{Pfrommer} et~al.}{2017}]{Pfrommer2017}
{Pfrommer} C.,  {Pakmor} R.,   {et al., } 2017, \mn@doi [\mnras]
  {10.1093/mnras/stw2941}, \href
  {https://ui.adsabs.harvard.edu/abs/2017MNRAS.465.4500P} {465, 4500}

\bibitem[\protect\citeauthoryear{{Piatti}}{{Piatti}}{2023}]{piatti2023}
{Piatti} A.~E.,  2023, \mn@doi [\mnras] {10.1093/mnras/stad2786}, 526, 391

\bibitem[\protect\citeauthoryear{{Polzin}, {Kravtsov}, {Semenov}
  et~al.}{{Polzin} et~al.}{2024}]{Polzin2024}
{Polzin} A.,  {Kravtsov} A.~V.,  {Semenov} V.~A.,   et~al., 2024, \mn@doi
  [\apj] {10.3847/1538-4357/ad32cb}, \href
  {https://ui.adsabs.harvard.edu/abs/2024ApJ...966..172P} {966, 172}

\bibitem[\protect\citeauthoryear{{Ramachandran}, {Hamann}
  et~al.}{{Ramachandran} et~al.}{2019}]{Ramachandran2019}
{Ramachandran} V.,  {Hamann} W.~R.,   et~al., 2019, \mn@doi [\aap]
  {10.1051/0004-6361/201935365}, \href
  {https://ui.adsabs.harvard.edu/abs/2019A&A...625A.104R} {625, A104}

\bibitem[\protect\citeauthoryear{{Rathjen}, {Naab}  \& {et al., }}{{Rathjen}
  et~al.}{2021}]{silcc6}
{Rathjen} T.-E.,  {Naab} T.,   {et al., } 2021, \mn@doi [\mnras]
  {10.1093/mnras/stab900}, 504, 1039

\bibitem[\protect\citeauthoryear{{Rathjen}, {Naab}  \& {et al., }}{{Rathjen}
  et~al.}{2023}]{silcc7}
{Rathjen} T.-E.,  {Naab} T.,   {et al., } 2023, \mn@doi [\mnras]
  {10.1093/mnras/stad1104}, 522, 1843

\bibitem[\protect\citeauthoryear{{Rathjen}, {Walch}, {Naab}  et~al.}{{Rathjen}
  et~al.}{2025}]{silcc8}
{Rathjen} T.-E.,  {Walch} S.,  {Naab} T.,   et~al., 2025, \mn@doi [\mnras]
  {10.1093/mnras/staf792}, \href
  {https://ui.adsabs.harvard.edu/abs/2025MNRAS.540.1462R} {540, 1462}

\bibitem[\protect\citeauthoryear{{R\'emy-Ruyer}, {Madden, S. C.}, {Galliano,
  F.}  et~al.}{{R\'emy-Ruyer} et~al.}{2014}]{Remy2014}
{R\'emy-Ruyer} A.,  {Madden, S. C.} {Galliano, F.}  et~al., 2014, \mn@doi
  [A\&A] {10.1051/0004-6361/201322803}, 563, A31

\bibitem[\protect\citeauthoryear{Rubele, Kerber  \& Girardi}{Rubele
  et~al.}{2009}]{Rubele2009}
Rubele S.,  Kerber L.,   Girardi L.,  2009, MNRAS, 403, 1156

\bibitem[\protect\citeauthoryear{{Rubio}, {Lequeux}  \& {Boulanger}}{{Rubio}
  et~al.}{1993}]{Rubio1993}
{Rubio} M.,  {Lequeux} J.,   {Boulanger} F.,  1993, \aap, \href
  {https://ui.adsabs.harvard.edu/abs/1993A&A...271....9R} {271, 9}

\bibitem[\protect\citeauthoryear{{Sabbi} et~al.}{{Sabbi}
  et~al.}{2009}]{Sabbi2009}
{Sabbi} E.,  et~al., 2009, \mn@doi [arXiv e-prints] {10.48550/arXiv.0908.3500},
  \href {https://ui.adsabs.harvard.edu/abs/2009arXiv0908.3500S} {p.
  arXiv:0908.3500}

\bibitem[\protect\citeauthoryear{Salpeter}{Salpeter}{1955}]{salpeter1955}
Salpeter E.~E.,  1955, ApJ, 121, 161

\bibitem[\protect\citeauthoryear{{Sandstrom}, {Bolatto}, {Bot}
  et~al.}{{Sandstrom} et~al.}{2012}]{Sandstrom2012}
{Sandstrom} K.~M.,  {Bolatto} A.~D.,  {Bot} C.,   et~al., 2012, \mn@doi [\apj]
  {10.1088/0004-637X/744/1/20}, \href
  {https://ui.adsabs.harvard.edu/abs/2012ApJ...744...20S} {744, 20}

\bibitem[\protect\citeauthoryear{{Savaglio}, {Glazebrook}, {Le Borgne}
  et~al.}{{Savaglio} et~al.}{2005}]{Savaglio2005}
{Savaglio} S.,  {Glazebrook} K.,  {Le Borgne} D.,   et~al., 2005, \mn@doi
  [\apj] {10.1086/497331}, \href
  {https://ui.adsabs.harvard.edu/abs/2005ApJ...635..260S} {635, 260}

\bibitem[\protect\citeauthoryear{{Schruba}, {Leroy}, {Walter}
  et~al.}{{Schruba} et~al.}{2011}]{Schruba2011}
{Schruba} A.,  {Leroy} A.~K.,  {Walter} F.,   et~al., 2011, \mn@doi [\aj]
  {10.1088/0004-6256/142/2/37}, \href
  {https://ui.adsabs.harvard.edu/abs/2011AJ....142...37S} {142, 37}

\bibitem[\protect\citeauthoryear{{Searle}}{{Searle}}{1971}]{Searle1971}
{Searle} L.,  1971, \mn@doi [\apj] {10.1086/151090}, \href
  {https://ui.adsabs.harvard.edu/abs/1971ApJ...168..327S} {168, 327}

\bibitem[\protect\citeauthoryear{{Seifried}, {Walch}, {Girichidis}
  et~al.}{{Seifried} et~al.}{2017}]{Seifried2017}
{Seifried} D.,  {Walch} S.,  {Girichidis} P.,   et~al., 2017, \mn@doi [\mnras]
  {10.1093/mnras/stx2343}, \href
  {https://ui.adsabs.harvard.edu/abs/2017MNRAS.472.4797S} {472, 4797}

\bibitem[\protect\citeauthoryear{{Sembach}, {Howk}, {Ryans}  et~al.}{{Sembach}
  et~al.}{2000}]{sembach2000}
{Sembach} K.~R.,  {Howk} J.~C.,  {Ryans} R. S.~I.,   et~al., 2000, \mn@doi
  [\apj] {10.1086/308173}, 528, 310

\bibitem[\protect\citeauthoryear{{Sextl}, {Kudritzki}, {Burkert}
  et~al.}{{Sextl} et~al.}{2024}]{Sextl2024}
{Sextl} E.,  {Kudritzki} R.-P.,  {Burkert} A.,   et~al., 2024, \mn@doi [\apj]
  {10.3847/1538-4357/ad08b3}, \href
  {https://ui.adsabs.harvard.edu/abs/2024ApJ...960...83S} {960, 83}

\bibitem[\protect\citeauthoryear{{Sharda}, {Amarsi}  \& {et al., }}{{Sharda}
  et~al.}{2023}]{Sharda2023}
{Sharda} P.,  {Amarsi} A.~M.,   {et al., } 2023, \mn@doi [\mnras]
  {10.1093/mnras/stac3315}, 518, 3985

\bibitem[\protect\citeauthoryear{{Shaver}, {McGee}, {Newton}  et~al.}{{Shaver}
  et~al.}{1983}]{Shaver1983}
{Shaver} P.~A.,  {McGee} R.~X.,  {Newton} L.~M.,   et~al., 1983, \mn@doi
  [\mnras] {10.1093/mnras/204.1.53}, \href
  {https://ui.adsabs.harvard.edu/abs/1983MNRAS.204...53S} {204, 53}

\bibitem[\protect\citeauthoryear{{Smart}, {Haffner}, {Barger}  et~al.}{{Smart}
  et~al.}{2019}]{Smart2019}
{Smart} B.~M.,  {Haffner} L.~M.,  {Barger} K.~A.,   et~al., 2019, \mn@doi
  [\apj] {10.3847/1538-4357/ab4d58}, \href
  {https://ui.adsabs.harvard.edu/abs/2019ApJ...887...16S} {887, 16}

\bibitem[\protect\citeauthoryear{{Strong}, {Moskalenko}  \& {et al.,
  }}{{Strong} et~al.}{2007}]{Strong2007}
{Strong} A.~W.,  {Moskalenko} I.~V.,   {et al., } 2007, \mn@doi [Annual Review
  of Nuclear and Particle Science] {10.1146/annurev.nucl.57.090506.123011},
  \href {https://ui.adsabs.harvard.edu/abs/2007ARNPS..57..285S} {57, 285}

\bibitem[\protect\citeauthoryear{{Sz{\'e}csi}, {Agrawal}, {W{\"u}nsch}
  et~al.}{{Sz{\'e}csi} et~al.}{2022}]{szecsi2020}
{Sz{\'e}csi} D.,  {Agrawal} P.,  {W{\"u}nsch} R.,   et~al., 2022, \mn@doi
  [\aap] {10.1051/0004-6361/202141536}, 658, A125

\bibitem[\protect\citeauthoryear{{Tacchella}, {Johnson}  et~al.}{{Tacchella}
  et~al.}{2023}]{Tacchella2023}
{Tacchella} S.,  {Johnson} B.~D.,   et~al., 2023, \mn@doi [\mnras]
  {10.1093/mnras/stad1408}, 522, 6236

\bibitem[\protect\citeauthoryear{Tanaka, Tan, Zhang  et~al.}{Tanaka
  et~al.}{2018}]{Tanaka2018}
Tanaka K. E.~I.,  Tan J.~C.,  Zhang Y.,   et~al., 2018, ApJ, 861, 68

\bibitem[\protect\citeauthoryear{{Tielens}}{{Tielens}}{2005}]{Tielens2005}
{Tielens} A.~G.~G.~M.,  2005, {The Physics and Chemistry of the Interstellar
  Medium}

\bibitem[\protect\citeauthoryear{{Togi} \& {Smith}}{{Togi} \&
  {Smith}}{2016}]{Togi2016}
{Togi} A.,  {Smith} J.~D.~T.,  2016, \mn@doi [\apj]
  {10.3847/0004-637X/830/1/18}, \href
  {https://ui.adsabs.harvard.edu/abs/2016ApJ...830...18T} {830, 18}

\bibitem[\protect\citeauthoryear{{Tremonti}, {Heckman}, {Kauffmann}
  et~al.}{{Tremonti} et~al.}{2004}]{Tremonti2004}
{Tremonti} C.~A.,  {Heckman} T.~M.,  {Kauffmann} G.,   et~al., 2004, \mn@doi
  [\apj] {10.1086/423264}, 613, 898

\bibitem[\protect\citeauthoryear{{Turk}, {Smith}  \& {et al., }}{{Turk}
  et~al.}{2011}]{Turk2011}
{Turk} M.~J.,  {Smith} B.~D.,   {et al., } 2011, \mn@doi [\apjs]
  {10.1088/0067-0049/192/1/9}, \href
  {https://ui.adsabs.harvard.edu/abs/2011ApJS..192....9T} {192, 9}

\bibitem[\protect\citeauthoryear{Virtanen et~al.}{Virtanen
  et~al.}{2020}]{scipy}
Virtanen P.,  et~al., 2020, \mn@doi [Nature Methods]
  {10.1038/s41592-019-0686-2}, \href {https://rdcu.be/b08Wh} {17, 261}

\bibitem[\protect\citeauthoryear{Walch, Wünsch  \& {et al., }}{Walch
  et~al.}{2011}]{Walch_2011}
Walch S.,  Wünsch R.,   {et al., } 2011, \mn@doi [ApJ]
  {10.1088/0004-637x/733/1/47}, 733, 47

\bibitem[\protect\citeauthoryear{{Walch}, {Whitworth}  \& {et al., }}{{Walch}
  et~al.}{2012}]{walch2012}
{Walch} S.~K.,  {Whitworth} A.~P.,   {et al., } 2012, \mn@doi [\mnras]
  {10.1111/j.1365-2966.2012.21767.x}, 427, 625

\bibitem[\protect\citeauthoryear{{Walch}, {Girichidis}  \& {et al., }}{{Walch}
  et~al.}{2015}]{silcc1}
{Walch} S.,  {Girichidis} P.,   {et al., } 2015, \mn@doi [\mnras]
  {10.1093/mnras/stv1975}, 454, 238

\bibitem[\protect\citeauthoryear{{Walter}, {Cannon}, {Roussel}
  et~al.}{{Walter} et~al.}{2007}]{Walter2007}
{Walter} F.,  {Cannon} J.~M.,  {Roussel} H.,   et~al., 2007, \mn@doi [\apj]
  {10.1086/514807}, \href
  {https://ui.adsabs.harvard.edu/abs/2007ApJ...661..102W} {661, 102}

\bibitem[\protect\citeauthoryear{{Wei{\ss}}, {Grishunin}, {Czubkowski}
  et~al.}{{Wei{\ss}} et~al.}{2023}]{Weiss2023}
{Wei{\ss}} A.,  {Grishunin} K.,  {Czubkowski} N.,   et~al., 2023, in
  {Ossenkopf-Okada} V.,  {Schaaf} R.,  {Breloy} I.,   {Stutzki} J.,  eds,
  Physics and Chemistry of Star Formation: The Dynamical ISM Across Time and
  Spatial Scales. p.~72

\bibitem[\protect\citeauthoryear{{Welty}, {Xue}  \& {Wong}}{{Welty}
  et~al.}{2012}]{Welty2012}
{Welty} D.~E.,  {Xue} R.,   {Wong} T.,  2012, \mn@doi [\apj]
  {10.1088/0004-637X/745/2/173}, \href
  {https://ui.adsabs.harvard.edu/abs/2012ApJ...745..173W} {745, 173}

\bibitem[\protect\citeauthoryear{{Whitcomb}, {Smith}, {Sandstrom}
  et~al.}{{Whitcomb} et~al.}{2024}]{Whitcomb2024}
{Whitcomb} C.~M.,  {Smith} J. D.~T.,  {Sandstrom} K.,   et~al., 2024, \mn@doi
  [\apj] {10.3847/1538-4357/ad66c8}, \href
  {https://ui.adsabs.harvard.edu/abs/2024ApJ...974...20W} {974, 20}

\bibitem[\protect\citeauthoryear{{Whitworth}, {Smith}  \& {et al.,
  }}{{Whitworth} et~al.}{2022}]{Whitworth2022}
{Whitworth} D.~J.,  {Smith} R.~J.,   {et al., } 2022, \mn@doi [\mnras]
  {10.1093/mnras/stab3622}, 510, 4146

\bibitem[\protect\citeauthoryear{{Wolfire}, {Hollenbach}  \& {et al.,
  }}{{Wolfire} et~al.}{1995}]{Wolfire1995}
{Wolfire} M.~G.,  {Hollenbach} D.,   {et al., } 1995, \mn@doi [\apj]
  {10.1086/175510}, 443, 152

\bibitem[\protect\citeauthoryear{{Wong} \& {Blitz}}{{Wong} \&
  {Blitz}}{2002}]{Wong2002}
{Wong} T.,  {Blitz} L.,  2002, \mn@doi [\apj] {10.1086/339287}, \href
  {https://ui.adsabs.harvard.edu/abs/2002ApJ...569..157W} {569, 157}

\bibitem[\protect\citeauthoryear{{W{\"u}nsch}, {Walch}, {Dinnbier}
  et~al.}{{W{\"u}nsch} et~al.}{2018}]{wunsch2018}
{W{\"u}nsch} R.,  {Walch} S.,  {Dinnbier} F.,   et~al., 2018, \mn@doi [\mnras]
  {10.1093/mnras/sty015}, \href
  {https://ui.adsabs.harvard.edu/abs/2018MNRAS.475.3393W} {475, 3393}

\bibitem[\protect\citeauthoryear{{W{\"u}nsch}, {Walch}  \& {et al.,
  }}{{W{\"u}nsch} et~al.}{2021}]{wunsch2021}
{W{\"u}nsch} R.,  {Walch} S.,   {et al., } 2021, \mn@doi [\mnras]
  {10.1093/mnras/stab1482}, 505, 3730

\bibitem[\protect\citeauthoryear{{Wyse} \& {Silk}}{{Wyse} \&
  {Silk}}{1989}]{Wyse1989}
{Wyse} R. F.~G.,  {Silk} J.,  1989, \mn@doi [\apj] {10.1086/167329}, \href
  {https://ui.adsabs.harvard.edu/abs/1989ApJ...339..700W} {339, 700}

\bibitem[\protect\citeauthoryear{{Yoon}, {Langer}  \& {Norman}}{{Yoon}
  et~al.}{2006}]{yoon2006}
{Yoon} S.~C.,  {Langer} N.,   {Norman} C.,  2006, \mn@doi [\aap]
  {10.1051/0004-6361:20065912}, 460, 199

\bibitem[\protect\citeauthoryear{{Zaritsky}}{{Zaritsky}}{1992}]{Zaritsky1992}
{Zaritsky} D.,  1992, \mn@doi [\apjl] {10.1086/186375}, \href
  {https://ui.adsabs.harvard.edu/abs/1992ApJ...390L..73Z} {390, L73}

\bibitem[\protect\citeauthoryear{{Zwicky}}{{Zwicky}}{1966}]{Zwicky1966}
{Zwicky} F.,  1966, \mn@doi [\apj] {10.1086/148490}, 143, 192

\bibitem[\protect\citeauthoryear{{van der Marel}}{{van der
  Marel}}{2006}]{Marel2004}
{van der Marel} R.~P.,  2006, in {Livio} M.,  {Brown} T.~M.,  eds, ~ Vol. 17,
  The Local Group as an Astrophysical Laboratory. pp 47--71 (\mn@eprint {arXiv}
  {astro-ph/0404192}), \mn@doi{10.48550/arXiv.astro-ph/0404192}

\bibitem[\protect\citeauthoryear{van~der Walt, Colbert  \& Varoquaux}{van~der
  Walt et~al.}{2011}]{vanderWalt2011}
van~der Walt S.,  Colbert S.~C.,   Varoquaux G.,  2011, \mn@doi [Computing in
  Science \& Engineering] {10.1109/MCSE.2011.37}, \href
  {https://ui.adsabs.harvard.edu/abs/2011CSE....13b..22V/abstract} {13, 22}

\makeatother
\end{thebibliography}




\appendix

\section{Stellar feedback at different metallicities}
\label{sec:app_stellarmodels}
In the following paragraphs, we will express the metallicity of the stellar models in units of solar metallicity, assuming 1~Z$_\odot$~=~0.014. Regarding the BoOST models, we compute their metallicity in solar units taking the value of $Z$ reported in Table~1 from \citealt{szecsi2020} and dividing by 0.014.

\subsection{Massive star lifetime}

\begin{figure}
	\includegraphics[width=\columnwidth]{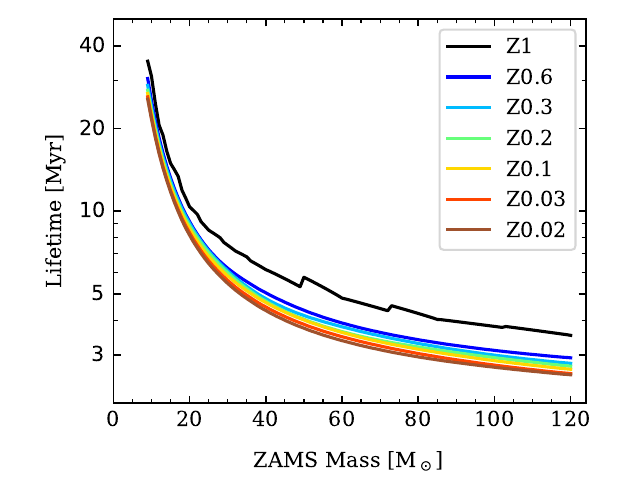}
    \caption{Lifetime of massive stars (in Myr) as a function of their initial mass (in M$_\odot$) for every considered
stellar grid. The Z1 line refers to the models at solar metallicity from \citealt{ekstrom2012}, whereas the others refer to the stellar models from \citealt{brott2011}, \citealt{szecsi2020} with respectively $Z$~=~0.64~Z$_\odot$ (Z0.6), 0.34~Z$_\odot$ (Z0.3), 0.15~Z$_\odot$ (Z0.2), 0.075~Z$_\odot$ (Z0.1), 0.03~Z$_\odot$ (Z0.03), 0.015~Z$_\odot$ (Z0.02). Here we note two facts: the lifetime of a star decreases as a function of its initial mass,
a well–known result from the theory of stellar evolution (see e.g. \citealt{Kippenhahnbook}). The second is that models with a lower
metallicity live for a shorter time.}
    \label{fig:lifetime_mass}
\end{figure}

In Fig.~\ref{fig:lifetime_mass}, we show how the lifetime of the stellar models depends on their initial (Zero Age Main Sequence, ZAMS) mass and metallicity. More massive stars have a briefer main sequence \citep{Kippenhahnbook}. A similar behaviour is observed for low–metallicity models, which are short-lived compared to metal-rich ones. Since metal-poor stars are more compact (\citealt{yoon2006}, \citealt{ekstrom2011}), and their convective core is larger and hotter, the region where nuclear burning is taking place is more expanded. Therefore, comparing two models with the same initial mass but different metallicity, the most metal–poor model will burn the same amount of fuel faster. However, the discrepancy seen between the Geneva and BoOST models is not only due to the different metallicities. In fact, BoOST models predict a larger convective core - due to a higher overshoot parameter - that brings more hydrogen from the outer layers into the burning regions. Hence, stars in the BoOST models exhaust their fuel sooner.

\subsection{Bolometric and wind luminosity}

\begin{figure}
	\includegraphics[width=\columnwidth]{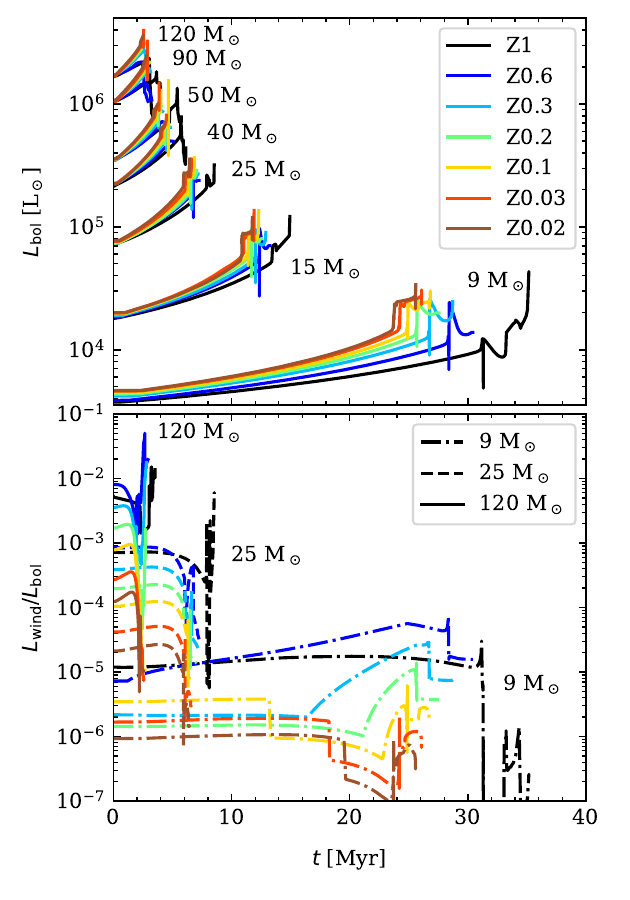}
    \caption{Evolution of the bolometric luminosity $L_\text{bol}$ in time (upper panel) and ratio of the wind luminosity $L_\text{wind}$ over $L_\text{bol}$ in time (bottom panel) for some representative models at all considered metallicities. In the bottom panel we use solid lines for models with an initial mass of 120~M$_\odot$, dashed lines for 25~M$_\odot$ and dash-dotted lines for 9~M$_\odot$.
    In the top panel we note that $L_\text{bol}$ depends both on the initial mass -- the larger, the brighter -- as well as on the metallicity, for which we have more luminous stars in low–metallicity conditions. In the bottom panel we observe that the ratio $L_\text{wind}$/$L_\text{bol}$ scales with the metallicity, as metal-poor stars have weaker winds and higher bolometric luminosities. Regarding the dependence $L_\text{wind}$/$L_\text{bol}$ with initial mass, this is dominated by the wind luminosity.}
    \label{fig:lbol_time}
\end{figure}

In the top panel of Fig.~\ref{fig:lbol_time} the time evolution of the bolometric luminosity is shown for some representative models at the metallicities considered. The luminosity has a positive correlation with the initial mass of the stars, as dictated by the mass-luminosity relation \citep{Kippenhahnbook} for which $L_\text{bol} \sim M^{3.37}$. We also note that the bolometric luminosity increases with decreasing metallicity. As mentioned above, in metal–poor stars, nuclear burning takes place in a more extended region. The direct consequence is that the luminosity due to nuclear burning is higher, as well as the total luminosity radiated by the surface. 

In the bottom panel of Fig.~\ref{fig:lbol_time}, the ratio $L_\text{wind}/L_\text{bol}$ of the wind luminosity to the bolometric luminosity is represented. The wind luminosity is the energy lost per unit time because of stellar winds, and is defined as
\begin{equation}
    L_\text{wind} = \frac{1}{2} \dot{M}_\text{wind} \times v^2_\text{wind}
    \label{eq:wind_lum}
\end{equation}
where $\dot{M}_\text{wind}$ is the mass loss rate due to the wind and $v_\text{wind}$ is the terminal velocity of the wind. As seen above, the bolometric luminosity increases with decreasing metallicity, and also the strength of winds is lower for metal-poor stars \citep[][]{Leitherer1992}. From this follows that the ratio $L_\text{wind}/L_\text{bol}$ scales with metallicity at fixed initial mass of the models. Moreover, this ratio has a positive correlation with initial mass, meaning that the wind luminosity dependence on the initial mass is stronger than the mass-luminosity relation.

\subsection{Ionizing radiation}

\begin{figure}
	\includegraphics[width=\columnwidth]{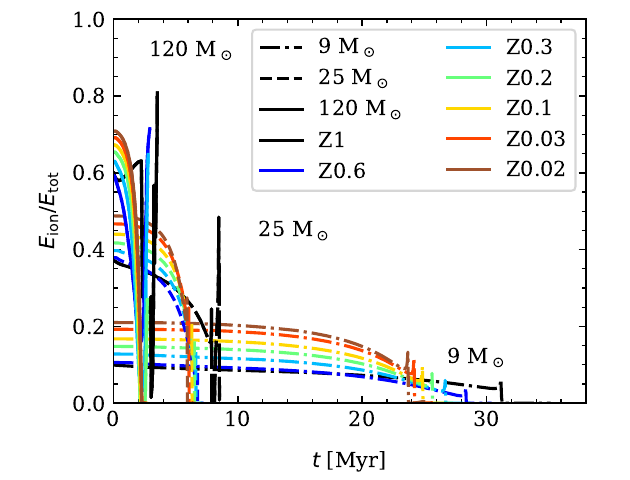}
    \caption{Ratio of the ionising radiation energy $E_\text{ion}$ over the total radiation energy $E_\text{tot}$ emitted by stars as a function of time. We use solid lines for the stellar models with initial mass of 120~M$_\odot$, dashed lines for 25~M$_\odot$ and dash-dotted lines for 9~M$_\odot$ models. We note that the fraction of ionising radiation increases with initial mass at the same metallicity, and scales with decreasing metallicity for fixed initial mass.}
    \label{fig:ion_en_frac}
\end{figure}

Another quantity that is of interest when studying the effects of stellar feedback is the amount of radiation emitted by stars that is able to ionise the hydrogen atoms of the surrounding ISM. We assume the stars to be black bodies, such that their spectrum can be described using Planck's law. We can rewrite the latter as a function of $E$~=~$h\nu$, where $h$ is the Planck constant and $\nu$ is the frequency of the emitted photons:
\begin{equation}
    B_E = \frac{2 E^3}{c^2 h^2} \frac{1}{\exp{\frac{E}{k_\mathrm{B} T}} - 1}
\end{equation}
where $c$ is the speed of light, k$_B$ the Boltzmann constant and $T$ the effective temperature of the star at a given time. Given that the ionisation energy for hydrogen is 13.6~eV, we can compute the fraction of ionising radiation as
\begin{equation}
    \frac{E_\mathrm{ion}}{E_\mathrm{tot}} = \frac{\int_{13.6}^{\infty} B_E dE}{\int_{0}^{\infty} B_E dE}
    \label{eq:ionisation_e_ratio}
\end{equation}
To be noted in this calculation is that the only parameter that actually depends on a given stellar model is the effective temperature at the surface of the star. Calculating the ratio from Eq.~\ref{eq:ionisation_e_ratio} for every timestep, we obtain its time evolution as shown in Fig.~\ref{fig:ion_en_frac}. We notice that low-metallicity models have a higher ionisation energy fraction compared to metal-rich models at the same initial mass and that the ionisation energy fraction scales with initial mass when fixing the metallicity. Both behaviours are a consequence of how the effective temperature of a star varies with metallicity and initial mass.    
In the first case, the higher effective temperature is due to lower opacities in metal-poor conditions \citep{Kippenhahnbook}. In the second case, the effective temperature of a star during the main sequence correlates with its mass \citep[again, see][]{Kippenhahnbook}{}{}. 

\section{Gas fragmentation as function of metallicity}
\label{appendix:dendro}
\begin{figure*}
	\includegraphics[width=0.99\textwidth]{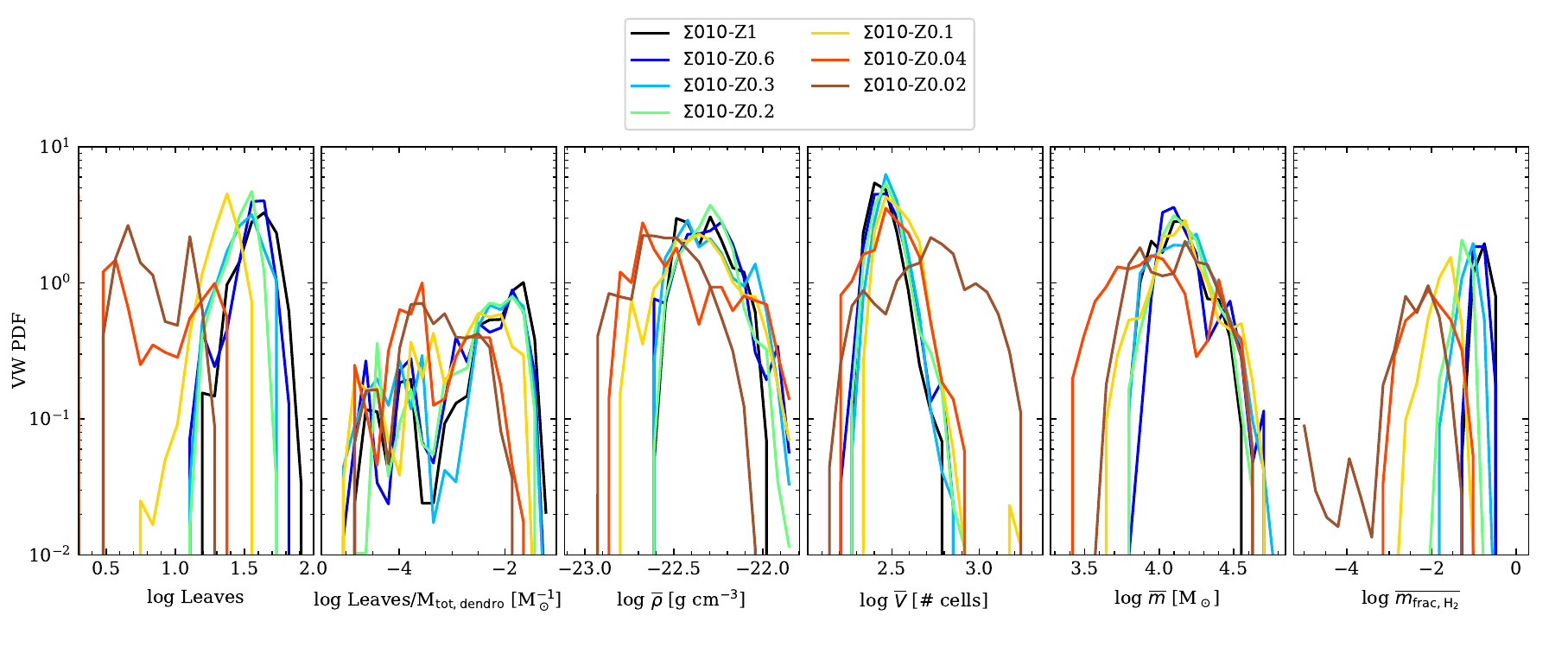}
    \caption{Same plot as Fig.~\ref{fig:dendro_2runs}, but showing the distributions computed for all runs.}
    \label{fig:dendro_allruns}
\end{figure*}

We show in Fig.~\ref{fig:dendro_allruns} the distributions for all runs that have already been discussed in Sec.~\ref{sec:gas_frag}. We note that the shown distributions behave similarly for metallicities larger than 0.1~Z$_\odot$, whereas they show a different behaviour for lower metallicities. 

\section{Gas scale height}
\begin{figure}
	\includegraphics[width=\columnwidth]{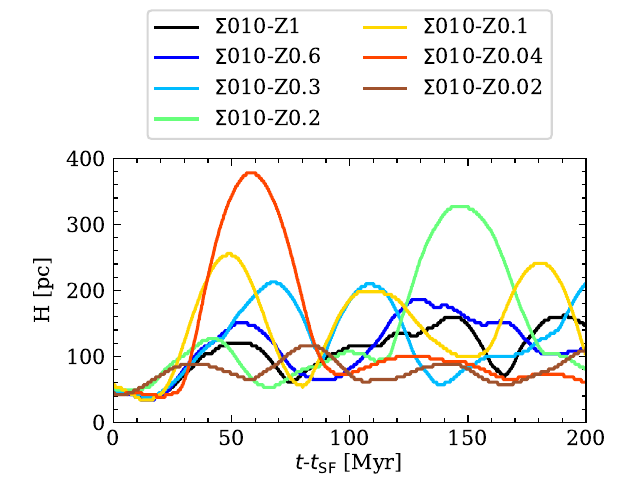}
    \caption{Scale height $H$ of the disc as a function of time, for every run. $H$ is defined as the height $\pm z$ around the midplane that encompasses 68\% of the total mass in the box. }
    \label{fig:scaleheight}
\end{figure}

In Fig.~\ref{fig:scaleheight} we show the evolution in time of the scale height of the disc, defined as the height $\pm z$, below and above the midplane, which contains 68\% of the total mass of the gas present in the box. We note that the peaks in the scale height of the disc correspond to the peaks in the star formation rate  surface density (see Fig.~\ref{fig:sfr}) for each run, but delayed by a few tens of Myr. This delay is due to the fact that the first SNe explode in a cold and dense environment; therefore, they have a lower efficiency. As more SNe go off, the gas near the midplane becomes warmer and stellar feedback is more effective in lifting the gas, therefore increasing the scale height of the disc. We notice that at low metallicity the disc becomes thicker because of the fact that a higher amount of warm gas is present (see Fig.~\ref{fig:vff}). The exception is the $\Sigma$010-Z0.02 run, where the scale height remains around 90 -- 100 pc due to the lack of a sufficient number of SNe. 

\bsp	
\label{lastpage}
\end{document}